\documentstyle[prd,epsbox,eqsecnum,aps]{revtex}

\begin{document}

\draft
\title{Preheating with non-minimally coupled scalar fields
         \\in higher-curvature inflation models }
\author{Shinji Tsujikawa$^1$\thanks{electronic
address:shinji@gravity.phys.waseda.ac.jp}, and
~Kei-ichi Maeda$^{1,2}$\thanks{electronic
address:maeda@gravity.phys.waseda.ac.jp}}
\address{$^1$ Department of Physics, Waseda University,
Shinjuku, Tokyo 169-8555, Japan\\[.3em]
$^2$ Advanced Research Institute for Science and Engineering,\\
Waseda University, Shinjuku, Tokyo 169-8555, Japan\\[.3em]}
\author{Takashi Torii \thanks{electronic
address:torii@th.phys.titech.ac.jp}
}
\address{Department of Physics, Tokyo
Institute of Technology, Meguro, Tokyo 152, Japan\\~}
\date{\today}
\maketitle
\begin{abstract}
In higher-curvature inflation
models ($R+\alpha_n R^n$), we study a parametric preheating of a
scalar field $\chi$ coupled non-minimally to a spacetime
curvature $R$ ($\xi R \chi^2$).
In the case of $R^2$-inflation model, efficient preheating becomes
possible for rather small values of $\xi$, i.e. $|\xi| \mbox{\raisebox{-1.ex}{$\stackrel
     {\textstyle<}{\textstyle \sim}$}}$
several.  
Although the maximal fluctuation $\sqrt{\langle \chi^2
\rangle}_{max} \approx 2 \times10^{17}$ GeV for
$\xi \approx -4$ is almost the same as the chaotic
 inflation model with a non-minimally coupled
$\chi$ field, the growth rate of the fluctuation becomes
much larger and efficient preheating is realized.
We also investigate preheating
for $R^4$ model  and find that the maximal fluctuation is
$\sqrt{\langle \chi^2 \rangle}_{max} \approx 8 \times
10^{16}$ GeV for $\xi \approx -35$.
\end{abstract}


 \baselineskip = 24pt

%%%%%%%%%%%%%%%%%%%%%%%%%%%%%%%%%%%%%%%%%%%%%%%%%%
%
%                                                 %
\section{Introduction}                            %
%                                                 %
%%%%%%%%%%%%%%%%%%%%%%%%%%%%%%%%%%%%%%%%%%%%%%%%%%
The inflation is one of the most promising models for the early stage of
the Universe in modern cosmology \cite{GS}.
It not only gives a natural explanation for
the horizon, flatness, and monopole problem but also provides us
density perturbations as seeds for a large scale structure
in the Universe\cite{Kolb}. So far, there are several
models for inflation, i.e. a new inflation proposed by Albrecht and
Steinhardt\cite{AS}, and Linde\cite{Linde1}, a chaotic inflation by
Linde\cite{Linde2}, and a higher-curvature inflation by
Starobinsky\cite{Starobinsky}. The first two models require a scalar
field which is called {\it inflaton} to drive inflation. The chaotic
inflation scenario would be natural in the sense  that   initial
conditions for inflation are not restricted so much as
compared with a new inflation. On the other hand, the
third model is based on adding the higher-curvature terms to the
Einstein-Hilbert action. This model is remarkable in the sense that
we have an inflationary solution without an
inflaton scalar field. That is, inflation can be realized by purely
gravitational coupling. Several authors\cite{higher_curvature}
studied some  constraints of
coupling with higher-curvature and investigated its various aspects
such as the  cosmic no-hair theorem.

An inflationary period will end when  kinetic energy of an
inflaton  dominates over its potential energy.
At this stage, the inflaton field begins to oscillate coherently around
the minimum of its potential.
The energy of the inflaton field will eventually be transferred to
radiation, and this  process is called reheating.
The original version of the reheating scenario was considered in Ref.
\cite{DA} just after the  new inflation model is proposed.
This old scenario   is based on perturbation theory, adding
a phenomenological decay term to the inflaton equation.
Since the reheating temperature is constrained to be relatively small such
as $T_{r}~\mbox{\raisebox{-1.ex}{$\stackrel
     {\textstyle<}{\textstyle \sim}$}}~10^9$ GeV and a production of GUT scale gauge
boson is kinematically impossible, the GUT scale baryogenesis does
not work well in this scenario.

Recently, it has been pointed out that an inflaton decay will begin
in a much more explosive process, called {\it preheating}
\cite{TB,KLS1,Boy,Kaiser,Son,Yoshi} before the perturbative
decay.
At this stage, the fluctuation of scalar particles can grow
quasi-exponentially by parametric resonance.
The importance of this non-perturbative stage was recognized
in ref.~\cite{TB} in the context of the new inflation scenario.
As for the chaotic inflation scenario, Kofman, Linde, and Starobinsky
first investigated the preheating structure with both massive
inflaton and a self-interacting potential\cite{KLS1}.
In the case with a self-interacting potential $\lambda\phi^4/4$
of a massive inflaton, the inflaton particles are produced
by parametric  resonance due to the self coupling of the inflaton
field. In this model, the evolution of inflaton quanta was
investigated
\cite{Boy,Kaiser} by
using a closed time path formalism\cite{CTP,Jor,Cal} and also studied
by fully non-linear lattice simulations\cite{KT1}.
If we adopt the value of $\lambda \sim 10^{-12}$ which is
constrained  by density perturbations, it was found that the
fluctuation
$\langle\delta\phi^2\rangle \sim 10^{-7} M_{\rm PL}^2$
will be produced in the preheating stage.
In the case of massive inflaton without a self-coupling, however,
we do not expect a parametric resonance.  Hence one may need
another scalar field
$\chi$ coupled to the inflaton field, where coupling makes
a resonant production of the $\chi$-particle possible. 
Especially when the coupling constant $g$ gets
large, a parametric resonance  will turn on in the broad resonance
regime. Consequently, although the expansion of the Universe reduces
the amplitude of inflaton field adiabatically, the
fluctuation of the $\chi$ field  increases quasi-exponentially.
As for the $\chi$-particle production,
several numerical works have been done by making use of
Hartree mean field approximations\cite{KT2}
and fully non-linear calculations\cite{KT3,PR}.
Analytic investigations including the backreaction and rescattering
effects are performed in ref.\cite{KLS2,Green}.
Although the structure of resonance shows a  complicated feature
which is  called $stochastic$, the analytic estimations based on the
Mathieu equation are found to be reliable to analyze the preheating
dynamics.

In the higher-curvature inflation model, a reheating process  in the
old scenario was  considered by Suen and Anderson\cite{SA}.
In the case of the $R^2$ inflation model, an effective scalar field
that appears by conformal transformation can be approximately written as
a massive scalar field in the reheating phase.
Therefore, the preheating stage is absent, since the inflaton
fluctuation cannot be amplified by a parametric resonance.
In the massive inflaton model, however, a new scenario  was recently
proposed, in which  the non-minimally coupled $\chi$ field   is
enhanced by a parametric resonance\cite{BL}. Especially when the
coupling $\xi$ is negative, the fluctuation of
the $\chi$-particle grows rapidly due to negative coupling instability.
The final fluctuation  $\langle\chi^2\rangle \sim 3 \times
10^{-4} M_{\rm PL}^2$ in the case of $\xi \approx -4$
is larger than the minimally coupled massive inflaton 
case with $g \approx 1 \times 10^{-3}$
\cite{TMT}.
However, in this model, rather large values of $|\xi|$
are required to have a comparable growth rate with previous 
existing preheating models.
Then a natural question arises.
When one considers the non-minimally coupled $\chi$ field in the
context of the higher-curvature  inflation model, how is the $\chi$
fluctuation amplified? If it turns out to be larger as compared with
the previous  cases, it would affect the baryogenesis in GUT scale
\cite{baryogenesis}, a non-thermal phase transition\cite{phasetransition}
and topological defect formation
\cite{defect}.
We show in this paper that in the higher-curvature inflation model,
especially in the $R^2$ inflation model, significant particle
production is expected only  due to  a non-minimal coupling.
As compared with the previous model of chaotic inflation with the
non-minimally coupled scalar field $\chi$,
efficient preheating becomes possible
even for rather small  $|\xi|$.

This paper is organized as follows.
In the next section, we briefly review our model of $R^n$ inflation
with the non-minimally coupled scalar field $\chi$.
In Sec.~III, the preheating in the
$R^2$ inflation model is investigated both by analytic approach
and numerical integration. We show that analytical estimation based
on  the Mathieu equation coincides well with numerical results.
In Sec.~IV, the $R^4$ inflation case is briefly discussed.
We give our conclusions and discussions in the final section.

%%%%%%%%%%%%%%%%%%%%%%%%%%%%%%%%%%%%%%%%%%%%%%%%%%
%%%%%%%%%%
%                                                          %
\section{Basic equations}    %
%                                                          %
%%%%%%%%%%%%%%%%%%%%%%%%%%%%%%%%%%%%%%%%%%%%%%%%%%
%%%%%%%%%%

%%%%%%%%%%%%%
We propose a model with a  higher-curvature term and a
non-minimally coupled scalar field $\chi$,
%%%%%%%%%%%%%
\begin{eqnarray}
{\cal L} = \sqrt{-g} \left[ \frac{1}{2\kappa^2}R
   +\alpha_n R^n
   -\frac12 \xi R \chi^2
   -\frac{1}{2}(\nabla \chi)^2
   -\frac12 m_{\chi}^2 \chi^2
    \right] ,
\label{B1}
\end{eqnarray}
%%%%%%%%%%%%%
where $\kappa^{2}/8\pi \equiv G =M_{\rm PL}^{-2} $ is Newton's
gravitational constant, $\alpha_n$ and $\xi$ are
coupling constants, and
$m_{\chi}$ is a mass of the $\chi$ field.
Although this system contains only one scalar field $\chi$,
it is reduced to a system described
by Einstein-Hilbert action with two scalar fields
by making  conformal transformation to the
Einstein frame \cite{conformal}.
We make the conformal transformation as follows:
%%%%%%%%%%%%%%
\begin{eqnarray}
\hat{g}_{\mu \nu}=e^{2\omega} g_{\mu \nu},
\label{B2}
\end{eqnarray}
%%%%%%%%%%%%%%
where
%%%%%%%%%%%%%%
\begin{eqnarray}
\omega=\frac{\kappa}{\sqrt{6}}\phi
\equiv \frac12 {\rm ln}
\left[ 1+2\kappa^2 n \alpha_n R^{n-1} -\xi \kappa^2 \chi^2 \right].
\label{B3}
\end{eqnarray}
%%%%%%%%%%%%%%
Then we obtain the following equivalent Lagrangian:
%%%%%%%%%%%%%%%
\begin{eqnarray}
{\cal L} = \sqrt{-\hat{g}} \left[ \frac{1}{2\kappa^2} \hat{R}
   -\frac{1}{2}(\hat{\nabla} \phi)^2
   -\frac12 e^{-\frac{\sqrt{6}}{3} \kappa \phi}
    (\hat{\nabla} \chi)^2
 -U_n(\phi,\chi) \right],
\label{B4}
\end{eqnarray}
%%%%%%%%%%%%%%%
%%%%%%%%%%%%%%%
where the potential $U_n(\phi,\chi)$ is defined by
%%%%%%%%%%%%%%%
\begin{eqnarray}
U_n(\phi,\chi)
\equiv e^{-\frac{2\sqrt{6}}{3} \kappa \phi }
\left[
\frac{1}{2 \kappa^2} R e^{\frac{\sqrt{6}}{3} \kappa \phi}
- \frac{1}{2\kappa^2} R- \alpha_n R^n
+\frac12 \xi R \chi^2+\frac12 m_{\chi}^2 \chi^2 \right].
\label{B5}
\end{eqnarray}
%%%%%%%%%%%%%%%
Eliminating the Ricci scalar term by using
Eq.~$(\ref{B3})$, $U_n(\phi,\chi)$ can be rewritten as
%%%%%%%%%%%%%%%
\begin{eqnarray}
U_n(\phi,\chi)
=e^{-\frac{2\sqrt{6}}{3} \kappa \phi } \left[ \lambda_n
\left( e^{\frac{\sqrt{6}}{3} \kappa \phi}
-1 +\xi\kappa^2\chi^2 \right)
^{\frac{n}{n-1}}+\frac12
m_{\chi}^2\chi^2 \right],
\label{B6}
\end{eqnarray}
%%%%%%%%%%%%%%%
where
%%%%%%%%%%%%%%%
\begin{eqnarray}
\lambda_n \equiv (n-1) \left[ \frac{1}{\alpha_n (2n\kappa^2)^n}
\right]^{\frac{1}{n-1}}.
\label{B30}
\end{eqnarray}
%%%%%%%%%%%%%%%
The scalar field $\phi$ plays the roll of an inflaton field
and the coupling constant $\alpha_n$
(or $\lambda_n$) is constrained by (i) e-folding
by the inflation and (ii) the density perturbation
produced in the inflationary stage as follows.

Since we need not take the $\chi$ field into account
in an inflationary stage,
the potential $U_n$ can be approximated as,
%%%%%%%%%%%%%%%
\begin{eqnarray}
V_n(\phi) \equiv U_n(\phi,0) =
\lambda_n e^{-\frac{2\sqrt{6}}{3} \kappa \phi }
\left( e^{\frac{\sqrt{6}}{3} \kappa \phi} -1 \right)
^{\frac{n}{n-1}}.
\label{potential}
\end{eqnarray}
%%%%%%%%%%%%%%%
The form of the potential is quite different between the
$n=2$ case and others. Since the potential has a plateau
for $\phi~\mbox{\raisebox{-1.ex}{$\stackrel
     {\textstyle>}{\textstyle\sim}$}}~M_{\rm PL}$ in the $n=2$ case
(See Fig.~1), the inflaton rolls down very slowly on
this plateau and can give the efficient inflation
without making  fine tuning on the initial condition
of the inflaton field. This is one of the important properties
of the $R^2$ inflation model.
In the case of $n>2$, however, $V_n(\phi)$ has a local maximum at
%%%%%%%%%%%%%%%
\begin{eqnarray}
\phi_{*}
=\sqrt{\frac{3}{16\pi}} {\rm log} \left[\frac{2(n-1)}{n-2}\right]
 M_{\rm PL},
\label{Lmaximum}
\end{eqnarray}
%%%%%%%%%%%%%%%
and inflation becomes  difficult to realize without
fine tuning of the initial value because the inflaton
easily falls down to the minimum of the potential.
Let us investigate the initial and final value of $\phi$ in the
inflationary stage.
First, the slow-roll parameters are described by the potential as
%%%%%%%%%%%%%%%
\begin{eqnarray}
\epsilon_1
\equiv \frac{1}{2\kappa^2}
\left(\frac{V_n'}{V_n}\right)^2
=\frac13 \left(
1-e^{\frac{\sqrt{6}}{3} \kappa \phi}  \right)^{-2}
\left(2-\frac{n-2}{n-1} e^{\frac{\sqrt{6}}{3} \kappa \phi}
\right)^2,
\label{slowroll1}
\end{eqnarray}
%%%%%%%%%%%%%%%
\begin{eqnarray}
\epsilon_2
\equiv \frac{1}{\kappa^2} \frac{V_n''}{V_n}
=\frac23  \left(
1-e^{\frac{\sqrt{6}}{3} \kappa \phi}  \right)^{-2}
\left[4-\frac{5n-8}{n-1} e^{\frac{\sqrt{6}}{3} \kappa \phi}
+\left(\frac{n-2}{n-1} \right)^2
 e^{\frac{2\sqrt{6}}{3} \kappa \phi}  \right].
\label{slowroll2}
\end{eqnarray}
%%%%%%%%%%%%%%%
The inflationary stage ends when
$\max$\{$\epsilon_1$, $|\epsilon_2|$\} becomes
of order unity.
One can easily find that $\epsilon_1$ increases to unity faster than
$|\epsilon_2|$, and the epoch when inflation ends can be estimated as
%%%%%%%%%%%%%%%
\begin{eqnarray}
\phi_f
=\sqrt{\frac{3}{16\pi}} {\rm log}
\left[ \frac{n-1}{2n^2-2n-1} \left\{
\left(1+\sqrt{3}\right)n+1 \right\} \right]  M_{\rm PL}.
\label{phiend}
\end{eqnarray}
%%%%%%%%%%%%%%%
For example, $\phi_f=0.188 M_{\rm PL} \approx M_{\rm PL}/5$
for $n=2$, and $\phi_f=0.108 M_{\rm PL}
\approx M_{\rm PL}/9$ for $n=4$.
Next, let us consider the value of $\phi$~($=\phi_{i}$)
at the epoch of horizon exit when
physical scales crossed outside the Hubble radius 60 e-folding
before the end of inflation.
By calculating e-folding number
%%%%%%%%%%%%%%%
\begin{eqnarray}
N=-\kappa^2 \int_{\phi_i}^{\phi_e}
\frac{V_n}{V_n'} d\phi,
\label{efolding}
\end{eqnarray}
%%%%%%%%%%%%%%%
and setting $N=60$, we obtain the value of $\phi_i$ as
%%%%%%%%%%%%%%%
\begin{eqnarray}
\phi_i =\cases{
             1.08 M_{\rm PL}, & ($n=2$) \cr & \cr
              \sqrt{\frac{3}{16\pi}} {\rm log}
              \left[ \frac{2(n-1)}{n-2}-
              \left\{ \frac{2(n-1)}{n-2}-
              e^{\frac{\sqrt{6}}{3} \kappa \phi_f}
\right\} 
              \left\{ \frac{2(n-1)}{n-2}
              e^{-\frac{\sqrt{6}}{3} \kappa \phi_f}
              \right\}^{-\frac{n-2}{n}}
              e^{-\frac{80(n-2)}{n}} \right] M_{\rm PL}.   &
              ($n\geq 4$) \cr
           } 
\label{phii}
\end{eqnarray}
%%%%%%%%%%%%%%%
Note that $\phi_i$ must be extremely close to  the
local maximum value
$\phi_{\ast}$ for $n>2$.

Next, we calculate the density perturbation
produced in the inflationary stage
and give the constraint of $\alpha_n$ by
comparing it with COBE data.
The amplitude of density perturbation can be calculated as
\cite{Lyth}
%%%%%%%%%%%%%%%
\begin{eqnarray}
\delta_H &=& \sqrt{\frac{32}{75} \frac{V_n(\phi_{i})}{M_{\rm PL}^4}
\epsilon_1^{-1} (\phi_i)} \nonumber \\
&=& \frac{2\sqrt{2 \lambda_n}}{5 M_{\rm PL}^2}
 e^{-\frac{\sqrt{6}}{3} \kappa \phi_i}
\left(e^{\frac{\sqrt{6}}{3} \kappa \phi_i}-1\right)
^{\frac{3n-2}{2(n-1)}}
\left(1-\frac{n-2}{2(n-1)}
e^{\frac{\sqrt{6}}{3} \kappa \phi_i} \right)^{-1}.
\label{B33}
\end{eqnarray}
%%%%%%%%%%%%%%%
The COBE data requires $\delta_H~\mbox{\raisebox{-1.ex}{$\stackrel
     {\textstyle<}{\textstyle \sim}$}}~10^{-5}$, hence
$\alpha_n$ is constrained to be
%%%%%%%%%%%%%%%
\begin{eqnarray}
         \cases{
           \alpha_2~\mbox{\raisebox{-1.ex}{$\stackrel
     {\textstyle>}{\textstyle\sim}$}}~10^9, & ($n=2$) \cr
           \alpha_n = \gamma_n \delta_{H}^{-2(n-1)}~
           \mbox{\raisebox{-1.ex}{$\stackrel
     {\textstyle>}{\textstyle\sim}$}}~ 10^{10(n-1)}\gamma_n, 
     & ($n\geq 4$) \cr
           }
\label{constraint}
\end{eqnarray}
%%%%%%%%%%%%%%%
where
%%%%%%%%%%%%%%%
\begin{eqnarray}
\gamma_n=\frac{M_{\rm PL}^{2(2-n)}}{(16\pi n)^n}
\left\{ \frac{8(n-1)}{25
\left(e^{\frac{\sqrt{6}}{3} \kappa \phi_i}
- e^{\frac{\sqrt{6}}{3} \kappa \phi_*} \right)^2} \right\}^{n-1}
\left(\frac{n}{n-2}\right)^{3n-2}.
\label{gamma}
\end{eqnarray}
%%%%%%%%%%%%%%%
Note that $\alpha_n$ needs to take a very large value especially
in the $n\geq 4$ case.
This is because the inflation field $\phi$ must be very near to
the local
maximum of its potential initially in order to cause the sufficient
inflation.
For example, in the case of $n=4$,
the dimensionless coupling constant
$\beta_4 \equiv \alpha_4 M_{\rm PL}^4$ is restricted to be
%%%%%%%%%%%%%%%
\begin{eqnarray}
\beta_4~\mbox{\raisebox{-1.ex}{$\stackrel
     {\textstyle>}{\textstyle\sim}$}}~10^{126}.
\label{B36}
\end{eqnarray}
%%%%%%%%%%%%%%%
As the increase of $n$, we have to choose larger
values of the coupling constant to fit the COBE data.
However, since we are interested in preheating after inflation,
it is worth investigating how $\chi$-particles are amplified
apart from the constraint of the coupling constant even for the
$n\geq 4$ case.

After the inflationary stage ended, the universe enters the oscillating phase of the $\phi$ field
and  $\chi$-particles can be created because the $\chi$ field is
non-minimally coupled with the spacetime curvature
which is oscillated by the $\phi$ field.
Let us obtain the basic equations of the $R^n$ inflation model
in the preheating stage.
Since the $\chi$ field is treated as a quantum field
on the classical background of spacetime and the $\phi$ field,
the conformal factor in Eq.~$(\ref{B2})$ includes a quantum
variable.
However, in order not to discuss quantum gravity,
this term should be replaced with an expectation value
$\langle \chi^2 \rangle$,
which may correspond to the number density of the
$\chi$-particle.
Then we regard the conformal factor $\Omega^2$ as
$1-\eta$, where $\eta \equiv \xi \kappa^2 \langle\chi^2\rangle$
\cite{comment}.

Since we assume that the spacetime and the inflaton field are
spatially homogeneous, we adopt the flat
Friedmann-Robertson-Walker metric as the background spacetime;
%%%%%%%%%%%%%%%
\begin{eqnarray}
d\hat{s}^2 = -dt^2 + \hat{a}^2(t) d {\bf x^2}.
\label{B12}
\end{eqnarray}
%%%%%%%%%%%%%%%
Hereafter we drop a caret since we argue only in the Einstein frame.
From the Lagrangian $(\ref{B4})$, the evolution of the scale factor
is described by
%%%%%%%%%%%%%%%
\begin{eqnarray}
\left(\frac{\dot{a}}{a}\right)^2 =
   \frac{\kappa^2}{3}
     \left[ \frac{1}{2} \dot{\phi}^2
   +\frac{1}{2}
   e^{-\frac{\sqrt{6}}{3} \kappa \phi}
   \langle (\nabla \chi)^2 \rangle
  e^{-\frac{2\sqrt{6}}{3} \kappa \phi }
  \left\{ \lambda_n \left( e^{\frac{\sqrt{6}}{3} \kappa \phi}
   -1 +\eta
  \right)^{\frac{n}{n-1}}
  +\frac12 m_{\chi}^2 \langle\chi^2\rangle \right\}
 \right], 
\label{B13}
\end{eqnarray}
%%%%%%%%%%%%%%%
where a dot denotes a derivative with respect to 
time coordinate $t$.

The evolution of the homogeneous inflaton field $\phi$ yields
%%%%%%%%%%%%%%%
\begin{eqnarray}
\ddot{\phi} &+& \frac{3\dot{a}}{a} \dot{\phi}
-\frac{2\sqrt{6}}{3}\kappa
e^{-\frac{2\sqrt{6}}{3} \kappa \phi}  
 \left[\frac14 e^ {\frac{\sqrt{6}}{3} \kappa \phi}
\langle (\nabla \chi)^2 \rangle \right. \nonumber \\
 &+& \left. 
 \lambda_n
\left( e^{\frac{\sqrt{6}}{3} \kappa \phi}-1+\eta \right)
^{\frac{1}{n-1}}
\left\{ \frac{n-2}{2(n-1)} e^{\frac{\sqrt{6}}{3} \kappa \phi}
-1+\eta \right\}
+\frac12 m_{\chi}^2
\langle\chi^2 \rangle \right]  =0. 
\label{B14}
\end{eqnarray}
%%%%%%%%%%%%%%%
Since we postulate that the $\phi$ field is  spatially homogeneous,
the fluctuation $\delta \phi$ is not considered.
In the present model, the $\phi$ field is a product of purely
gravitational origin.
Hence  considering the fluctuation $\delta \phi$ is equivalent to
taking into account the perturbation of the metric.
Recently, Bassett et al. \cite{mperturbation4} investigated
the evolution of metric perturbation in a two-field model
of a massive inflaton and a massless $\chi$ field
interacting with the inflaton, and found that metric fluctuation is
resonantly amplified in preheating phase.
There will be a possibility that  $\delta \phi$ grows
by a parametric resonance also in the present model.
Moreover, the rescattering effect of the $\chi$-particle and the
 $\delta \phi$ field are expected to make the $\chi$ field be
amplified.
Although we do not consider the
$\phi$ fluctuation in this paper,
we should take into account
this effect for a complete study of preheating.
It is under consideration.

The equation of the $\chi$  field is expressed as
%%%%%%%%%%%%%%%
\begin{eqnarray}
\ddot{\chi}
   + \left( \frac{3\dot{a}}{a}-
   \frac{\sqrt{6}}{3} \kappa \dot{\phi}
   \right) \dot{\chi}-\partial_i\partial^i \chi
   + e^{-\frac{\sqrt{6}}{3} \kappa \phi}
   \frac{\partial}{\partial \chi}
   \left[ \lambda_n
   \left( e^{\frac{\sqrt{6}}{3} \kappa \phi}-1 +
   \xi\kappa^2\chi^2 \right)^{\frac{n}{n-1}}
    +\frac12 m_{\chi}^2 \chi^2 \right] =0, 
\label{B18}
\end{eqnarray}
%%%%%%%%%%%%%%%
where an index with a roman character denotes space coordinates.
In order to study a quantum particle creation of the $\chi$ field,
we make the following
mean field approximation with respect to $\chi$, which provides us a
linearized equation for the quantum field $\chi$:
%%%%%%%%%%%%%%%
\begin{eqnarray}
\ddot{\chi}
   + \left(  \frac{3\dot{a}}{a}- \frac{\sqrt{6}}{3} \kappa
   \dot{\phi}
   \right) \dot{\chi}-\partial_i\partial^i \chi 
  + e^{-\frac{\sqrt{6}}{3} \kappa \phi}~\nabla_{\chi}^2
   \left[ \lambda_n
   \left(e^{\frac{\sqrt{6}}{3} \kappa \phi}-1 +\eta \right)^
    {\frac{n}{n-1}}
    +\frac12 m_{\chi}^2 \langle\chi^2\rangle \right] \chi=0.
\label{B19}
\end{eqnarray}
%%%%%%%%%%%%%%%
where $\nabla_{\chi}=2\sqrt{\langle \chi^2\rangle} ~\partial / \partial
\langle \chi^2\rangle$.
Expanding the  scalar field $\chi$ as
%%%%%%%%%%%%%%%
\begin{eqnarray}
\chi=\frac{1}{(2\pi)^{3/2}} \int \left(a_k \chi_k(t)
 e^{-i {\bf k} \cdot {\bf x}}+a_k^{\dagger} \chi_k^{*}(t)
 e^{i {\bf k} \cdot {\bf x}} \right) d^3{\bf k},
\label{B20}
\end{eqnarray}
%%%%%%%%%%%%%%%
where $a_k$ and $a_k^{\dagger}$ are the annihilation and
creation operators, respectively,
$\chi_k$ obeys the following equation of motion:
%%%%%%%%%%%%%%%
\begin{eqnarray}
\ddot{\chi}_k
   &+& \left( \frac{3\dot{a}}{a}- \frac{\sqrt{6}}{3}
  \kappa \dot{\phi}
   \right) \dot{\chi}_k 
  + \left[ \frac{k^2}{a^2} +e^{-\frac{\sqrt{6}}{3} \kappa \phi}
  \right. \nonumber \\
  &\times& \left.
    \left\{ \frac{2n}{n-1} \lambda_n \xi \kappa^2
   \left(e^{\frac{\sqrt{6}}{3} \kappa \phi}-1+\eta \right)
   ^{-\frac{n-2}{n-1}}
   \left( e^{\frac{\sqrt{6}}{3} \kappa \phi}-1+\frac{n+1}{n-1}
   \eta \right)
  +m_{\chi}^2 \right\} \right] \chi_k=0. 
\label{B21}
\end{eqnarray}
%%%%%%%%%%%%%%%
The expectation value of $\chi^2$ is expressed with $\chi_k$ as
%%%%%%%%%%%%%%%
\begin{eqnarray}
\langle \chi^2 \rangle = \frac1{2\pi^2} \int k^2|\chi_k|^2 dk.
\label{B22}
\end{eqnarray}
%%%%%%%%%%%%%%%

Let us obtain the approximate equations of the $\phi$ and $\chi$
fields in order to estimate the preheating phenomenon analytically.
First we rewrite Eqs.~$(\ref{B13})$ and $(\ref{B14})$
by  following two approximations:

%\vspace{0.2cm}
\begin{enumerate}
\item{The value of $\phi$ is much smaller than $M_{\rm PL}$,
      namely
%%%%%%%%%%%%%%%
\begin{eqnarray}
\phi \ll M_{\rm PL}.
\label{approx1}
\end{eqnarray}
%%%%%%%%%%%%%%%
}
%\vspace{0.2cm}
\item{The backreaction effect by the $\chi$-particles is negligible,
     namely
%%%%%%%%%%%%%%%
\begin{eqnarray}
\eta \ll 1~~~{\rm and}~~~\langle (\nabla \chi)^2 \rangle
\ll M_{\rm PL}^4.
\label{approx2}
\end{eqnarray}
%%%%%%%%%%%%%%%
}
\end{enumerate}
%\vspace{0.2cm}
Then, the evolution equation of the scale factor $(\ref{B13})$
is written by
%%%%%%%%%%%%%%%
\begin{eqnarray}
\left(\frac{\dot{a}}{a}\right)^2
= \frac{\kappa^2}{3}
\left[\frac12 \dot{\phi}^2+ K_n \phi^{\frac{n}{n-1}} \right],
\label{scale}
\end{eqnarray}
%%%%%%%%%%%%%%%
where
%%%%%%%%%%%%%%%
\begin{eqnarray}
K_n \equiv \lambda_n \left(\frac23 \kappa^2\right)
^{\frac{n}{2(n-1)}},
\label{Kn}
\end{eqnarray}
%%%%%%%%%%%%%%%
is a constant.
On the other hand, the $\phi$ field equation $(\ref{B14})$
can be written by
%%%%%%%%%%%%%%%
\begin{eqnarray}
\ddot{\phi}+\frac{3\dot{a}}{a}\dot{\phi}
+\frac{n}{n-1} K_n \phi^{\frac{1}{n-1}} = 0.
\label{phi}
\end{eqnarray}
%%%%%%%%%%%%%%%
Note that the potential of
the $\phi$ field  in reheating phase is
represented as $V_n(\phi) \approx K_n \phi^{\frac{n}{n-1}}$.
For $n=2$, the potential becomes
$V_2(\phi) \approx K_2 \phi^2$, and the $\phi$ field
behaves as a massive inflaton.  For $n\geq 4$, the potential
is not regarded as the mass term  and the
curvature of the potential around its bottom becomes larger
than that in the $n=2$ case (See Fig~1).

Making use of a time averaged relation: $\langle\dot{\phi}^2\rangle
=\frac{n}{n-1} \langle V_n(\phi) \rangle$, we can easily find
the time evolution of the scale factor from
the relations $(\ref{scale})$ and $(\ref{phi})$ as
%%%%%%%%%%%%%%%
\begin{eqnarray}
a \approx \left(\frac{t}{t_0}
\right)^{\frac{3n-2}{3n}},
\label{scale2}
\end{eqnarray}
%%%%%%%%%%%%%%%
where $t_0$ is the initial time of the coherent oscillation.
Then, introducing a new time variable
%%%%%%%%%%%%%%%
\begin{eqnarray}
\tau \equiv \frac{n}{2(n-1)} t_0 \left(\frac{t}{t_0}\right)
^{\frac{2(n-1)}{n}}=
\frac{n}{2(n-1)} t_0 a^{\frac{6(n-1)}{3n-2}},
\label{tau}
\end{eqnarray}
%%%%%%%%%%%%%%%
and a new scalar field
%%%%%%%%%%%%%%%
\begin{eqnarray}
\varphi \equiv \left(\frac{t}{t_0}\right)^{\frac{2(n-1)}{n}} \phi
=a^{\frac{6(n-1)}{3n-2}} \phi,
\label{varphi}
\end{eqnarray}
%%%%%%%%%%%%%%%
the $\phi$ field equation $(\ref{phi})$ is rewritten as
%%%%%%%%%%%%%%%
\begin{eqnarray}
\frac{d^2\varphi}{d\tau^2}+
\frac{n}{n-1} K_n \varphi^{\frac{1}{n-1}}=0.
\label{varphi2}
\end{eqnarray}
%%%%%%%%%%%%%%%
This equation is easy to be integrated and we find
a conserved quantity,
%%%%%%%%%%%%%%%
\begin{eqnarray}
E \equiv \frac12 \left(\frac{d\varphi}{d\tau}\right)^2
+K_n \varphi^{\frac{n}{n-1}}.
\label{energy}
\end{eqnarray}
%%%%%%%%%%%%%%%
Integrating Eq.~$(\ref{energy})$
by the variable $\zeta \equiv \left(\frac{K_n}{E}
\right)^{\frac{1}{n}} \varphi^{\frac{1}{n-1}}$, $\tau$ is
represented as
%%%%%%%%%%%%%%%
\begin{eqnarray}
\tau=\pm \frac{n-1}{\sqrt{2E}} \left(\frac{E}{K_n}\right)^
{\frac{n-1}{n}} \int \frac{\zeta^{n-2}}{\sqrt{1-\zeta^n}} d\zeta.
\label{tau2}
\end{eqnarray}
%%%%%%%%%%%%%%%
For example, when $n=2$, $\tau$ is written as
$\tau=(1/\sqrt{2K_2})\cos^{-1} \zeta +\tau_0$.
Hence the oscillation of the $\varphi$ field is sinusoidal:
%%%%%%%%%%%%%%%
\begin{eqnarray}
\varphi=\sqrt{\frac{E}{K_2}} \cos
\left[ \sqrt{2K_2} (\tau-\tau_0) \right].
\label{sin}
\end{eqnarray}
%%%%%%%%%%%%%%%
For $n>2$, however, the $\varphi$ field does not oscillate
sinusoidally but its behavior  is rather
complicated since the integral in Eq~(\ref{tau2}) is
expressed by an elliptic or hyperelliptic integral.
We investigate this issue for the $n=4$ case in Sec. IV.

Next we rewrite the $\chi$ field equation $(\ref{B21})$ with
the new time variable $\tau$.
It yields
%%%%%%%%%%%%%%%
\begin{eqnarray}
& & \frac{d^2\chi_k}{d\tau^2} +
 2b \frac{d\chi_k}{d\tau} +a^{-\frac{6(n-2)}{3n-2}}
 \left[\frac{k^2}{a^2}+e^{-\frac{\sqrt{6}}{3} \kappa \phi}
 \right. \nonumber \\
&\times& \left.
 \left\{ \frac{2n}{n-1} \lambda_n \xi \kappa^2
   \left(e^{\frac{\sqrt{6}}{3} \kappa \phi}-1+\eta\right)
   ^{-\frac{n-2}{n-1}}
   \left(e^{\frac{\sqrt{6}}{3} \kappa \phi}-1+\frac{n+1}{n-1}
   \eta \right)
  +m_{\chi}^2 \right\} \right] \chi_k=0,
\label{chi2}
\end{eqnarray}
%%%%%%%%%%%%%%%
where
%%%%%%%%%%%%%%%
\begin{eqnarray}
b \equiv
\frac{6(n-1)}{3n-2} \frac{1}{a} \frac{da}{d\tau}
  -\frac{\kappa }{\sqrt{6}}
  \frac{d\phi}{d\tau}.
\label{b}
\end{eqnarray}
%%%%%%%%%%%%%%%
Transforming the scalar field $\chi_k$ as
%%%%%%%%%%%%%%%
\begin{eqnarray}
X_k \equiv e^{\int b d\tau } \chi_k
=a^{\frac{6(n-1)}{3n-2}}
e^{-\frac{\kappa}{\sqrt{6}} \kappa\phi} \chi_k,
\label{X}
\end{eqnarray}
%%%%%%%%%%%%%%%
Eq.~$(\ref{chi2})$ becomes
%%%%%%%%%%%%%%%
\begin{eqnarray}
\frac{d^2 X_k}{d\tau^2}+\omega_k^2 X_k=0,
\label{chi3}
\end{eqnarray}
%%%%%%%%%%%%%%%
where the time dependent frequency $\omega_k^2$ is
%%%%%%%%%%%%%%%
\begin{eqnarray}
\omega_k^2 &\equiv&
a^{-\frac{6(n-2)}{3n-2}}
 \left[\frac{k^2}{a^2}+
 e^{-\frac{\sqrt{6}}{3} \kappa \phi}
 \left\{ \frac{2n}{n-1} \lambda_n \xi \kappa^2
   \left(e^{\frac{\sqrt{6}}{3} \kappa \phi}-1+\eta \right)
   ^{-\frac{n-2}{n-1}}
   \left(e^{\frac{\sqrt{6}}{3} \kappa \phi}-1
  +\frac{n+1}{n-1} \eta \right)
  +m_{\chi}^2  \right\} \right]
  \nonumber  \\
  & & - \frac{6(n-1)}{3n-2}
  \left\{ \frac{1}{a} \left(\frac{d^2a}{d\tau^2} \right)
   -\frac{1}{a^2} \left(\frac{da}{d\tau} \right)^2 \right\}
  +\frac{\kappa}{\sqrt{6}} \frac{d^2\phi}{d\tau^2}
  - \left\{ \frac{6(n-1)}{3n-2} \frac{1}{a} \frac{da}{d\tau}
  -\frac{\kappa}{\sqrt{6}} \frac{d\phi}{d\tau}
   \right\}^2.
\label{omega}
\end{eqnarray}
%%%%%%%%%%%%%%%
Since the frequency of the $X_k$ field includes the inflaton
field which oscillates coherently under two approximations
$(\ref{approx1})$ and $(\ref{approx2})$, the
parametric resonance may be expected due to these terms
even in the $n\geq 4$ case.
We investigate  $\chi$-particle production by the
oscillating background field $\phi$ especially when
$n=2$ and $n=4$.
We follow the semiclassical picture of preheating which is
first considered in Ref.~\cite{KT1}, where the
$\chi$-particles are created by quantum fluctuation initially,
and they are treated as a classical field later as the
$\chi$-particles are
produced.   We give an initial distribution
of $X_k$-state in our previous paper~\cite{TMT}.
In the next section, we analyze the structure of preheating
for the $n=2$ case and compare it with numerical results.

%%%%%%%%%%%%%%%%%%%%%%%%%%%%%%%%%%%%%%%%%%%%%%%%%%
%
%                                                 %
\section{preheating by $R^2$ inflation model}                            %
%                                                 %
%%%%%%%%%%%%%%%%%%%%%%%%%%%%%%%%%%%%%%%%%%%%%%%%%%
In the $n=2$ case, the reheating phase starts at $\phi \approx
 M_{\rm PL}/5$ and the $\phi$ field begins to oscillate coherently
around the minimum of its potential after $\phi$ drops down to
$\phi~\mbox{\raisebox{-1.ex}{$\stackrel
     {\textstyle<}{\textstyle \sim}$}}~M_{\rm PL}/20$.
At this stage, as is found from Eqs.~$(\ref{scale2})$-$(\ref{varphi})$,
the scale factor evolves as $a\approx
(t/t_0)^{2/3}$ and $\tau$, $\varphi$ are defined
by $\tau=t$ and $\varphi=a^{3/2}\phi$.
Since the
$\varphi$ field oscillates sinusoidally as Eq.~$(\ref{sin})$
with the mass
%%%%%%%%%%%%%%%
\begin{eqnarray}
m \equiv \sqrt{2K_2} =\frac{M_{\rm PL}}
{\sqrt{96\pi\alpha_2}},
\label{mass}
\end{eqnarray}
%%%%%%%%%%%%%%%
the evolution of the $\phi$ field for
$\phi~\mbox{\raisebox{-1.ex}{$\stackrel
     {\textstyle<}{\textstyle \sim}$}}~M_{\rm PL}/20$ is sinusoidal with decreasing
amplitude $\Phi$ \cite{KLS2}:
%%%%%%%%%%%%%%%
\begin{eqnarray}
\phi=\Phi \sin mt,~~~{\rm with}~~~\Phi
=\frac{M_{\rm PL}}{\sqrt{3\pi} mt}
=\frac{M_{\rm PL}}{2\pi\sqrt{3\pi}\bar{t}},
\label{E1}
\end{eqnarray}
%%%%%%%%%%%%%%%
which is equivalent to the case of the massive inflaton.
Here we introduced the dimensionless time
variable $\bar{t}\equiv mt/2\pi$, which represents the number
of the oscillation of the inflaton field.
Note that Eq.~$(\ref{E1})$ is valid for $\bar{t}~\mbox{\raisebox{-1.ex}{$\stackrel {\textstyle>}{\textstyle\sim}$}}~1$.
As the $\chi$-particles are produced, the backreaction effect
by $\eta$ and $\langle (\nabla \chi)^2 \rangle$ is expected to
appear.
However, numerical calculations shows that $\eta$ and
$\langle (\nabla \chi)^2 \rangle/m^2M_{\rm PL}^2$ do not exceed
0.1 even at the final stage of preheating,
and the $\varphi$ field oscillates almost coherently
during the whole stage of preheating.

Let us investigate the evolution of  $\langle\chi^2\rangle$
both analytically and numerically.
First, $\langle\chi^2\rangle$ is normalized as
%%%%%%%%%%%%%%%
\begin{eqnarray}
\langle\bar{\chi}^2\rangle \equiv
\frac{\langle\chi^2\rangle}{M_{\rm PL}^2}
=\left( \frac{m}{M_{\rm PL}} \right)^2
 \frac1{2\pi^2} \int \bar{k}^2|\bar{\chi}_k|^2 d\bar{k}
=\frac{1}{96\pi\alpha_2}
 \frac1{2\pi^2} \int \bar{k}^2|\bar{\chi}_k|^2 d\bar{k},
\label{E40}
\end{eqnarray}
%%%%%%%%%%%%%%%
where $\bar{k}=k/m$ and $\bar{\chi}_k=\sqrt{m} \chi_k$.
Hence $\langle\bar{\chi}^2\rangle$
depends on the coupling constant $\alpha_2$.
Here
we adopt the lowest value $\alpha_2=2 \times 10^9$
which is determined
by density perturbation.
Note that in this case,
$m \sim 10^{-6} M_{\rm PL} \sim 10^{13}$ GeV,
which is almost the same value as that in the
chaotic inflation model.
Next, let us rewrite Eq.~$(\ref{chi3})$ to the form of
Mathieu equation.
Since the scale factor decreases so slow as compared with
the time scale of the
oscillation of the inflaton field for $\phi~\mbox{\raisebox{-1.ex}{$\stackrel
     {\textstyle<}{\textstyle \sim}$}}~M_{\rm PL}/20$,
we can approximate the frequency as
%%%%%%%%%%%%%%%
\begin{eqnarray}
\omega_k^2 &\approx &
\frac{k^2}{a^2}+4\lambda_2\xi\kappa^2
\left\{1-(1-3\eta)\left(1-\frac{\sqrt{6}}{3}\kappa\phi
\right) \right\}
+m_{\chi}^2 \left(1-\frac{\sqrt{6}}{3}\kappa\phi \right)
+\frac{\kappa}{\sqrt{6}} \frac{d^2\phi}{d\tau^2}
\nonumber \\
&\approx& \frac{k^2}{a^2} +m_{\chi}^2 +9\xi\eta m^2
+\frac{2}{3\pi \bar{t}}
\left[ 3\xi m^2 (1-3\eta)
-m_{\chi}^2-\frac12 m^2 \right] \sin mt.
\label{E3}
\end{eqnarray}
%%%%%%%%%%%%%%%
Here we did not neglect $\eta$. Since $\eta$ may increase up
to 0.1, we should consider the effect from
the backreaction terms $\xi\eta$ and $3\eta$ in the
$X_k$ field equation.
Setting $mt=2z-\pi/2$, Eq.~$(\ref{chi3})$ is reduced to
the Mathieu equation:
%%%%%%%%%%%%%%%
\begin{eqnarray}
\frac{d^2}{dz^2} X_k +\left(A_k -2q \cos 2z \right) X_k=0,
\label{E4}
\end{eqnarray}
%%%%%%%%%%%%%%%
where
%%%%%%%%%%%%%%%
\begin{eqnarray}
A_k=4 \left(\frac{\bar{k}^2}{a^2}+\bar{m}_{\chi}^2
+9\xi\eta \right),
\label{E5}
\end{eqnarray}
%%%%%%%%%%%%%%%
\begin{eqnarray}
q=\frac{4}{3\pi \bar{t}} \left[ 3\xi (1-3\eta)-
\bar{m}_{\chi}^2 -\frac12 \right].
\label{E6}
\end{eqnarray}
%%%%%%%%%%%%%%%
$\bar{m}_{\chi}$ is normalized as $\bar{m}_{\chi}={m}_{\chi}/m$.
Since
we can find from Eq.~$(\ref{E5})$ that $A_k$ always takes
positive value even in the negative $\xi$ case [Remember
$\eta = \xi \kappa^2 \langle \chi^2 \rangle$],
the resonance bands are narrower than those in the model discussed
in Ref.~\cite{TMT} when $\xi <0$ case.
However, since $A_k$ is not limited by $A_k \ge 2q$,
the resonance bands can be broader than those in
the model of $\frac12 m^2\phi^2+\frac12 g^2\phi^2\chi^2$.
Moreover $q$ decreases as $1/\bar{t}$ by the expansion
of the universe.
This is slower than in the massive inflaton plus coupled
scalar field case, where $q \sim 1/\bar{t}^2$\cite{TMT}.
This is due to the fact that the dominant oscillating term in
Eq.~$(\ref{E3})$ is not the $\phi^2$ term but the $\phi$ term,
hence the period of the preheating becomes rather longer
and we can expect an efficient resonance even if $|\xi|$ is
not very large.
Below, we consider the case of $\xi>0$ and $\xi<0$ separately.

%%%%%%%%%%%%%%%%%%%%%%%%%%%
\subsection{Case of $\xi\geq 0$}
%%%%%%%%%%%%%%%%%%%%%%%%%%%
%%%%%%%%%%%%%%%%%%%%%%%%%%%%%%%%%%
\subsubsection{massless $\chi$-particle case}
%%%%%%%%%%%%%%%%%%%%%%%%%%%%%%%%%%
First, we investigate the production of the
massless $\chi$-particle for the case $\xi>0$.
In the initial stage of preheating, we can neglect the $\xi\eta$
term. Hence $A_k$ is almost constant ($A_k \approx 0$)
if we consider modes with small momentum $\bar{k}$, which may be the
dominant
ingredients of the particle production independently
on $\xi$. When the $\phi$ field begins
to oscillate coherently ($\bar{t}\approx 1$),
i.e., the analysis by using the Mathieu equation becomes valid,
the
initial value of $q$ ($=q_i$) is determined by $\xi$. After
this, $q$ decreases as $1/\bar{t}$ by the expansion of the
universe. Hence the Mathieu parameters trace the line near the
$q$-axis horizontally on the Mathieu chart. When the $chi$ field
passes across the first instability band,
the resonance terminates. What determines the efficiency of
the preheating is the initial value of $q$. When $q_i$ is large,
the Floquet index is large and  the growth rate of
$\langle \bar{\chi}^2 \rangle$ is expected to be large.
Moreover the period when the Mathieu parameters stay
in the instability band becomes long, which
means that the total abundance of $\langle \bar{\chi}^2 \rangle$
may be large. As a result, we can conclude that the large $\xi$
gives the efficient preheating as long as the backreaction
effect can be neglected.
Let us examine the evolution with the concrete values
of $\xi$ numerically.

For small $\xi$ ($\xi~\mbox{\raisebox{-1.ex}{$\stackrel
     {\textstyle<}{\textstyle \sim}$}}~1$),
$q_{i}~\mbox{\raisebox{-1.ex}{$\stackrel
     {\textstyle<}{\textstyle \sim}$}}~1$.
Hence, parametric resonance turns on in the narrow resonance
regime and $\chi$-particles are hardly produced because
$q$ decreases and the resonant range becomes narrower
and narrower by the expansion of the universe.
On the other hand, for $\xi~\mbox{\raisebox{-1.ex}{$\stackrel
     {\textstyle>}{\textstyle\sim}$}}~1$, the resonance can  be expected.
Let us consider several cases to compare the differences of the resonance
structures.
For $\xi=3$, $q_{i} \sim 3.61$, and the resonance band is rather broad
at the initial stage. In Fig.~3(a), we depict the evolution of
$\langle\bar{\chi}^2\rangle \equiv \langle\chi^2\rangle/M_{\rm PL}^2$
as a function of $\bar{t}$ for $\xi=3$.
After $\bar{t} \approx 1$,
$\langle\bar{\chi}^2\rangle$ increases almost exponentially
by the parametric
resonance, and  it reaches its maximum value
$\langle\bar{\chi}^2\rangle_f=2.21 \times 10^{-8} $
when $\bar{t}_f=5.61$. [We summarize the data of numerical
calculations in Table~I.]
After that, $\langle\bar{\chi}^2\rangle$ decreases monotonically
by the adiabatic damping of the Hubble expansion.
Since the total amount of created particles is not so large, $\xi\eta$
is much smaller than unity during  the whole stage and we can neglect the
backreaction effect.
In this case, the factor which
terminates the resonance is the expansion of the universe.
$A_k$ and $q$ both decrease and the resonance stops when $q$ drops
down to $q_f=0.64$ ($\bar{t}_f=5.61$).
This evolution is illustrated by the solid curve in Fig.~2.
For $\xi=5$, $q_{i} \sim 6.15$, and more efficient resonance
can occur. Actually, the final value of $\langle\bar{\chi}^2\rangle_f=1.74
\times 10^{-4}$ (when $\bar{t}_f=6.27$ and
$q_f=0.98$) is
much larger than the case of $\xi=3$ (Fig.~3(b)).
 Since the term $36\xi\eta$ in
Eq.~$(\ref{E5})$ is about 3.94 at $\bar{t}_f=6.27$.
Hence we can not neglect the backreaction in this case.
The backreaction mainly affects the
behavior of $A_k$. It makes $A_k$ increase rapidly independently
on the momentum $k$.
Hence the Mathieu parameter deviates from the first instability
band and the resonant band becomes very narrow. By this
effect the parametric resonance terminates.
For $\xi=5$, however,
this deviation occurs when $q$ drops down to about
unity, hence we can say that the final value of
$\langle\bar{\chi}^2\rangle$
is mainly determined by the decrease of $q$.
The case of $\xi \approx 5$ is the marginal case, whether the value of
$\langle\bar{\chi}^2\rangle_f$ is determined by the decrease of
$q$ due to the expansion of the
universe, or by the increase of $A_k$ due to the backreaction.

We can find that  $\langle\bar{\chi}^2\rangle$ increases with
the same period as the oscillation of the inflaton field. This is
different from the resonance in the chaotic inflation model with
the massive inflation, where the $\langle\bar{\chi}^2\rangle$
increases twice during one oscillation of the inflation \cite{KLS2}.
In the chaotic inflation model, the Mathieu parameters are
constrained as $A_k \geq 2q$, and the square of the time
dependent frequency of the $\chi$ field is always positive.
And the resonance occurs when the non-adiabatic condition:
$\omega_k^2 \ll d\omega_k/dt$ is satisfied.
Since this condition is realized when $\phi \approx 0$,
the resonance occurs twice during each oscillation
of the inflaton field. In the present case, however, $\omega_k^2$
is not constrained to be positive but
can become a negative value. In this case $\chi$-particles
are produced mainly by negative coupling instability \cite{BL}.
As we can see from Eq~(\ref{E3}), $\omega_k^2$ becomes
negative once for each oscillation of $\phi$. As a result,
$\langle\bar{\chi}^2\rangle$ increases once during each
oscillation of the $\phi$ field. This picture is independent on the sign
of the coupling constant $\xi$.

For $\xi=10$, $q_{i} \sim 12.52$, and the final values are
$t_f=4.28$, $q_f=2.71$, and $\langle\bar{\chi}^2\rangle_f =
9.95 \times 10^{-5}$ (Fig.~3(c)).
In this case, the value of $\langle\bar{\chi}^2\rangle_f$
decreases slightly compared with the $\xi=5$ case.
As the $\chi$-particles are created, $A_k$ increases because of
the $\xi\eta$ term and the $\chi$ field deviates from
instability bands before $q$
drops down under unity.
This behavior is illustrated by the dashed curve in Fig.~2.
Namely, in this case, the created $\chi$-particles themselves terminate
the resonance. For $\xi~\mbox{\raisebox{-1.ex}{$\stackrel
     {\textstyle>}{\textstyle\sim}$}}~5$, the mechanism to stop the resonance
is the same as in the $\xi=10$ case. One can estimate the final abundance
$\langle\bar{\chi}^2\rangle_f$ by making use of this property.
The maximal momentum to cause the negative coupling
instability is given by Eq~(\ref{E3}) as
%%%%%%%%%%%%%%%
\begin{eqnarray}
\frac{\bar{k}_{max}^2}{a^2} = \frac{2}{3\pi \bar{t}}
\left[ 3\xi (1-3\eta)-m_{\chi}^2
-\frac12 \right]
-9\xi\eta -\bar{m}_{\chi}^2,
\label{E11}
\end{eqnarray}
%%%%%%%%%%%%%%%
where $\bar{k}_{max} \equiv k_{max}/m$ and $\bar{m}_{\chi}
\equiv m_{\chi}/m$.
The relation $(\ref{E11})$ suggests that $\bar{k}_{max}^2/a^2$
decreases
and approaches zero as the preheating  proceeds.
Actually, numerical calculation shows  that
$\bar{k}_{max}^2/a^2$ reaches  zero at $\bar{t}_f$ in the
case of $\xi~\mbox{\raisebox{-1.ex}{$\stackrel
     {\textstyle>}{\textstyle\sim}$}}~5$.
For this reason, it is enough to set $\bar{k}^2/a^2=0$ in Eq.~$(\ref{E5})$
in order to obtain the analytical value of $\langle\bar{\chi}^2\rangle_f$.
For $\xi~\mbox{\raisebox{-1.ex}{$\stackrel
     {\textstyle>}{\textstyle\sim}$}}~5$, parametric resonance terminates when
$A_k$ begins to increase
rapidly as the increase of $\xi\eta$.
Since the resonance bands are very narrow for the region $A_k~\mbox{\raisebox{-1.ex}{$\stackrel
     {\textstyle>}{\textstyle\sim}$}}~3q$,
we adopt the criterion:
When the Mathieu parameters pass across the
$A_k =3q$ line, the preheating terminates.
Then, we find the final abundance of the $\chi$-particles as
%%%%%%%%%%%%%%%
\begin{eqnarray}
\langle\bar{\chi}^2\rangle_f
\approx \frac{1}{24\pi\xi(\pi\bar{t}_f+1)}\left(1-\frac{1}{6\xi}
\right).
\label{E7}
\end{eqnarray}
%%%%%%%%%%%%%%%
Note that $\langle\bar{\chi}^2\rangle_f$ includes two variables:
$\xi$ and $\bar{t}_f$. Generally, as $\xi$ increases ($\xi~\mbox{\raisebox{-1.ex}{$\stackrel
     {\textstyle>}{\textstyle\sim}$}}~5$),
$\bar{t}_f$ decreases because the production rate of $\chi$-particles
becomes large and the $\chi$ field deviates from the 
instability bands  in the early stage.
Numerically, $\langle\bar{\chi}^2\rangle_f$ is the slowly decreasing
function of $\xi$ for $\xi~\mbox{\raisebox{-1.ex}{$\stackrel
  {\textstyle>}{\textstyle\sim}$}}~5$ and this suggests that the 
$\xi$ effect plays a more important role.
The analytic estimation by Eq.~$(\ref{E7})$ shows good agreement with
numerical calculation.
For example, when $\xi=10$, the estimated value
is $\langle\bar{\chi}^2\rangle_f =9.02 \times 10^{-5}$
and the numerical value is $\langle\bar{\chi}^2\rangle_f =9.95 \times
10^{-5}$. For $\xi=100$,  the analytic value is
$\langle\bar{\chi}^2\rangle_f=2.87 \times 10^{-5}$,
since $\bar{t}_f=1.15$.
This is close to the numerical value
$\langle\bar{\chi}^2\rangle_f =2.51 \times 10^{-5}$.
This coincidence shows the validity of the criterion and the
picture which we adopted.
For $\xi~\mbox{\raisebox{-1.ex}{$\stackrel
     {\textstyle>}{\textstyle\sim}$}}~100$, $\bar{t}_f$
approaches a constant value $\bar{t}_f \sim 1$, and
$\langle\bar{\chi}^2\rangle_f$ decreases as $1/\xi$.
Other examples are listed in Table I.
We also show the final value of
$\langle\bar{\chi}^2\rangle_f$ as a function of $\xi$ in Fig.~4.
For $1~\mbox{\raisebox{-1.ex}{$\stackrel
     {\textstyle<}{\textstyle \sim}$}}~\xi~\mbox{\raisebox{-1.ex}{$\stackrel {\textstyle<}{\textstyle \sim}$}}~5$, $\langle\bar{\chi}^2\rangle_f$ is the
increasing function of $\xi$ and is determined by the decrease
of $q$ due to the Hubble expansion.
On the other hand, $\langle\bar{\chi}^2\rangle_f$ is a
slowly decreasing function of $\xi$ for $\xi~\mbox{\raisebox{-1.ex}{$\stackrel {\textstyle>}{\textstyle\sim}$}}~5$,
where the backreaction effect becomes important.

We can conclude that
$\langle\bar{\chi}^2\rangle_f$ takes maximal value
$\langle\bar{\chi}^2\rangle_{max}=1.74 \times 10^{-4}$
when $\xi \approx
5$ for $\xi>0$ for the massless $\chi$-particle case.

%%%%%%%%%%%%%%%%%%%%%%%%%%%%%%%%%%
\subsubsection{massive $\chi$-particle case}
%%%%%%%%%%%%%%%%%%%%%%%%%%%%%%%%%%
Whether the $\chi$-particle with a mass heavier than the
GUT scale can be produced or not is one of the most
important issues of the preheating, which is related to
GUT scale baryogenesis. One can easily obtain the properties of
the preheating by the massive $\chi$ field and difference from
the massless case from Eqs.~$(\ref{E5})$ and Eq.~$(\ref{E6})$.
Since the $\chi$-field couples with the inflaton field by $\xi$
and $m_{\chi}$ terms as we can see from Eq.~$(\ref{B21})$,
one may consider that
parametric resonance would occur by only the mass term
without the non-minimal coupling.
This, however, is not the case.
In the case of $\xi=0$, $A_k \ge 4\bar{m}_{\chi}^2$ and
$|q_i| \approx 0.4 \left(\bar{m}_{\chi}^2+\frac12\right)$.
This means that $A_k$ is at least 10 times larger than $q$
when $\bar{t} \approx 1$, in which case the resonance bands
are very
narrow. Moreover, even if the $\chi$ field stays in an instability band
initially, $A_k$ and $q$ decrease by the expansion
of the universe and soon
deviate from the instability bands.
As a result, it is difficult to produce the massive
$\chi$-particle enough without the non-minimal coupling.
We confirmed this by numerical calculation.

For the $m_{\chi} \ne 0$ case, the existence of the
$\bar{m}_{\chi}^2$ term constrains
$A_k \geq 4\bar{m}_{\chi}^2$ and
decreases the value of $q$, hence the  resonance is suppressed
as compared with the massless case.
First, consider the case of $m_{\chi}=m$ ($\bar{m}_{\chi}=1$).
$A_k$ takes  the value more than 4 due to the mass effect, and
the modes with the momentum which contribute to the resonance
mainly exist in the second instability band of the
Mathieu chart  (See Fig.~2).
Since there is no contribution from the first resonance band,
which has larger Floquet index than the second one with
the same $q$ value, the larger
value of $\xi$ is expected to be
needed for the $\chi$-particle production to occur.
Numerically, when $\xi~\mbox{\raisebox{-1.ex}{$\stackrel
     {\textstyle<}{\textstyle \sim}$}}~3$, the parametric resonance
does not come about.
In the case of $\xi=5$, the
resonance occurs, but
this process is very weak (Fig.~5(a)).
For $\bar{t}~\mbox{\raisebox{-1.ex}{$\stackrel
     {\textstyle>}{\textstyle\sim}$}}~1$, $\langle\bar{\chi}^2\rangle$ increases, but
$\langle\bar{\chi}^2\rangle_f$ is smaller than the initial value.
When $\xi=10$, the parametric resonance after $\bar{t} \approx 1$
is rather effective (Fig.~5(b)).
The final values are $\bar{t}_f=9.22$, $q_f=1.31$, and
$\langle\bar{\chi}^2\rangle_f =1.17 \times 10^{-5}$
(See Table~II).
Since the contribution from the backreaction term
$36\xi\eta$  in Eq.~$(\ref{E5})$
takes 1.06 as a final value, the deviation from the instability band
by the increase of $A_k$ begins to be important.
When $\xi=20$, the final values are $\bar{t}_f=3.23$, $q_f=7.07$,
and $\langle\bar{\chi}^2\rangle_f =5.19 \times 10^{-5}$ (Fig.~5(c)).
In this case, the backreaction effect by the  $\xi\eta$ term
completely determines the final
$\chi$-particle abundance because $q_f=7.07$ when resonance ends.
Numerically, $\langle\bar{\chi}^2\rangle_f$ takes the maximal value of
$\langle\bar{\chi}^2\rangle_{max} \approx
5 \times 10^{-5}$ for $\xi=20 \sim 30$.
For $\xi~\mbox{\raisebox{-1.ex}{$\stackrel
 {\textstyle>}{\textstyle\sim}$}}~30$, $\langle\bar{\chi}^2\rangle_f$ is a slowly
decreasing function of $\xi$.
Generally, when the $\xi\eta$ suppression effect plays the
relevant role, one can estimate $\langle\bar{\chi}^2\rangle_f$
in the same way as the massless $\chi$-particle case.
By adopting the criterion: $A_k=3q$ when the resonance
terminates and
setting $\bar{k}^2/a^2=0$ in Eq.~$(\ref{E5})$, we obtain
%%%%%%%%%%%%%%%
\begin{eqnarray}
\langle\bar{\chi}^2\rangle_f
\approx \frac{1}{72\pi\xi^2} \left[\frac{6\xi-1}{2(\pi\bar{t}_f+1)}
-\bar{m}_{\chi}^2 \right].
\label{E12}
\end{eqnarray}
%%%%%%%%%%%%%%%
Apparently, the mass effect suppresses the final
$\chi$-particle abundance.
Let us compare this estimation
with numerical results.
For example, when $\bar{m}_{\chi}=1$ and $\xi=20$,
the numerical value is
$\langle\bar{\chi}^2\rangle_f=5.19 \times 10^{-5}$
(at $\bar{t}_f=3.23$) and the analytic one is
$\langle\bar{\chi}^2\rangle_f =4.80 \times 10^{-5}$
; for $\xi=100$, the numerical one is
$\langle\bar{\chi}^2\rangle_f=2.78 \times 10^{-5}$
(at $\bar{t}_f=1.18$) and the
analytic one is $\langle\bar{\chi}^2\rangle_f=2.76 \times 10^{-5}$.
When $\bar{m}_{\chi}=1$, the mass term in Eq.~$(\ref{E12})$
can not be  neglected as compared with the former term for
the $\xi~\mbox{\raisebox{-1.ex}{$\stackrel
     {\textstyle<}{\textstyle \sim}$}}~50$ case.
In this case, the mass effect is one of the
important factors to determine the final
abundance of the $\chi$ particles.
On the other hand, for $\xi~\mbox{\raisebox{-1.ex}{$\stackrel
     {\textstyle>}{\textstyle\sim}$}}~50$, the mass effect
can be neglected and does not play
a relevant role for the final abundance, although the growth rate
and other properties during the resonance are
different from the massless case.
In Table II, we show the analytically estimated and numerical
values of $\langle\bar{\chi}^2\rangle_f$ in various parameters.
We can see that the
analytic results give good agreement with the numerical
ones.

For the large $\xi$ where the backreaction becomes
important, the right hand side (r.h.s.) in
Eq.~$(\ref{E12})$ must be a positive value, hence $\bar{m}_{\chi}$ is
constrained to be
%%%%%%%%%%%%%%%
\begin{eqnarray}
\bar{m}_{\chi}^2~\mbox{\raisebox{-1.ex}{$\stackrel
     {\textstyle<}{\textstyle \sim}$}}~\frac{6\xi-1}{2(\pi\bar{t}_f+1)}.
\label{E13}
\end{eqnarray}
%%%%%%%%%%%%%%%
Since $\bar{t}_f$ approaches to the constant value of
$\bar{t}_f \approx 1$ as $\xi$ increases, we can obtain the constraint
of the mass of the created $\chi$-particle for large $\xi$ as
%%%%%%%%%%%%%%%
\begin{eqnarray}
\bar{m}_{\chi}~\mbox{\raisebox{-1.ex}{$\stackrel
     {\textstyle<}{\textstyle \sim}$}}~\sqrt{\xi}.
\label{E14}
\end{eqnarray}
%%%%%%%%%%%%%%%
From this constraint we find that if $\xi$
is more than 100, $\chi$-particle
whose mass is of order $m_{\chi}=10m \sim 10^{14}$ GeV
can be produced by the
non-minimal coupling in the $R^2$ inflation model.
In order to produce the GUT scale gauge boson whose mass is about
$m_{\chi}=10^3 m \sim 10^{16}$ GeV, $\xi$ is needed as 
at least $10^6$.
Namely, unless $\xi$ takes a very large value, the GUT scale boson
is hard to be created.
Moreover, even if $\xi$ takes a very large value,
since the Mathieu parameters $A_k$ and $q$
deviate from instability band due to mass effect from the very
beginning, the final abundance of the $\chi$-particles must be
very small as estimated by Eq.~(\ref{E12}) as
$\langle\bar{\chi}^2\rangle_f \propto  1/\xi$.
Finally, we show the  value of
$\langle\bar{\chi}^2\rangle_f$ as a function
of $\bar{m}_{\chi}$ in the case of $\xi=100$ in Fig.~6.
$\langle\bar{\chi}^2\rangle_f$ is the quadratically
decreasing function of $\bar{m}_{\chi}$ as is expected from
Eq.~$(\ref{E12})$.
We find that the $\chi$-particle whose mass is
more than $m_{\chi}~\mbox{\raisebox{-1.ex}{$\stackrel
     {\textstyle>}{\textstyle\sim}$}}~10m$ can not be produced,
justifying the estimation $(\ref{E14})$.

%%%%%%%%%%%%%%%%%%%%%%%%%%%
\subsection{Case of $\xi<0$}    %
%%%%%%%%%%%%%%%%%%%%%%%%%%%
%%%%%%%%%%%%%%%%%%%%%%%%%%%%%%%%%%
\subsubsection{massless $\chi$-particle case}
%%%%%%%%%%%%%%%%%%%%%%%%%%%%%%%%%%
Let us investigate the $\xi<0$ case for the massless $\chi$-particle.
In the $\xi<0$ case, $q$ takes negative value as is found by
Eq.~$(\ref{E6})$.
Since the Mathieu chart is symmetrical with respect to
$A_k$-axis, it is enough to consider the absolute value of $q$:
%%%%%%%%%%%%%%%
\begin{eqnarray}
|q|=\frac{4}{3\pi \bar{t}} \left[ 3 |\xi| (1-3\eta)
+ \frac12 \right].
\label{E8}
\end{eqnarray}
%%%%%%%%%%%%%%%
Note that $|q|$ is larger than  in the positive
$\xi$ case because of the last term $+1/2$, which makes the
resonant more efficient.
Even if $\xi=-1$ ($q_i \approx 1.5$),  parametric resonance
evidently occurs, although the final abundance is very small:
$\langle\bar{\chi}^2\rangle_f=2.63 \times 10^{-11}$ (see Fig.~7(a)).
For $\xi=-3$ ($q_i \approx 4.0$),  the final values are
$\bar{t}_f=7.02$, $q_f=0.57$ and $\langle\bar{\chi}^2\rangle_f=
8.71 \times 10^{-5}$.
Hence parametric resonance is much more efficient than in the
$\xi=3$ case.
One characteristic property in the $\xi<0$ case is that
$\langle\bar{\chi}^2\rangle$ can increase before the $\varphi$ field
enters the coherent oscillation stage.
This is because  the time dependent frequency
$\omega_k^2$ of Eq.~$(\ref{omega})$  takes negative
value initially for $k=0$, and $\langle\bar{\chi}^2\rangle$ grows due
to negative coupling instability.
For example, in the $\xi=-3$ case, $\langle\bar{\chi}^2\rangle$
decreases slightly initially by the expansion of the universe, but
begins to increase after that (Fig.~7(b)).
This is another factor where the negative $\xi$ case gives
more efficient preheating than the positive one.
After $\bar{t} \approx 1$,
the approximation~(\ref{approx1}) is valid and the
analytical estimation by the
Mathieu equation is relevant.
The final value $q_f$ is small for $\xi >-3$, we
can neglect the back
reaction effect and the resonance terminates by
passing  across the first resonant band.
For $\xi=-5$, however, the final values are
$\bar{t}_f=3.80$, $q_f=1.57$ and $\langle\bar{\chi}^2\rangle_f=
2.63 \times 10^{-4}$ (Fig.~7(c)).
Since $q_f$ is larger than unity, the backreaction by the
$\xi\eta$ term in
Eq.~$(\ref{E5})$  becomes important and one can estimate
the final abundance $\langle\bar{\chi}^2\rangle_f$ for $\xi~\mbox{\raisebox{-1.ex}{$\stackrel
     {\textstyle<}{\textstyle \sim}$}}-4$
by the same method as the $\xi>0$ case.
By using the criterion $A_k=3|q|$ for Eqs.~$(\ref{E5})$  and
$(\ref{E8})$, $\langle\bar{\chi}^2\rangle_f$ is
estimated as
%%%%%%%%%%%%%%%
\begin{eqnarray}
\langle\bar{\chi}^2\rangle_f
\approx \frac{1}{24\pi|\xi|(\pi\bar{t}_f-1)}\left(1+\frac{1}{6|\xi|}
\right).
\label{E50}
\end{eqnarray}
%%%%%%%%%%%%%%%
This shows that the maximal value of $\langle\bar{\chi}^2\rangle_f$
becomes larger than that of the $\xi>0$ case in Eq.~$(\ref{E6})$.
Actually, the value of $\langle\bar{\chi}^2\rangle_f$ in the range
$-15~\mbox{\raisebox{-1.ex}{$\stackrel
     {\textstyle<}{\textstyle \sim}$}}~\xi~\mbox{\raisebox{-1.ex}{$\stackrel
     {\textstyle<}{\textstyle \sim}$}}~-4$  is larger than the maximal value for
$\xi>0$ case~(See Fig.~4).
For $\xi =-10$, $\langle\bar{\chi}^2\rangle_f$ estimated by
Eq.~$(\ref{E50})$ is $\langle\bar{\chi}^2\rangle_f=1.74 \times
10^{-4}$ (at $t_f=2.78$) and the numerical value is
$\langle\bar{\chi}^2\rangle_f=1.78 \times 10^{-4}$.
The analytic estimation
$(\ref{E50})$ gives quite good agreement with the numerical results
for $-20~\mbox{\raisebox{-1.ex}{$\stackrel
     {\textstyle<}{\textstyle \sim}$}}~\xi~\mbox{\raisebox{-1.ex}{$\stackrel
     {\textstyle<}{\textstyle \sim}$}}~-4$ (See Table III).

For $\xi~\mbox{\raisebox{-1.ex}{$\stackrel
     {\textstyle<}{\textstyle \sim}$}}~-20$, however, the final point $\bar{t}_f$ becomes
less than unity, and the estimation $(\ref{E50})$ based on the Mathieu
equation is broken.
In this case, since $\omega_k^2$
takes negative value for most $k$ modes,
$\langle\bar{\chi}^2\rangle$ and $|\eta|$ increase rapidly.
Then, the term $e^{\frac{\sqrt{6}}{3}\kappa\phi}-1+3\eta$
in Eq.~$(\ref{omega})$
decreases and becomes negative while $\omega_k^2$ becomes
positive  for any $\phi $, which means the negative coupling
instability does not occur any more.
Note that the $\phi$ field
equation $(\ref{B14})$ includes a similar term
: $e^{\frac{\sqrt{6}}{3}\kappa\phi}-1+\eta$.
Although the $\eta$ term is expected to work as the backreaction on
the $\phi$ field, the effect on the $\chi$
field equation is dominant
due to the difference by factor 3. After the inflaton
field enters the coherent oscillation stage, $A_k$ is much
larger than $|q|$ and the further resonance does not come about.
By making use of this picture, we can estimate the final abundance of
$\langle\bar{\chi}^2\rangle_f$ for $\xi~\mbox{\raisebox{-1.ex}{$\stackrel
     {\textstyle<}{\textstyle \sim}$}}~-20$ as
%%%%%%%%%%%%%%%
\begin{eqnarray}
\langle\bar{\chi}^2\rangle_f
\approx \frac{1}{24\pi|\xi|}
\left( e^{\frac{\sqrt{6}}{3} \kappa \phi_f} -1\right),
\label{E41}
\end{eqnarray}
%%%%%%%%%%%%%%%
where $\phi_f=\phi(t_f)$.
For example, when $\xi=-50$, numerical values are
$\bar{t}_f=0.58$,
$\langle\bar{\chi}^2\rangle_f=2.82 \times 10^{-4}$
and $\phi_f=0.128M_{\rm PL}$.
Then, $\langle\bar{\chi}^2\rangle_f$ estimated by
Eq.~$(\ref{E41})$ is
$\langle\bar{\chi}^2\rangle_f= 1.83 \times 10^{-4}$.
In spite of the naive estimation,
it gives a value close to the numerical one.
We show the final abundance $\langle\bar{\chi}^2\rangle_f$
for $\xi <0$
in Fig.~4.
We can find that
$\langle\bar{\chi}^2\rangle_f$ takes its maximum value
$\langle\bar{\chi}^2\rangle_{max} \approx 3 \times 10^{-4}$
for $\xi \approx -4$ (and $\xi \approx -40$),
which is larger than $\xi>0$ case.
For $\xi~\mbox{\raisebox{-1.ex}{$\stackrel
     {\textstyle<}{\textstyle \sim}$}}~-40$, $\langle\bar{\chi}^2\rangle_f$ begins to
decrease by the factor: $1/|\xi|$.

%%%%%%%%%%%%%%%%%%%%%%%%%%%%%%%%%%
\subsubsection{massive $\chi$-particle case}
%%%%%%%%%%%%%%%%%%%%%%%%%%%%%%%%%%
We investigate the $\xi<0$, $m_{\chi} \ne 0$ case
as the final case in this section.
Since $A_k$ and $|q|$ are expressed as Eq.~$(\ref{E5})$ and
%%%%%%%%%%%%%%%
\begin{eqnarray}
|q|=\frac{4}{3\pi \bar{t}} \left[ 3 |\xi| (1-3\eta)
+\bar{m}_{\chi}^2 + \frac12 \right],
\label{E60}
\end{eqnarray}
%%%%%%%%%%%%%%%
the mass effect makes both $A_k$ and $|q|$ larger than the
massless case.
Although the large $q$ value gives an efficient preheating,
the increase of $A_k$ is more significant than the increase of $|q|$,
and $\chi$-particle production is suppressed by taking
the mass effect into account.
For example, when $\bar{m}_{\chi}=1$, the resonance of $\chi$-particles
can not be seen for $-3~\mbox{\raisebox{-1.ex}{$\stackrel
     {\textstyle<}{\textstyle \sim}$}}~\xi~\mbox{\raisebox{-1.ex}{$\stackrel
     {\textstyle<}{\textstyle \sim}$}}~0$.
Consider the case of $\xi=-5$ and $\bar{m}_{\chi}=1$
(See Fig.~8(a)).
Although the growth rate is small, $\langle\bar{\chi}^2\rangle$
is slowly increasing up to
$\langle\bar{\chi}^2\rangle_f=3.24 \times 10^{-8}$
(at $\bar{t}_f=7.74$).
In this case, since $q_f=0.90$ and $36\xi\eta \ll 1$, the resonance
terminates by the expansion of the universe.
When $\xi=-10$ and $\bar{m}_{\chi}=1$,
the resonance is efficient and the final values are
$\langle\bar{\chi}^2\rangle_f=1.13 \times 10^{-4}$,
$\bar{t}_f=2.69$ and $q_f=4.61$ (Fig.~8(b)).
In this case, the increase of $A_k$ due to $\chi$-particle
production determines the final abundance.
By using the same criterion as in the previous cases, we can
estimate the value of $\langle\bar{\chi}^2\rangle_f$ as
%%%%%%%%%%%%%%%
\begin{eqnarray}
\langle\bar{\chi}^2\rangle_f
\approx \frac{1}{72\pi\xi^2} \left[\frac{6|\xi|+1}{2(\pi\bar{t}_f-1)}
-\bar{m}_{\chi}^2 \right].
\label{E61}
\end{eqnarray}
%%%%%%%%%%%%%%%
In the $\xi=-10$ case, the estimated
value  is
$\langle\bar{\chi}^2\rangle_f=1.37 \times 10^{-4}$, which is close to
the numerical value.
Eq.~$(\ref{E61})$ is relevant for $-30~\mbox{\raisebox{-1.ex}{$\stackrel
     {\textstyle<}{\textstyle \sim}$}}~\xi~\mbox{\raisebox{-1.ex}{$\stackrel
     {\textstyle<}{\textstyle \sim}$}}~-10$
because $\bar{t}_f$ is larger than unity.
However, for $\xi~\mbox{\raisebox{-1.ex}{$\stackrel
     {\textstyle<}{\textstyle \sim}$}}~-30$, where the resonance
terminates before the inflaton field begins to oscillate coherently,
we should use the estimation
%%%%%%%%%%%%%%%
\begin{eqnarray}
\langle\bar{\chi}^2\rangle_f
\approx \frac{1}{24\pi|\xi|}
\left( e^{\sqrt{\frac23} \kappa \phi_f} -1
-\frac{\bar{m}_{\chi}^2}{3|\xi|} \right),
\label{E62}
\end{eqnarray}
%%%%%%%%%%%%%%%
which can be obtained similarly to Eq.~$(\ref{E41})$.
Note that Eq.~$(\ref{E62})$ is valid when $\bar{m}_{\chi}^2$
is not as large as $|\xi|$.
If $\bar{m}_{\chi}^2$ is large as compared with $|\xi|$,
$\omega_k^2$ becomes positive  at the initial stage by
Eq.~$(\ref{omega})$ and can not give the negative
coupling instability.
For example, when $\xi=-50$ and $\bar{m}_{\chi}=1$, the 
numerical values
are $\langle\bar{\chi}^2\rangle_f=2.79 \times 10^{-4}$,
$\bar{\phi}_f=0.128 M_{\rm PL}$ at $\bar{t}_f=0.58$,
while the estimated
value by Eq.~$(\ref{E62})$ is $\langle\bar{\chi}^2\rangle_f= 1.81
\times 10^{-4}$
(See Table IV  for other examples).
In this case, as is found by Eq.~$(\ref{E62})$, the mass term does
not affect a significant role.
However, if $\bar{m}_{\chi}$ is large, $\langle\bar{\chi}^2\rangle_f$
is significantly suppressed by a mass effect
(Compare with $\xi=-50$,
$\bar{m}_{\chi}=7$ case).
Since $\chi$-particle production is possible naively when the r.h.s. in
Eq.~$(\ref{E62})$ takes positive value,
the produced mass of the $\chi$-particle is bounded as
%%%%%%%%%%%%%%%
\begin{eqnarray}
\bar{m}_{\chi}~\mbox{\raisebox{-1.ex}{$\stackrel
     {\textstyle<}{\textstyle \sim}$}}~
\sqrt{3|\xi| \left( e^{\sqrt{\frac23} \kappa \phi_f} -1\right)}.
\label{E63}
\end{eqnarray}
%%%%%%%%%%%%%%%
 $\phi_f$ is smaller than the initial value $M_{\rm PL}/5$ for
large $\xi$. This means that
$\bar{m}_{\chi}$ is constrained to be
%%%%%%%%%%%%%%%
\begin{eqnarray}
\bar{m}_{\chi}~\mbox{\raisebox{-1.ex}{$\stackrel
     {\textstyle<}{\textstyle \sim}$}}~2 \sqrt{|\xi|}.
\label{E64}
\end{eqnarray}
%%%%%%%%%%%%%%%
The massive $\chi$-particle whose mass is of the order
$m_{\chi} \sim 10^{13}$ GeV can be generated even if $\xi \sim -5$,
but with the increase of $m_{\chi}$, the $\chi$-particle is hard to 
produce because $m_{\chi}$ is strongly restricted by
Eq.~$(\ref{E64})$. In particular,
$|\xi|$ still needs to take a very large value
$|\xi|~\mbox{\raisebox{-1.ex}{$\stackrel
     {\textstyle>}{\textstyle\sim}$}}~10^5$ in order to produce the GUT scale gauge boson:
$m_{\chi} \sim 10^{16}$GeV.
As a result, it seems difficult to produce
the GUT scale particle by parametric resonance
in the $R^2$ inflation model in both the $\xi>0$ and $\xi<0$
cases.

%%%%%%%%%%%%%%%%%%%%%%%%%%%%%%%%%%%%%%%%%%%%%%%%%%
%
%                                                 %
\section{preheating by $R^4$ inflation model}                            %
%                                                 %
%%%%%%%%%%%%%%%%%%%%%%%%%%%%%%%%%%%%%%%%%%%%%%%%%%
In this section, we consider the $R^4$ inflation model as the next
example.
In the $R^4$ inflation model, inflation is hardly realized because
a fine tuning of the initial value of the inflaton field
 is needed and the coupling constant
$\alpha_4$ is constrained to be very large\cite{R4}.
It is worth, however, investigating whether a preheating stage
exists or not and how the preheating process proceeds if it occurs.
In this case, the value of the inflaton when the reheating process
turns on is $\phi\approx M_{\rm PL}/9$.
After $\phi$ decreases under $\phi \approx M_{\rm PL}/20$,
the scale factor
evolves as $a\approx \left(t/t_0\right)^{5/6}$, where
$\tau$ and
$\varphi$ are defined as $\tau=2t_0/3
\left(t/t_0\right)^{3/2}$ and $\varphi=a^{9/5}\phi$,
respectively
by  Eqs.~$(\ref{scale2})$-$(\ref{varphi})$.
As for the $\varphi$ field, integrating the r.h.s. in
Eq.~$(\ref{tau2})$, we find that $\tau$ is
expressed by the combination of
the complete elliptic integrals of the  first kind
$F\left( \theta, 1/\sqrt{2} \right)$ and the second
kind $E\left( \theta, 1/\sqrt{2} \right)$ \cite{elliptic}:
%%%%%%%%%%%%%%%
\begin{eqnarray}
F \left(\theta, \frac{1}{\sqrt{2}} \right)-
2E \left(\theta, \frac{1}{\sqrt{2}} \right)
=\pm \frac23 \left(\frac{K_4^{\; 3}}{E}\right)^{1/4}
(\tau-\tau_0),
\label{elliptic}
\end{eqnarray}
%%%%%%%%%%%%%%%
where
%%%%%%%%%%%%%%%
\begin{eqnarray}
\theta=\cos^{-1} \left[
\left(\frac{K_4}{E} \right)^{1/4} \varphi^{1/3} \right].
\label{theta}
\end{eqnarray}
%%%%%%%%%%%%%%%
Since $F \left( \theta, 1/\sqrt{2} \right)$ and
$E \left(\theta, 1/\sqrt{2} \right)$ are well
approximated as
$F \left(\theta, 1/\sqrt{2} \right)
\approx 1.1803 \theta$ and
$E \left(\theta, 1/\sqrt{2}  \right)
\approx 0.8598 \theta$,
we obtain the evolution of the $\varphi$ field as
%%%%%%%%%%%%%%%
\begin{eqnarray}
\varphi^{\frac13} \approx
\varphi_0^{\frac13}
\cos \left[ \frac{c\sqrt{K_4}}{\varphi_0^{1/3}}
(\tau-\tau_0) \right],
\label{phiapprox}
\end{eqnarray}
%%%%%%%%%%%%%%%
where $c=1.2360$ is a constant and
$\varphi_0 \equiv \left(E/K_4 \right)^{\frac34}$ represents
the amplitude of the $\varphi$ field.
Note that the variable
%%%%%%%%%%%%%%%
\begin{eqnarray}
m_4 \equiv  \frac{c\sqrt{K_4}}{\varphi_0^{1/3}}
\label{m4}
\end{eqnarray}
%%%%%%%%%%%%%%%
plays the role of the ``inflaton mass" for the $R^4$ inflation model
when the inflaton field oscillates coherently.

Next, consider the evolution of the $\chi$ field. The
time dependent frequency in the present model is expressed as
%%%%%%%%%%%%%%%
\begin{eqnarray}
\omega_k^2 =
a^{-\frac{6}{5}}
 \left[\frac{k^2}{a^2}+
 e^{-\frac{\sqrt{6}}{3} \kappa \phi}
 \left\{ \frac{8}{3} \lambda_4 \xi \kappa^2
   \left(e^{\frac{\sqrt{6}}{3} \kappa \phi}-1+\eta \right)
   ^{-\frac{2}{3}}
   \left(e^{\frac{\sqrt{6}}{3} \kappa \phi}-1
  +\frac{5}{3} \eta \right)
  +m_{\chi}^2  \right\} \right] 
  + \frac{\kappa}{\sqrt{6}} \frac{d^2\phi}{d\tau^2}
  +\frac{\kappa^2}{6} \left(\frac{d\phi}{d\tau}\right)^2,
\label{omega22}
\end{eqnarray}
%%%%%%%%%%%%%%%
where we used the approximation
$\kappa \frac{d\phi}{d\tau} \gg \frac{da}{d\tau}$.
Further, adopting the approximations:
Eqs.~$(\ref{approx1})$ and $(\ref{approx2})$,
the frequency  can be written by
%%%%%%%%%%%%%%%
\begin{eqnarray}
\omega_k^2 \approx
a^{-\frac65} \left(\frac{k^2}{a^2}+m_{\chi}^2\right)
+\left(\frac23 \right)^{\frac32} a^{-\frac95}
(6\xi-1)\kappa K_4 \varphi_0^{\frac13}
\cos \left[m_4
(\tau-\tau_0) \right],
\label{omega2}
\end{eqnarray}
%%%%%%%%%%%%%%%
where we used Eq.~$(\ref{phiapprox})$.
Setting $2z=m_4 (\tau-\tau_0)+\pi$, Eq.~$(\ref{chi3})$
is reduced to the Mathieu equation
$(\ref{E4})$, with
%%%%%%%%%%%%%%%
\begin{eqnarray}
A_k=4 a^{-\frac65}
\left(\frac{\bar{k}^2}{a^2}+\bar{m}_{\chi}^2 \right),
\label{Ak}
\end{eqnarray}
%%%%%%%%%%%%%%%
\begin{eqnarray}
q=c' \left( \xi-\frac16 \right)
a^{-\frac95}
\bar{\varphi}_0,
\label{q}
\end{eqnarray}
%%%%%%%%%%%%%%%
where $c' \equiv \frac{32}{c^2}\sqrt{\frac{\pi}{3}} =21.4338$,
$\bar{k} \equiv k/{m_4}$, $\bar{m}_{\chi} \equiv m_{\chi}/m_4$
and $\bar{\varphi}_0 \equiv \varphi_0/M_{\rm PL}$.
Both of the Mathieu parameters decrease by the
expansion of the universe in the $R^4$ model.
Note that this relation can only be used for $\eta \ll 1$.
As $\chi$-particles are produced, we have to take into account
the suppression effect by the increase of $|\eta |$.
As the $R^2$ inflation model, the difference
between the $\xi>0$ and $\xi<0$ cases appears in  the sign of the
term $1/6$
in Eq.~$(\ref{q})$. This difference, however, is not important
for $|\xi| \gg1$.

The expectation value $\langle\chi^2\rangle$ is normalized as
%%%%%%%%%%%%%%%
\begin{eqnarray}
\langle\bar{\chi}^2\rangle \equiv
\frac{\langle\chi^2\rangle}{M_{\rm PL}^2}
=\left( \frac{m_4}{M_{\rm PL}} \right)^2
 \frac1{2\pi^2} \int \bar{k}^2|\bar{\chi}_k|^2 d\bar{k}
=\frac{c^2}{16(12\pi \bar{\varphi}_0)^{\frac23}\beta^{\frac13} }
 \frac1{2\pi^2} \int \bar{k}^2|\bar{\chi}_k|^2 d\bar{k},
\label{normalise}
\end{eqnarray}
%%%%%%%%%%%%%%%
where $\bar{\chi}_k \equiv \sqrt{m_4} \chi_k$.
Since $\beta$ is strongly constrained
as Eq.~$(\ref{B36})$, the normalized  value of
$\langle\bar{\chi}^2\rangle$
is very small initially.
However, since $\langle\bar{\chi}^2\rangle$ can grow exponentially by
the parametric resonance, we can expect a considerable amount
of the $\chi$-particles at the final stage of the preheating.
We use a dimensionless time parameter defined by $\tilde{t}
\equiv M_{\rm PL} \beta^{-\frac16} t$ in numerical calculations.
Counting time from $\phi \approx M_{\rm PL}/9$, the
evolution of the $\varphi$ field is described by Eq.~$(\ref{phiapprox})$
with
$\varphi_0 \approx 0.05 M_{\rm PL}$ (at
$\tilde{t}_0 \approx 20$). Then Eq.~$(\ref{q})$ is rewritten as
%%%%%%%%%%%%%%%
\begin{eqnarray}
q \approx 1.0716 \left(\xi-\frac16 \right)
\left(\frac{\tilde{t}}{20}\right)^{-\frac32}.
\label{q2}
\end{eqnarray}
%%%%%%%%%%%%%%%
We find from this relation that $q$ decreases faster as compared
with the $R^2$ inflation model.

First, consider the $\xi>0$ case for the massless $\chi$-particles.
For $0~\mbox{\raisebox{-1.ex}{$\stackrel
     {\textstyle<}{\textstyle \sim}$}}~\xi~\mbox{\raisebox{-1.ex}{$\stackrel
     {\textstyle<}{\textstyle \sim}$}}~3$,  $\langle\bar{\chi}^2\rangle$ does
not increase because the initial value of $q$ is small.
For $\xi~\mbox{\raisebox{-1.ex}{$\stackrel
     {\textstyle>}{\textstyle\sim}$}}~5$, however,  we can see that
$\langle\bar{\chi}^2\rangle$ grows by the
parametric resonance.
Consider the case of $\xi=10$ (See Fig.~9(a)).
Numerically, the final values are $\tilde{t}_f=161$
and $\langle\bar{\chi}^2\rangle_f=5.89 \times 10^{-27}$
(See Table IV).
Although the total amount of created $\chi$-particles
is not large  because of the small
initial value, parametric resonance evidently
occurs due to the oscillating $\varphi$ field.
In this case, since  $\eta$ is much smaller than unity,
the backreaction effect due to $\chi$-particle
production does not affect the $\chi$-field
evolution. When the Mathieu parameters
with low momentum, which contribute the resonance
mainly, pass across the first instability band by Hubble
expansion, the resonance terminates.
In the case of $\xi=30$, the final values are $\tilde{t}_f=185$
and
$\langle\bar{\chi}^2\rangle=1.29 \times 10^{-5}$ (Fig.~9(b)).
In spite of the small initial fluctuation,
$\langle\bar{\chi}^2\rangle$ increases up to
$\langle\bar{\chi}^2\rangle_f \sim 10^{-5}$
by copious production of the $\chi$-particles.
In the case of $\xi=50$, the final values are $\tilde{t}_f=140$
and $\langle\bar{\chi}^2\rangle_f=
8.28 \times 10^{-6}$ (Fig.~9(c)).
Although the growth rate of $\langle\bar{\chi}^2\rangle$
becomes larger than the $\xi=30$ case because
of the large initial value of $q$, the final
abundance is almost the same.
This means that the suppression effect by the backreaction
is significant when the fluctuation
increases up to $\langle\bar{\chi}^2\rangle \sim 10^{-5}$.
In such cases, we can estimate the final abundance of
$\chi$-particles as follows.
When $\eta \ll 1$, the time dependent frequency (\ref{omega22})
can become negative for $\phi~\mbox{\raisebox{-1.ex}{$\stackrel
     {\textstyle<}{\textstyle \sim}$}}~0$ and the $\chi$-particle
can be produced by the negative coupling instability.
However as the $\chi$-particles are produced enough, the second term
in the square bracket of Eq.~(\ref{omega22}) becomes
positive for any value of $\phi$.
We can roughly estimate  that the resonance
finally terminates when $\eta$ grows up to
%%%%%%%%%%%%%%%
\begin{eqnarray}
 \frac53 \eta \approx e^{\frac{\sqrt{6}}{3} \kappa \Phi}-1
\approx \frac{\sqrt{6}}{3} \kappa \Phi
\approx \frac{\sqrt{6}}{3} \kappa
\left(\frac{t_f}{t_0}\right)^{-\frac32} \varphi_0,
\end{eqnarray}
%%%%%%%%%%%%%%%
where we used the relation $\Phi_f \approx
\left(\frac{t_f}{t_0}\right)^{-\frac32}
\varphi_0$. $\Phi$ represents the amplitude of $\phi$.
Namely,
%%%%%%%%%%%%%%%
\begin{eqnarray}
\langle\bar{\chi}^2\rangle_f
\approx \frac{1}{10\xi} \sqrt{\frac{3}{\pi}}
\left(\frac{\tilde{t}_f}{\tilde{t}_0}\right)^{-\frac32}
\bar{\varphi}_0.
\label{etaestimate}
\end{eqnarray}
%%%%%%%%%%%%%%%
Let us adopt the values of $\bar{\varphi}_0=0.05$ and $\tilde{t}_0
=20$. For the $\xi=30$ and $\xi=50$ cases,
$\langle\bar{\chi}^2\rangle_f$ estimated
by Eq.~$(\ref{etaestimate})$ are
$\langle\bar{\chi}^2\rangle_f=5.79 \times 10^{-6}$ and
$\langle\bar{\chi}^2\rangle_f=5.28 \times 10^{-6}$, respectively
(Other examples are listed in Table V).
In spite of the naive estimation of Eq.~$(\ref{etaestimate})$,
it gives fairly close values to the numerical ones.
We did not take into account the $d^2\phi/d\tau^2$ term in
Eq~(\ref{omega22}) for the estimation. This is one of the
reasons why the analytical values are smaller than the numerical
values.
Note that the r.h.s in Eq.~$(\ref{etaestimate})$ can be
regarded as the function of
two variables $\xi$ and $t_f$.
With the increase of $\xi$, numerical calculation shows that
$t_f$ decreases. For
$25~\mbox{\raisebox{-1.ex}{$\stackrel
     {\textstyle<}{\textstyle \sim}$}}~\xi~\mbox{\raisebox{-1.ex}{$\stackrel
     {\textstyle<}{\textstyle \sim}$}}~100$, the decreasing
rate $1/\xi$ of $\langle\bar{\chi}^2\rangle_f$  almost balances
with the increasing rate
$\left(t_f/t_0\right)^{-\frac32}$.
Hence $\langle\bar{\chi}^2\rangle_f$ takes almost constant value
$\langle\bar{\chi}^2\rangle_f \approx (1 \sim 3) \times 10^{-5}$
(See Fig.~10).
To be more precise, $\langle\bar{\chi}^2\rangle_f$ takes the
maximal value $\langle\bar{\chi}^2\rangle_f \approx 3 \times
10^{-5}$ for $\xi \approx 40$.
For $\xi~\mbox{\raisebox{-1.ex}{$\stackrel
     {\textstyle>}{\textstyle\sim}$}}~100$, the decreasing
rate $1/\xi$ surpasses the increasing rate
$\left(\frac{t_f}{t_0}\right)^{-\frac32}$ and
$\langle\bar{\chi}^2\rangle_f$ decreases with the increase of $\xi$.

Next, we consider the $\xi<0$ case for the massless $\chi$-particles.
In this case, since the absolute value of $q$ is represented as
%%%%%%%%%%%%%%%
\begin{eqnarray}
|q|=c' \left( |\xi|+\frac16 \right)
\left(\frac{t}{t_0}\right)^{-\frac32}
\bar{\varphi}_0,
\label{q3}
\end{eqnarray}
%%%%%%%%%%%%%%%
the initial value of $|q|$ is slightly larger than the $\xi>0$ case.
Moreover, since $\omega_k^2$ in Eq.~$(\ref{omega22})$ can take
negative value from the first stage of preheating,
$\langle\bar{\chi}^2\rangle$ grows
rather effectively.
Numerically, while
$\langle\bar{\chi}^2\rangle$ does not increase
for $-3~\mbox{\raisebox{-1.ex}{$\stackrel
     {\textstyle<}{\textstyle \sim}$}}~\xi~\mbox{\raisebox{-1.ex}{$\stackrel
     {\textstyle<}{\textstyle \sim}$}}~0$ at all,
the growth of $\langle\bar{\chi}^2\rangle$
is expected for $\xi~\mbox{\raisebox{-1.ex}{$\stackrel
     {\textstyle<}{\textstyle \sim}$}}~-3$.
For example, in the case of $\xi=-5$, $\langle\bar{\chi}^2\rangle$
increases exponentially and
reaches the final value $\langle\bar{\chi}^2\rangle_f=
1.95 \times 10^{-33}$ at $\tilde{t}_f=100$ (Fig.~11(a)).
At the initial stage, $\langle\bar{\chi}^2\rangle$ decreases
slightly
by the Hubble expansion, and then increases due to negative
instability from $\tilde{t} \approx 10$, which is earlier than
the $\xi>0$ case.
The smaller the value of $\xi$ we adopt,
the larger the final abundance
$\langle\bar{\chi}^2\rangle_f$ becomes,
and it reaches  a plateau
when $\xi \approx -20$.
In the case of $\xi=-20$, the final values are $\tilde{t}_f=156$ and
$\langle\bar{\chi}^2\rangle_f=3.08 \times 10^{-5}$ (Fig.~11(b)).
As $\xi$ further decreases under $\xi \approx -20$, the suppression
effect by $\chi$-particle production begins to work, and
$\langle\bar{\chi}^2\rangle_f$ does not increase while the
growth rate becomes large
as the decrease of $\xi$.
Numerically, $\langle\bar{\chi}^2\rangle_f$ takes almost constant
value $\langle\bar{\chi}^2\rangle_f \approx (2 \sim 5) \times
10^{-5}$ for the range $-100~\mbox{\raisebox{-1.ex}{$\stackrel
     {\textstyle<}{\textstyle \sim}$}}~\xi~\mbox{\raisebox{-1.ex}{$\stackrel
     {\textstyle<}{\textstyle \sim}$}}~-20$.
For example, in the case of $\xi=-50$,
the final values are $\tilde{t}_f=82$ and
$\langle\bar{\chi}^2\rangle_f=
2.18 \times 10^{-5}$ (Fig.~11(c)).
We can estimate $\langle\bar{\chi}^2\rangle_f$ in the
same way
as the $\xi>0$ case, which yields
%%%%%%%%%%%%%%%
\begin{eqnarray}
\langle\bar{\chi}^2\rangle_f
\approx \frac{1}{10|\xi|} \sqrt{\frac{3}{\pi}}
\left(\frac{t_f}{t_0}\right)^{-\frac32}
\bar{\varphi}_0.
\label{etaestimate2}
\end{eqnarray}
%%%%%%%%%%%%%%%
When $\xi=-50$, $\langle\bar{\chi}^2\rangle_f$
estimated by Eq.~$(\ref{etaestimate2})$ is
$\langle\bar{\chi}^2\rangle_f=1.18 \times 10^{-5}$, which
is close to the numerical value
$\langle\bar{\chi}^2\rangle_f=2.18 \times 10^{-5}$.
In the case of $\xi~\mbox{\raisebox{-1.ex}{$\stackrel
     {\textstyle<}{\textstyle \sim}$}}~-100$, the factor $1/|\xi|$  becomes
dominant and $\langle\bar{\chi}^2\rangle_f$ decreases with
the increase of $|\xi|$.
In Table VI, we show the numerical and  analytical values by
Eq.~$(\ref{etaestimate2})$ in various cases.
We also depict the numerical value of
$\langle\bar{\chi}^2\rangle_f$ as a function of $\xi$
in Fig.~10.
We find that
$\langle\bar{\chi}^2\rangle_f$ takes almost constant
value  for $-100~\mbox{\raisebox{-1.ex}{$\stackrel
     {\textstyle<}{\textstyle \sim}$}}~\xi~\mbox{\raisebox{-1.ex}{$\stackrel
     {\textstyle<}{\textstyle \sim}$}}~-20$.
The achieved maximal fluctuation
in the $R^4$ inflation model is $\langle\bar{\chi}^2\rangle_{max}
\approx 5 \times 10^{-5}$ at $\xi \approx -35$, 
which is smaller than that in the
$R^2$ inflation model.

Finally, we take the mass of the $\chi$-particles into account.
In the $R^2$ inflation model, $A_k$ is constrained as $A_k \geq
4\bar{m}_{\chi}^2$ throughout the preheating. In the
$R^4$ inflation model, although $A_k$ increases by the mass term
similarly, it can decrease due to the prefactor $a^{-6/5}$
and the system might give a resonance which is not inefficient
compared with the massless case. However, we have to note that $q$
value decreases by the Hubble expansion simultaneously. If $q$ decreases
faster than $A_k$, the
parametric
resonance is suppressed as in the $R^2$ inflation model.
Unfortunately, this is the case.
As one example, we show the
$\langle\bar{\chi}^2\rangle$ evolution in the case
of $\xi=-50$, $m_{\chi}=m_4$ in Fig.~12.
We find that the growth of  $\langle\bar{\chi}^2\rangle$
is strongly suppressed
by the mass effect.
Although $A_k$ and $q$ in Eq.~$(\ref{Ak})$ and $(\ref{q})$
are not written by the forms which include the term $\eta$,
we can roughly estimate the maximal value of $m_{\chi}$
by using the criterion: $A_k~\mbox{\raisebox{-1.ex}{$\stackrel
     {\textstyle<}{\textstyle \sim}$}}~3q$.
Considering the $k=0$ mode, $\bar{m}_{\chi}$ is constrained as
%%%%%%%%%%%%%%%
\begin{eqnarray}
\bar{m}_{\chi}~\mbox{\raisebox{-1.ex}{$\stackrel
     {\textstyle<}{\textstyle \sim}$}}~
\sqrt{\frac{\xi}{2}}
\left(\frac{\bar{t}}{20}\right)^{-\frac14}.
\label{massconstraint}
\end{eqnarray}
%%%%%%%%%%%%%%%
The above estimation gives a much lower upper bound
than that in the $R^2$ inflation case.
From Eq.~$(\ref{massconstraint})$,
the $\chi$-particle whose mass is of the order $m_{\chi} \sim
m_4 \sim 3 \times 10^{-22} M_{\rm PL}$ can be produced if
$|\xi|~\mbox{\raisebox{-1.ex}{$\stackrel
     {\textstyle>}{\textstyle\sim}$}}~100$.
We conclude that an unnaturally large value of
$|\xi|$ is needed even if GeV scale bosons are to be
 produced.

%%%%%%%%%%%%%%%%%%%%%%%%%%%%%%%%%%%%%%%%%%%%%%%%%%
%
%                                                 %
\section{conclusions and discussions}                            %
%                                                 %
%%%%%%%%%%%%%%%%%%%%%%%%%%%%%%%%%%%%%%%%%%%%%%%%%%
In this paper, we analyze preheating with non-minimally coupled
scalar field $\chi$ in the higher-curvature inflation model.
We have examined  properties of resonance, especially
for $R^2$- and $R^4$-inflation models.
In the $R^2$ model, $R^2$-term provides us an effective
inflaton field $\phi$, and  inflation is realized through the
flat plateau for $\phi~\mbox{\raisebox{-1.ex}{$\stackrel
     {\textstyle>}{\textstyle\sim}$}}~M_{\rm PL}$.
In the reheating phase, the $\phi$ field
behaves as a massive scalar field.
Although the evolutions of scale factor and $\phi$ field 
in the present model are
almost the same as those for a massive inflaton with 
$V(\phi)=\frac12 m^2\phi^2$ in the non-minimal coupling 
$\frac12 \xi R\chi^2$
case, the structure of resonance is different.
In the $R^2$ model, since the resonance parameter $q$ decreases as
$1/t$ due to the expansion of the universe, which is slower than 
that in the model with
$V(\phi, \chi)=\frac12 m^2\phi^2+\frac12 \xi R\chi^2$,
we do not need so large a value of $|\xi|$ as $|\xi|~\mbox{\raisebox{-1.ex}{$\stackrel
     {\textstyle>}{\textstyle\sim}$}}~10$
for an effective resonance.
In the $\xi>0$ case with a massless $\chi$-particle,
if $\xi$ is larger than $\xi~\mbox{\raisebox{-1.ex}{$\stackrel
     {\textstyle>}{\textstyle\sim}$}}~1$,
$\chi$-particle production is possible against the diluting
effect by cosmic expansion.
When $1~\mbox{\raisebox{-1.ex}{$\stackrel
     {\textstyle<}{\textstyle \sim}$}}~\xi~\mbox{\raisebox{-1.ex}{$\stackrel
     {\textstyle<}{\textstyle \sim}$}}~5$, the total amount of created 
$\chi$-particles
$\langle\chi^2\rangle_f$ grows
with the increase of $\xi$, because the initial
value of $q$ becomes larger.
However, for $\xi~\mbox{\raisebox{-1.ex}{$\stackrel
     {\textstyle>}{\textstyle\sim}$}}~5$, $\langle\chi^2\rangle_f$
slowly decreases as the increase of $\xi$
because of the suppression effect by $\chi$-particle
production.
Hence, in the positive $\xi$ case,
$\langle\chi^2\rangle_f$ takes the maximal value
$\sqrt{\langle\chi^2\rangle}_{max} \approx 1.5 \times
10^{17}$ GeV at $\xi \approx 5$.
When $\xi$ is negative, the resonance structure does not
change so much except for the small increase of the 
$|q|$ value.
In this case, the $\chi$-particle production becomes possible
for $\xi~\mbox{\raisebox{-1.ex}{$\stackrel
     {\textstyle<}{\textstyle \sim}$}}~-1$.
For $-4~\mbox{\raisebox{-1.ex}{$\stackrel
     {\textstyle<}{\textstyle \sim}$}}~\xi~\mbox{\raisebox{-1.ex}{$\stackrel
     {\textstyle<}{\textstyle \sim}$}}~-1$, $\langle\chi^2\rangle_f$
increases with the decrease of $\xi$, and $\langle\chi^2\rangle_f$
takes maximal value $\sqrt{\langle\chi^2\rangle}_{max}
\approx 2 \times 10^{17}$ GeV at $\xi \approx -4$.
For the case of $\xi~\mbox{\raisebox{-1.ex}{$\stackrel
     {\textstyle<}{\textstyle \sim}$}}~-20$, a $\chi$-particle fluctuation
rapidly grows with the passage of time and reaches the maximal
value $\langle\chi^2\rangle_f$ before the $\phi$-field begins to
oscillate sinusoidally.
This means that in the $\xi<0$ case, the effective $\chi$-particle
production occurs due to a negative coupling instability
irrespective of the coherent oscillation of $\phi$-field.
Numerically, $\langle\chi^2\rangle_f$ slowly increases
for $-40~\mbox{\raisebox{-1.ex}{$\stackrel
     {\textstyle<}{\textstyle \sim}$}}~\xi~\mbox{\raisebox{-1.ex}{$\stackrel
     {\textstyle<}{\textstyle \sim}$}}~-20$ with the decrease of $\xi$,
and the value of $\langle\chi^2\rangle_f$ in
the $\xi \approx -40$ case
is almost the same as the $\xi \approx -4$ case.
In summary, in the $R^2$ model, the achieved maximal value
of $\chi$-particle fluctuation is
$\sqrt{\langle\chi^2\rangle}_{max} \approx 2 \times 10^{17}$ GeV
for $\xi \approx -4$ (and $\xi \approx -40$).
As for the massive $\chi$-particle case, a parametric resonance
is suppressed by the mass effect.
We find the relation for the massive $\chi$-particle
production to occur: $\bar{m}_{\chi}~\mbox{\raisebox{-1.ex}{$\stackrel
     {\textstyle<}{\textstyle \sim}$}}~\sqrt{\xi}$
for $\xi>0$ and $\bar{m}_{\chi}~\mbox{\raisebox{-1.ex}{$\stackrel
     {\textstyle<}{\textstyle \sim}$}}~2\sqrt{|\xi|}$ for $\xi<0$.
Hence it is difficult to produce GUT scale gauge bosons
unless $|\xi|$ is as enormously large as $|\xi|~\mbox{\raisebox{-1.ex}{$\stackrel
     {\textstyle>}{\textstyle\sim}$}}~10^5$.

As for the $R^4$ model, inflation is difficult to realize
unless selecting the very large coupling constant $\alpha_4$
with spacetime curvature.
Nevertheless, we have examined the resonance structure, because
we are interested in whether a preheating stage generally 
exists or not for $n>2$.
Since the potential of the $\phi$ field is written as
$V(\phi) \sim \phi^{\frac43}$ for
$\phi~\mbox{\raisebox{-1.ex}{$\stackrel
     {\textstyle<}{\textstyle \sim}$}}~M_{\rm PL}/20$, the oscillation of the $\phi$ field is not exactly sinusoidal.
However, we find that the behavior of $\phi$ field is
well approximated as a sinusoidal oscillation, and the
equation of the $\chi$ field is reduced to the Mathieu
equation.
In the $R^4$ case, the resonance structure changes
and $q$ decreases as $q \sim t^{-\frac32}$, which is faster 
than the $R^2$ case.
For $\xi~\mbox{\raisebox{-1.ex}{$\stackrel
     {\textstyle>}{\textstyle\sim}$}}~5$ and $\xi~\mbox{\raisebox{-1.ex}{$\stackrel
     {\textstyle<}{\textstyle \sim}$}}~-3$, the fluctuation of
$\chi$-particles grows with the passage of time.
However, $\langle\bar{\chi}^2\rangle_f$ is rather
small for $|\xi|~\mbox{\raisebox{-1.ex}{$\stackrel
     {\textstyle<}{\textstyle \sim}$}}~20$, because a parametric resonance
is not so effective.
For $20~\mbox{\raisebox{-1.ex}{$\stackrel
     {\textstyle<}{\textstyle \sim}$}}~|\xi|~\mbox{\raisebox{-1.ex}{$\stackrel
     {\textstyle<}{\textstyle \sim}$}}~100$, $\langle\chi^2\rangle_f$
takes the almost constant value  $\sqrt{\langle\chi^2\rangle}_f
 \approx (4 \sim 8) \times10^{16}$ GeV, and for $|\xi|>100$,
$\langle\chi^2\rangle_f$ decreases with the increase of $|\xi|$.
The maximum value achieved in the $R^4$ model is
$\sqrt{\langle\chi^2\rangle}_f \approx 8 \times 10^{16}$ GeV
at $\xi \approx -35$.
Taking into account the mass effect of $\chi$-particles,
the final fluctuation is more strongly suppressed
as compared with the $R^2$ inflation model.

As for $R^n (n>4)$ model, the coupling constant $\alpha_n$ is
constrained to be very large as Eq.~$(\ref{constraint})$.
Although this model is rather unnatural, we should stress
that it is still possible to investigate the resonance
structure by the same method as in the $n=4$ case.
The oscillation of the $\phi$ field around the potential
Eq.~$(\ref{potential})$ is approximately written by the 
sinusoidal form, so the fluctuation of $\chi$-particles 
can grow by a parametric resonance.
Hence, in this sense, a preheating stage generically exists
in thw higher-curvature inflation model with
a non-minimal coupling, 
irrespective of the problem of the large coupling constant.

In a higher-curvature inflation model,
inflation is realized through the coupling
with spacetime curvature.
Because of this, the inflaton field is the product of a 
purely gravitational source.
Hence for the complete study of preheating, we should take into account
the metric perturbation.
It was pointed out in Ref.
\cite{mperturbation1,mperturbation2,mperturbation3,mperturbation4,mperturbation5,mperturbation6}
that the metric perturbation
undergoes the effect of parametric resonance in a reheating phase.
Although it is rather difficult to study preheating in the present model
by taking into account metric perturbations, it is of interest how the
preheating dynamics are modified  by this effect.

In our study of amplification of a scalar field $\chi$ coupled
non-minimally to a spacetime
curvature $R$ ($\frac12\xi R \chi^2$) in a higher-curvature inflation
model ($R+\alpha_n R^n$),
we have found that the value of $\xi$ does not need to be so large
as in the case of massive inflaton plus non-minimally
coupled scalar field $\chi$ to have an effective resonance.
In other gravitational theories such as the Brans-Dicke
theory\cite{BD} and  the induced gravity\cite{induced},
we may have different types of preheating dynamics.
Recently, Mazumdar and Mendes\cite{BDnew} studied preheating
in the Brans-Dicke theory with a model of massive inflaton plus a
minimally coupled scalar field $\sigma$ and found that
the Brans-Dicke field as well as the $\sigma$ field undergoes
the effect of parametric resonance.
It is of interest to investigate the amplification of
a non-minimally coupled scalar field in Generalized Einstein
Theories.
Furthermore, although we mainly concentrate
on the $\chi$-particle production in this paper, we may
discuss an inflaton particle production
with a model
of some inflaton potential and non-minimally coupled
scalar field $\phi$ as well.
These issues are under consideration.

%%%%%%%%%%%%%%%%%%%%%%%%%%%%%%%%%%%%%%%%%%%%%%%%%%
\section*{ACKOWLEDGEMENTS}
T. T. is thankful for financial support from the JSPS. This
work was partially supported by a Grant-in-Aid for  Scientific
Research Fund of the Ministry of Education, Science and Culture
(No. 09410217 and Specially Promoted Research No. 08102010), by a
JSPS Grant-in-Aid
(No. 094162), and by the Waseda University Grant for Special
Research Projects.

\newpage
%%%%%%%%%%%%%%%%%%%%%%%%%%%%%%%%%%%%%%%
%%%%%%%%%%%%%%%%%%%%%%%%%%%%%%%%%%%%%%%

\newpage

\begin{table}
\caption{The final values $\langle \bar{\chi}^2 \rangle_{f}$
obtained by the analytical estimation and by the numerical
calculation for $\xi>0, m_{\chi}=0$ in the case of the $R^2$
inflation model.
The analytical estimation  gives a good approximation
to the numerical results.
$\langle \bar{\chi}^2 \rangle_{numerical}$ takes
the maximal value  at $\xi \sim 5$. We also show the time
$\bar{t}_f$ when the resonance stops and the value of
$36\xi\eta$ which indicates a suppression effect.
For large values of $\xi$, the suppression effect by $\chi$-
particle production is significant.
}

\vskip .3cm
\noindent
\begin{tabular}{crc|cc|ccc|lc}
 ~& $\xi$ & ~& $\bar{t}_{f} $ &~&
         $\langle \bar{\chi}^2 \rangle_{analytic}$
         & $\langle \bar{\chi}^2 \rangle_{numerical}$
   &~& $36\xi \eta_{numerical}$ &~\\
        \hline
 ~&3 & ~& 5.61 & ~&--- & $2.21 \times 10^{-8}$ &~ & $1.80\times 10^{-4}$
&~\\
 ~& 5 & ~& 6.27 & ~& $1.24 \times 10^{-4}$ & $1.74 \times 10^{-4}$  &~ &
3.94  &~\\
 ~&10 & ~& 4.28 & ~& $9.02 \times 10^{-5}$ & $9.95 \times 10^{-5}$  &~ &
9.00  &~\\
~&30 & ~& 2.15 &~&
$5.67 \times 10^{-5}$   & $7.50 \times 10^{-5}$  &~ &
 $6.10 \times 10^{1}$ &~\\
 ~&50 & ~& 1.30 &~& $5.20 \times 10^{-5}$    & $2.90 \times
10^{-5}$ &~ & $6.56 \times 10^1$ &~\\
 ~&100 & ~&1.15 &~& $2.87 \times10^{-5}$  & $2.51 \times10^{-5}$
&~ & $2.27 \times 10^2$   &~\\
 ~&500 & ~&1.06 &~& $6.13 \times 10^{-6}$ & $7.96 \times 10^{-6}$
&~& $1.80 \times 10^3$  &~
\end{tabular}
\end{table}
%%%%%%%%%%%%%%%%
\vspace{1cm}
\begin{table}
\caption{The final values $\langle \bar{\chi}^2 \rangle_{f}$
obtained by
the analytical estimation and by the numerical
calculation for $\xi>0, m_{\chi} \ne 0$
in the case of the $R^2$ inflation model.
The mass effect suppresses the final $\chi$-particle
abundance, and massive $\chi$-particle which does not
satisfy the condition Eq.~(3.14)
can not be created.
}

\vskip .3cm
\noindent
\begin{tabular}{crc|cc|ccc|lc}
 ~& $\xi$ & ~& $\bar{m}_{\chi} $ &~&
         $\langle \bar{\chi}^2 \rangle_{analytic}$
         & $\langle \bar{\chi}^2 \rangle_{numerical}$
   &~& $\bar{t}_f$ &~\\
        \hline
 ~&5 & ~& 1 & ~&--- &$3.80 \times 10^{-12}$ &~ & 6.27 &~\\
 ~& 10 & ~& 1 & ~&--- & $1.17 \times 10^{-5}$  &~ & 9.22 &~\\
 ~&20 & ~& 1 & ~& $4.80 \times 10^{-5}$ & $5.19 \times 10^{-5}$  &~ &
3.23  &~\\
~&50 & ~& 2 &~& $2.80 \times 10^{-5}$   & $2.51 \times 10^{-5}$
 &~ & 2.08 &~\\
 ~&100 & ~& 3 &~& $2.36 \times 10^{-5}$    & $2.32 \times
10^{-5}$ &~ & 1.21 &~\\
 ~&500 & ~&10 &~& $3.87 \times 10^{-6}$ & $3.80 \times 10^{-6}$
&~& 1.18  &~
\end{tabular}
\end{table}
%%%%%%%%%%%%%%%%
\vspace{1cm}
\begin{table}
\caption{The final values $\langle \bar{\chi}^2 \rangle_{f}$
obtained by the analytical estimation and by the numerical
calculation for $\xi<0, m_{\chi}=0$
in the case of the $R^2$ inflation model.
The analytical estimation  gives a good approximation
to the numerical results.
We also show the time
$\bar{t}_f$ when the resonance stops and the value of
$36\xi\eta$ which indicates a suppression effect.
$\langle \bar{\chi}^2 \rangle_{numerical}$ takes
the maximal value at
$\xi\approx -4$ and $\xi \approx -40$.
}

\vskip .3cm
\noindent
\begin{tabular}{crc|cc|ccc|lc}
 ~& $\xi$ & ~& $\bar{t}_{f} $ &~&
         $\langle \bar{\chi}^2 \rangle_{analytic}$
         & $\langle \bar{\chi}^2 \rangle_{numerical}$
   &~& $36\xi \eta_{numerical}$ &~\\
        \hline
 ~&-3 & ~& 7.02 & ~&--- & $8.71 \times 10^{-5}$ &~ & 0.71 &~\\
 ~& -5 & ~& 3.80 & ~& $2.51 \times 10^{-4}$ & $2.63 \times 10^{-4}$  &~ &
5.95  &~\\
 ~&-10 & ~& 2.78  & ~& $1.74 \times 10^{-4}$ & $1.78 \times 10^{-4}$  &~ &
$1.61 \times 10^1$  &~\\
~&-30 & ~& 0.70 &~&
$1.71 \times 10^{-4}$   & $2.79 \times 10^{-4}$  &~ &
 $2.27 \times 10^2$ &~\\
 ~&-50 & ~& 0.58 &~& $1.83 \times 10^{-4}$    & $2.82 \times
10^{-4}$ &~ & $6.38 \times 10^2$ &~\\
 ~&-100 & ~&0.58 &~& $9.13 \times10^{-5}$  & $1.22 \times10^{-4}$
&~ & $1.10 \times 10^3$   &~\\
 ~&-500 & ~&0.37 &~& $3.19 \times 10^{-5}$ & $5.31 \times 10^{-5}$
&~& $1.20 \times 10^4$  &~
\end{tabular}
\end{table}
%%%%%%%%%%%%%%%%
\vspace{1cm}
\begin{table}
\caption{The final values $\langle \bar{\chi}^2 \rangle_{f}$
obtained by
the analytical estimation and by the numerical
calculation for $\xi<0, m_{\chi} \ne 0$
in the case of the $R^2$ inflation model.
The mass effect suppresses the final $\chi$-particle
abundance, and massive $\chi$-particle which does not
satisfy the condition Eq.~(3.22)
can not be created.
}

\vskip .3cm
\noindent
\begin{tabular}{crc|cc|ccc|lc}
 ~&  $\xi$ &~& $m_{\chi}$ &~
       &$\langle \bar{\chi}^2 \rangle_{analytic}$
         & $\langle \bar{\chi}^2 \rangle_{numerical}$
         &~& $\bar{t}_f$ &\\
        \hline
~&-5 & ~& 1 & ~&--- &$3.24 \times 10^{-8}$ &~ & 7.74 &~\\
 ~& -10 & ~& 1 & ~& $1.37 \times 10^{-4}$ & $1.13 \times 10^{-4}$  &~ &
2.69 &~\\
 ~&-30 & ~& 3 & ~& $9.46 \times 10^{-5}$ & $1.58 \times 10^{-4}$  &~ &
0.73  &~\\
~&-50 & ~& 7 &~& $3.45 \times 10^{-5}$   & $5.01 \times 10^{-5}$
 &~ & 0.67 &~\\
 ~&-100 & ~& 10 &~& $6.75 \times 10^{-5}$    & $8.67 \times
10^{-5}$ &~ & 0.52 &~\\
 ~&-500 & ~&30 &~& $1.59 \times 10^{-5}$ & $8.91 \times 10^{-6}$
&~& 0.37  &~
\end{tabular}
\end{table}
%%%%%%%%%%%%%%%%
%%%%%%%%%%%%%%%%
\vspace{1cm}
\begin{table}
\caption{The final values $\langle \bar{\chi}^2 \rangle_{f}$
obtained by
the analytical estimation and by the numerical
calculation for $\xi>0, m_{\chi}=0$
in the case of the $R^4$ inflation model.
}

\vskip .3cm
\noindent
\begin{tabular}{crc|cc|ccc}
 ~&  $\xi$ &~& $\bar{t}_f$ &~
       &$\langle \tilde{\chi}^2 \rangle_{analytic}$
         & $\langle \bar{\chi}^2 \rangle_{numerical}$
         &~ \\
        \hline
~&10 & ~& 161 & ~&--- &$5.89 \times 10^{-27}$ &~\\
 ~&30 & ~& 185 & ~& $5.79 \times 10^{-6}$ & $1.29 \times 10^{-5}$  &~\\
~&50 & ~& 140 & ~& $5.28 \times 10^{-6}$ & $8.28 \times 10^{-6}$  &~\\
 ~&70 & ~& 112 &~& $5.27 \times 10^{-6}$   & $1.09 \times 10^{-5}$
 &~\\
 ~&100 & ~& 88 &~& $5.29 \times 10^{-6}$    & $1.06 \times
10^{-5}$ &~\\
 ~&300 & ~&59 &~& $3.21 \times 10^{-6}$ & $5.37 \times 10^{-6}$
&~
\end{tabular}
\end{table}
%%%%%%%%%%%%%%%%
%%%%%%%%%%%%%%%%
\vspace{1cm}
\begin{table}
\caption{The final values $\langle \bar{\chi}^2 \rangle_{f}$
obtained by
the analytical estimation and by the numerical
calculation for $\xi<0, m_{\chi}=0$
in the case of the $R^4$ inflation model.
}

\vskip .3cm
\noindent
\begin{tabular}{crc|cc|ccc}
 ~&  $\xi$ &~& $\bar{t}_f$ &~
       &$\langle \tilde{\chi}^2 \rangle_{analytic}$
         & $\langle \bar{\chi}^2 \rangle_{numerical}$
         &~ \\
        \hline
~&-20 & ~& 156 & ~&$1.13 \times 10^{-5}$ &$3.08 \times 10^{-5}$ &~\\
 ~& -30 & ~& 130 & ~& $9.80 \times 10^{-6}$ & $2.24 \times 10^{-5}$  &~\\
~&-50 & ~& 82 & ~& $1.18 \times 10^{-5}$ & $2.18 \times 10^{-6}$  &~\\
~&-70 & ~& 65 &~& $1.19 \times 10^{-5}$   & $2.12 \times 10^{-5}$
 &~\\
 ~&-100 & ~& 51 &~& $1.20 \times 10^{-5}$    & $2.30 \times
10^{-5}$ &~\\
 ~&-300 & ~&36 &~& $6.74 \times 10^{-6}$ & $5.37 \times 10^{-6}$
&~
\end{tabular}
\end{table}
%%%%%%%%%%%%%%%%

\newpage
%%%%%%%%%%%%%%%%%%%%
%   figures
%%%%%%%%%%%%%%%%%%%%
\begin{flushleft}
{ Figure Captions}
\end{flushleft}
\noindent
%%%%%%%%
\parbox[t]{2cm}{FIG. 1:\\~}\ \
\parbox[t]{12cm}
{The potential in the equivalent system to the $R^n$
gravity thoery.  The solid  and dotted curve denote
the $n=2$ and $n=4$ cases respectively.
The potential of the $n=2$ case is flat enough for
$\phi~\mbox{\raisebox{-1.ex}{$\stackrel
     {\textstyle>}{\textstyle\sim}$}}~M_{\rm PL}$ so that inflation occurs, but for the $n>2$ case,
the potential has a local maximum at
Eq.~(2.9) and inflation is rather difficult to realize.
}\\[1em]
\noindent
%%%%%%%%
\parbox[t]{2cm}{FIG. 2:\\~}\ \
\parbox[t]{12cm}
{The schematic diagram of the Mathieu chart and the typical
paths for two types of resonance.  The lined regions
denote  the  instability bands.
Generally, The growth of $\chi$ fluctuation
terminates in two distinct ways.
In the first case, when $|\xi|$ is not so large, resonance ceases
by universe expansion before $\chi$-particles are
sufficiently produced.
In this case, the increase of $A_k$ due to $\chi$-particle
production is negligible, and resonance stops when $q$ drops
down under unity (solid curve).
In the second case, when $|\xi|$ is as large as $|\xi|~
\mbox{\raisebox{-1.ex}{$\stackrel {\textstyle>}{\textstyle\sim}$}}~10$,
the growth of $\chi$ fluctuation stops because of the
increase of $A_k$ due to $\chi$-particle production
(dotted curve).
In this case, resonance ends when the $\chi$ field reaches 
the line $A_k \approx 3q$ because few instability bands exist
for $A_k~\mbox{\raisebox{-1.ex}{$\stackrel
     {\textstyle>}{\textstyle\sim}$}}~3q$.
 }\\[1em]
\noindent
%%%%%%%%
%%%%%%%%
\parbox[t]{2cm}{FIG. 3:\\~}\ \
\parbox[t]{12cm}
{The evolution of $\langle \bar{\chi}^2 \rangle$ as a function of
$\bar{t}$ for $\xi>0, m_{\chi}=0$ in the case of
the $R^2$ inflation model ((a)~$\xi=3$,
(b)~$\xi=5$, and (c)~$\xi=10$).
We find that the
parametric resonance occurs by the positive $\xi$-coupling.
$\langle \bar{\chi}^2 \rangle$ increases
quasi-exponentially and reaches  its final value
$\langle \bar{\chi}^2 \rangle_f$. After then, it decreases by the
adiabatic expansion.
$\langle \bar{\chi}^2 \rangle_f$ takes the maximal value when
$\xi \approx 5$. For larger $\xi$ ($\xi~\mbox{\raisebox{-1.ex}{$\stackrel
     {\textstyle>}{\textstyle\sim}$}}~100$), although
the growth rate becomes large, the final value
$\langle \bar{\chi}^2 \rangle_f$ is suppressed by
$\eta$ term.
}\\[1em]
\noindent
%%%%%%%%
%%%%%%%%
\parbox[t]{2cm}{FIG. 4:\\~}\ \
\parbox[t]{12cm}
{The final value of $\langle \bar{\chi}^2 \rangle$ as a function
of $\xi$ for $m_{\chi}=0$ in the case of the $R^2$ inflation model.
When $\xi$ is positive, $\langle \bar{\chi}^2 \rangle_f$
takes its maximal value  $\langle \bar{\chi}^2 \rangle_f
=1.74 \times 10^{-4}$ when $\xi \approx 5$.
For $\xi<0$, $\langle \bar{\chi}^2 \rangle_f$
takes its maximal value $\langle\bar{\chi}^2\rangle_{max}
\approx 3 \times 10^{-4}$
when  $\xi \approx -4$ and $\xi \approx -40$,
which is  larger than the $\xi>0$ case.
}\\[1em]
\noindent
%%%%%%%%
%%%%%%%%
\parbox[t]{2cm}{FIG. 5:\\~}\ \
\parbox[t]{12cm}
{The evolution of $\langle \bar{\chi}^2 \rangle$ as a function of
$\bar{t}$ for $\xi>0, m_{\chi} = m$ in the case of the 
$R^2$ inflation model ((a)~$\xi=5$,
(b)~$\xi=10$, and (c)~$\xi=20$).
The growth of $\langle \bar{\chi}^2 \rangle$ is suppressed by
the mass of $\chi$-particles, especially when $\xi$ is small.
However, massive $\chi$-particles of the order
$m_{\chi} \sim 10^{13}$ GeV can be created sufficiently when
$\xi~\mbox{\raisebox{-1.ex}{$\stackrel
     {\textstyle>}{\textstyle\sim}$}}~10$.
}\\[1em]
\noindent
%%%%%%%%
%%%%%%%%
\parbox[t]{2cm}{FIG. 6:\\~}\ \
\parbox[t]{12cm}
{The final value of $\langle \bar{\chi}^2 \rangle$ as a function
of $m_{\chi}$ for $\xi=100$ in the case of the $R^2$ inflation model.
$\langle \bar{\chi}^2 \rangle_f$ is a quadratically decreasing
function of $m_{\chi}$. $\chi$-particles whose mass is
$m_{\chi}~\mbox{\raisebox{-1.ex}{$\stackrel
     {\textstyle>}{\textstyle\sim}$}}~10m \sim 10^{14}$ GeV can not be produced,
as is found by Eq.~$(\ref{E14})$.
}\\[1em]
\noindent
%%%%%%%%
%%%%%%%%
\parbox[t]{2cm}{FIG. 7:\\~}\ \
\parbox[t]{12cm}
{The evolution of $\langle \bar{\chi}^2 \rangle$ as a function of
$\bar{t}$ for $\xi<0, m_{\chi}=0$ in the case of the 
$R^2$ inflation model ((a)~$\xi=-1$,
(b)~$\xi=-3$, and (c)~$\xi=-5$).
Since the value of $|q|$ is larger as compared with the same
absolute value of $\xi$, parametric resonance occurs
even if $\xi=-1$.
The value of $\langle \bar{\chi}^2 \rangle_f=
2.63 \times 10^{-4}$ in the $\xi=-5$ case is larger than the maximal
value $\langle \bar{\chi}^2 \rangle_f=
1.74 \times 10^{-4}$ in the positive $\xi$ case.
In the negative $\xi$ case, $\langle \bar{\chi}^2 \rangle_f$
takes maximal value $\langle \bar{\chi}^2 \rangle_f \approx
3 \times 10^{-4}$ at $\xi \approx -4$ and $\xi \approx -40$.
}\\[1em]
\noindent
%%%%%%%%
%%%%%%%%
\parbox[t]{2cm}{FIG. 8:\\~}\ \
\parbox[t]{12cm}
{The evolution of $\langle \bar{\chi}^2 \rangle$ as a function of
$\bar{t}$ for $\xi<0, m_{\chi}=m$ in the case of the 
$R^2$ inflation model ((a)~$\xi=-5$ and
(b)~$\xi=-10$).
When $\xi=-5$, the mass effect significantly suppresses
 the final abundance.
However, when $\xi=-10$, the mass effect of the order $m_{\chi}=m$
does not play a relevant role.
We find that massive $\chi$-particles of order
$m_{\chi} \sim 10^{13}$ GeV can be created sufficiently when
$\xi~\mbox{\raisebox{-1.ex}{$\stackrel
     {\textstyle<}{\textstyle \sim}$}}~-10$.
 }\\[1em]
\noindent
%%%%%%%%
%%%%%%%%
\parbox[t]{2cm}{FIG. 9:\\~}\ \
\parbox[t]{12cm}
{The evolution of $\langle \bar{\chi}^2 \rangle$ as a function of
$\tilde{t}$ for $\xi>0, m_{\chi}=0$ in the case of the 
$R^4$ inflation model ((a)~$\xi=10$,
(b)~$\xi=30$ and (c)~$\xi=50$).
We find that the
parametric resonance occurs by the positive $\xi$-coupling.
Since the initial fluctuation is small, the final value of
$\langle \bar{\chi}^2 \rangle$ is rather small for $\xi=10$.
However, in the case of $\xi=30$ and $\xi=50$,
the final value of $\langle \bar{\chi}^2 \rangle$ reaches 
the value $\langle \bar{\chi}^2 \rangle_f \sim 10^{-5}$.
For $\xi~\mbox{\raisebox{-1.ex}{$\stackrel
     {\textstyle>}{\textstyle\sim}$}}~25$, $\langle \bar{\chi}^2 \rangle_f$ does
not increase because of the suppression effect due to
$\chi$-particle production.
As a result, in the $\xi>0$ case,
$\langle \bar{\chi}^2 \rangle_f$ takes almost constant
value: $(1 \sim 3) \times 10^{-5}$ for $25~\mbox{\raisebox{-1.ex}{$\stackrel
     {\textstyle<}{\textstyle \sim}$}}~\xi~\mbox{\raisebox{-1.ex}{$\stackrel
     {\textstyle<}{\textstyle \sim}$}}~100$.
}\\[1em]
\noindent
%%%%%%%%
%%%%%%%%
\parbox[t]{2cm}{FIG. 10:\\~}\ \
\parbox[t]{12cm}
{The final value of $\langle \bar{\chi}^2 \rangle$ as a function
of $\xi$ for $m_{\chi}=0$ in the case of the $R^4$ inflation model.
$\langle \bar{\chi}^2 \rangle_f$
takes almost constant value  $\langle \bar{\chi}^2 \rangle_f
\approx (1 \sim 5) \times 10^{-5}$ for $20~\mbox{\raisebox{-1.ex}{$\stackrel
     {\textstyle<}{\textstyle \sim}$}}~|\xi|~\mbox{\raisebox{-1.ex}{$\stackrel
     {\textstyle<}{\textstyle \sim}$}}~100$.
The maximal value of $\langle \bar{\chi}^2 \rangle_f$ is
$\langle \bar{\chi}^2 \rangle_{max} \approx 5 \times 10^{-5}$
at $\xi \approx -35$.
}\\[1em]
\noindent
%%%%%%%%
%%%%%%%%
\parbox[t]{2cm}{FIG. 11:\\~}\ \
\parbox[t]{12cm}
{The evolution of $\langle \bar{\chi}^2 \rangle$ as a function of
$\tilde{t}$ for $\xi<0, m_{\chi}=0$ in the case of the 
$R^4$ inflation model ((a)~$\xi=-5$, (b)~$\xi=-20$
(c)~$\xi=-50$ ).
When $\xi=-5$, parametric resonance evidently occurs, but
the final value of $\langle \bar{\chi}^2 \rangle$ is rather small.
However, in the case of $\xi~\mbox{\raisebox{-1.ex}{$\stackrel
     {\textstyle<}{\textstyle \sim}$}}~-20$,
$\chi$-particles are sufficiently produced.
For $-100~\mbox{\raisebox{-1.ex}{$\stackrel
     {\textstyle<}{\textstyle \sim}$}}~\xi~\mbox{\raisebox{-1.ex}{$\stackrel
     {\textstyle<}{\textstyle \sim}$}}~-20$, $\langle \bar{\chi}^2 \rangle_f$ takes
almost constant value $\langle \bar{\chi}^2 \rangle_f
\approx (2 \sim 5) \times 10^{-5}$.
}\\[1em]
\noindent
%%%%%%%%
%%%%%%%%
\parbox[t]{2cm}{FIG. 12:\\~}\ \
\parbox[t]{12cm}
{The evolution of $\langle \bar{\chi}^2 \rangle$ as a function of
$\tilde{t}$ for $\xi=-50, m_{\chi}=m_4$ in the case of the 
$R^4$ inflation model.
As compared with the massless case, $\langle \bar{\chi}^2 \rangle$
is strongly suppressed by the mass effect.
Generally, massive $\chi$-particles which do not satisfy the
condition Eq.~(4.13) can not be created.
 }\\[1em]
\noindent
%%%%%%%%

\newpage

\begin{figure}
\begin{center}
\psbox[hscale=1.0,vscale=1.0]{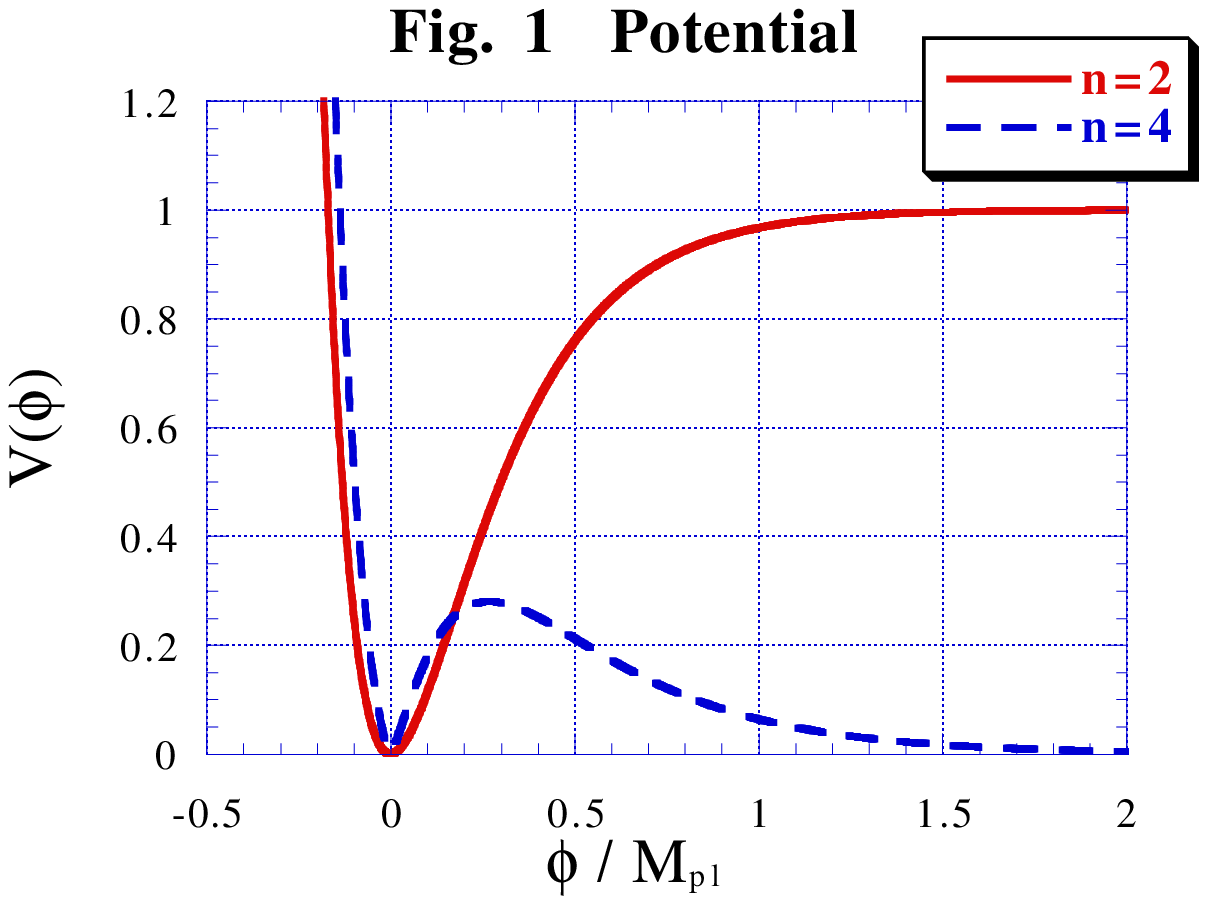}
\end{center}
\end{figure}

\newpage
\begin{figure}
\begin{center}
\psbox[hscale=0.8,vscale=0.8]{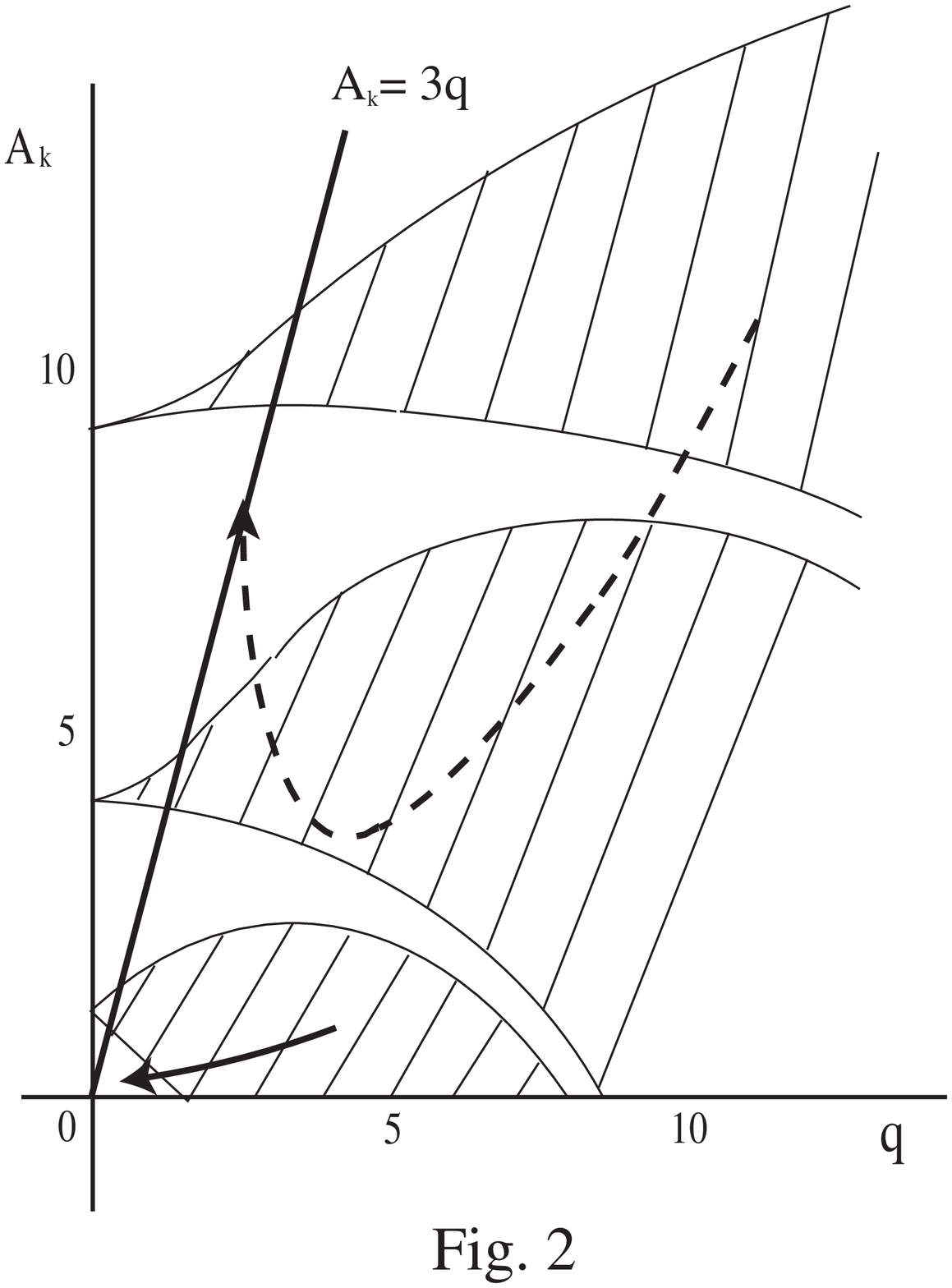}
\end{center}
\end{figure}

\newpage
\begin{figure}
\begin{center}
\psbox[hscale=0.7,vscale=0.7]{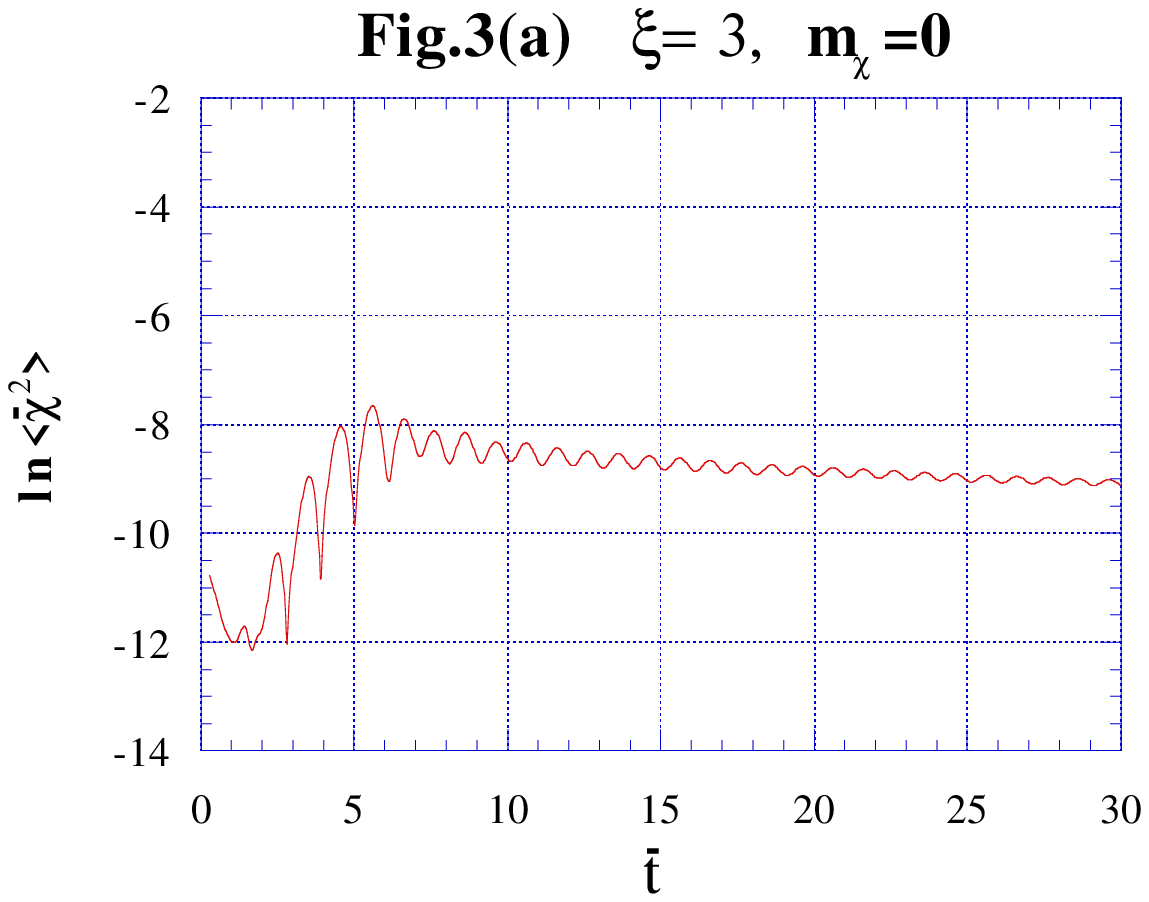}
\end{center}
\end{figure}

\begin{figure}
\begin{center}
\psbox[hscale=0.7,vscale=0.7]{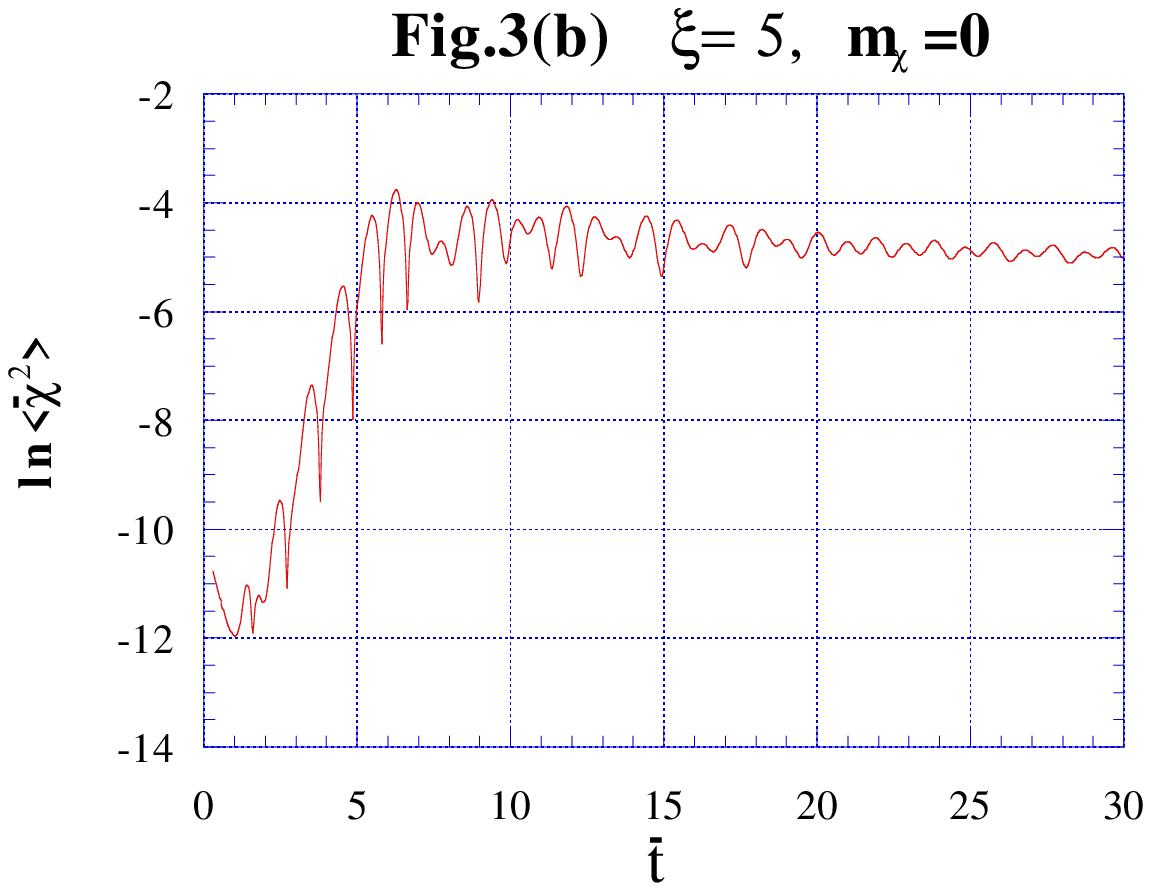}
\end{center}
\end{figure}

\begin{figure}
\begin{center}
\psbox[hscale=0.7,vscale=0.7]{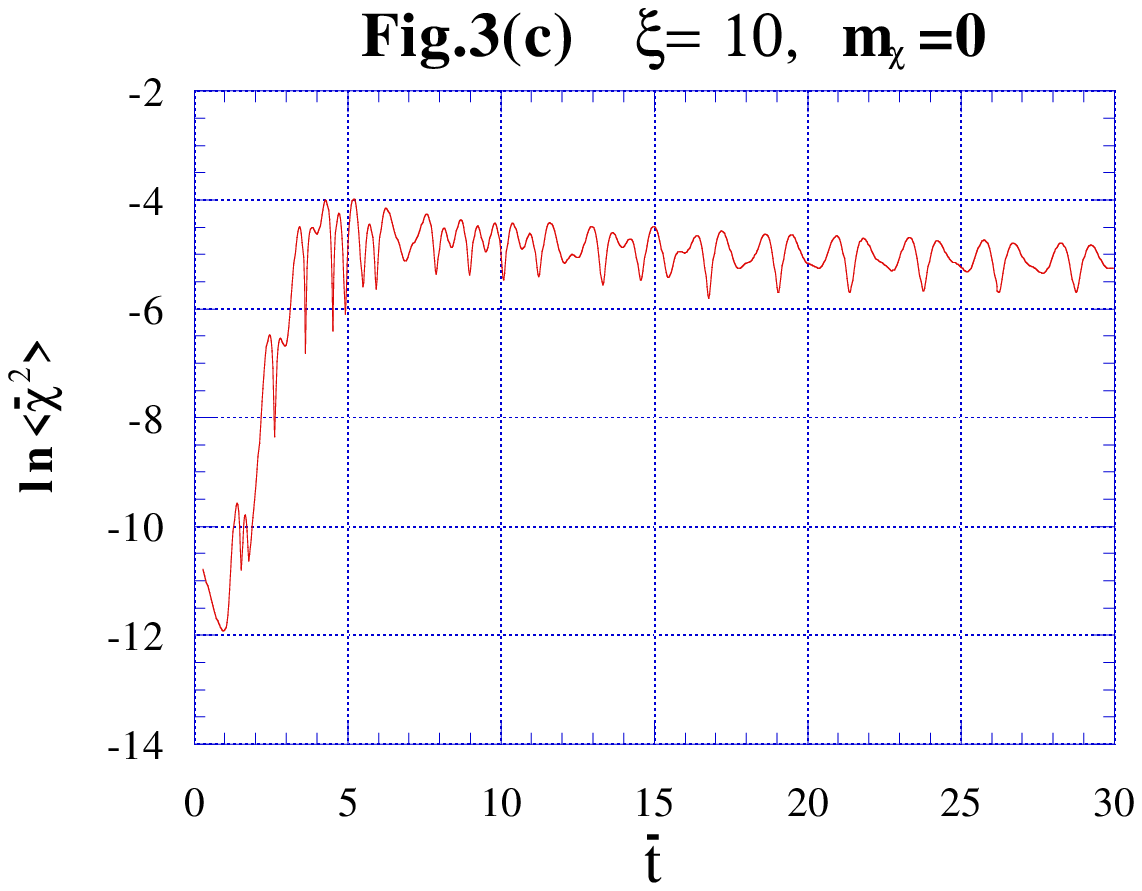}
\end{center}
\end{figure}

\begin{figure}
\begin{center}
\psbox[hscale=0.9,vscale=0.9]{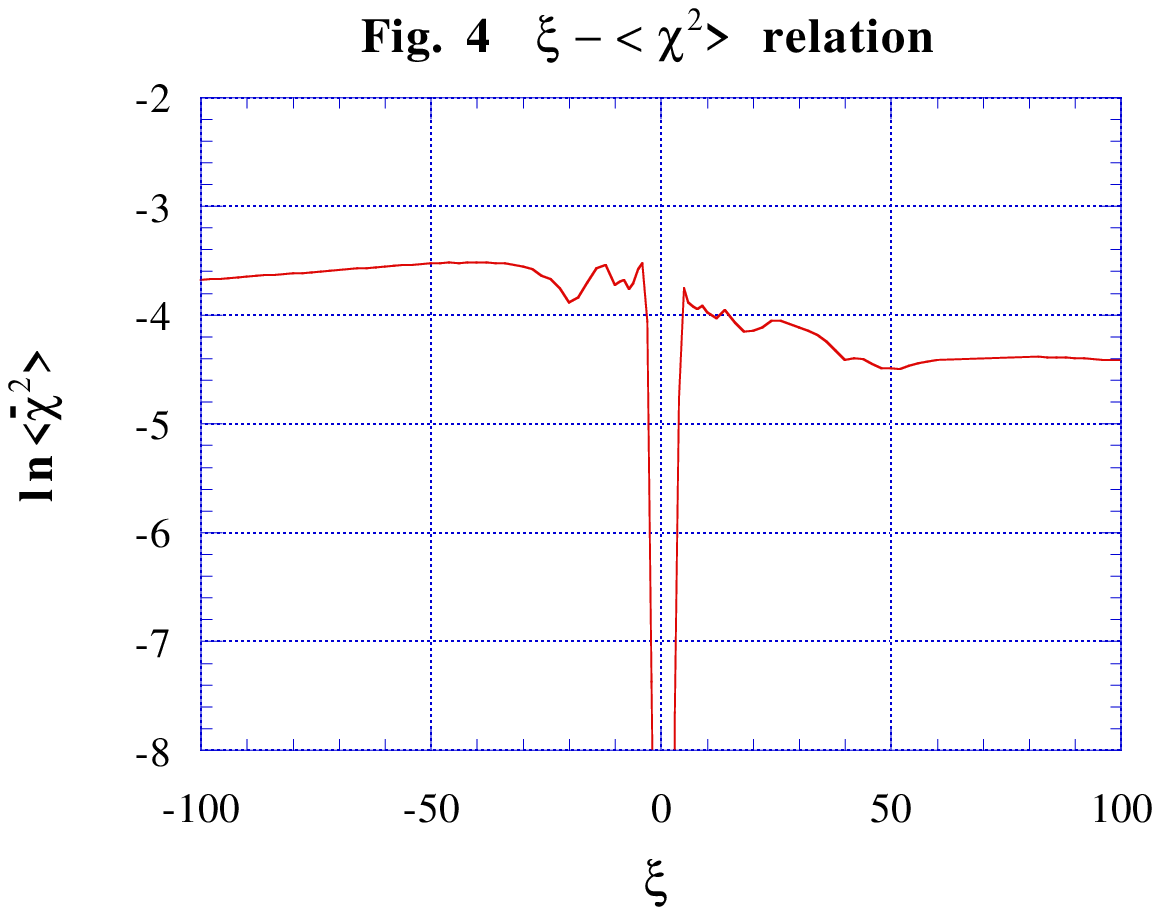}
\end{center}
\end{figure}

\newpage
\begin{figure}
\begin{center}
\psbox[hscale=0.6,vscale=0.6]{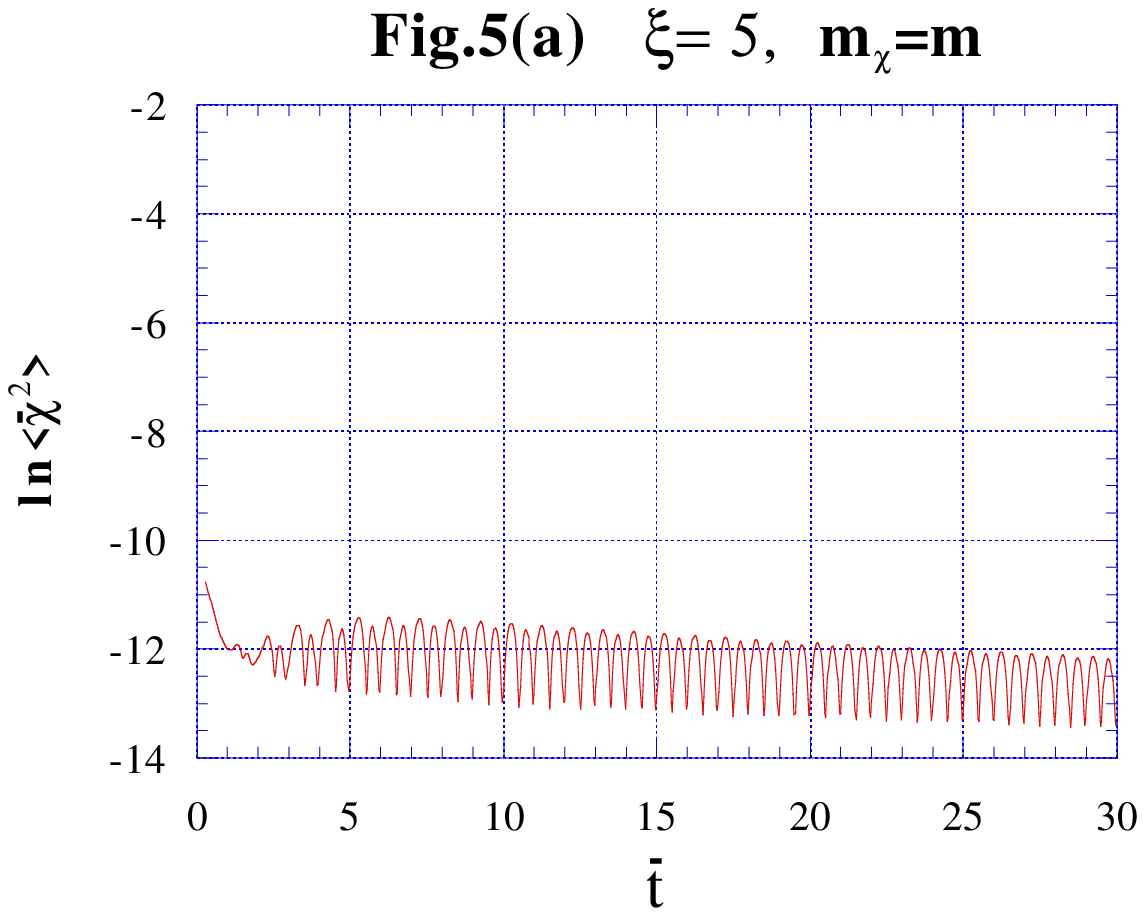}
\end{center}
\end{figure}

\begin{figure}
\begin{center}
\psbox[hscale=0.6,vscale=0.6]{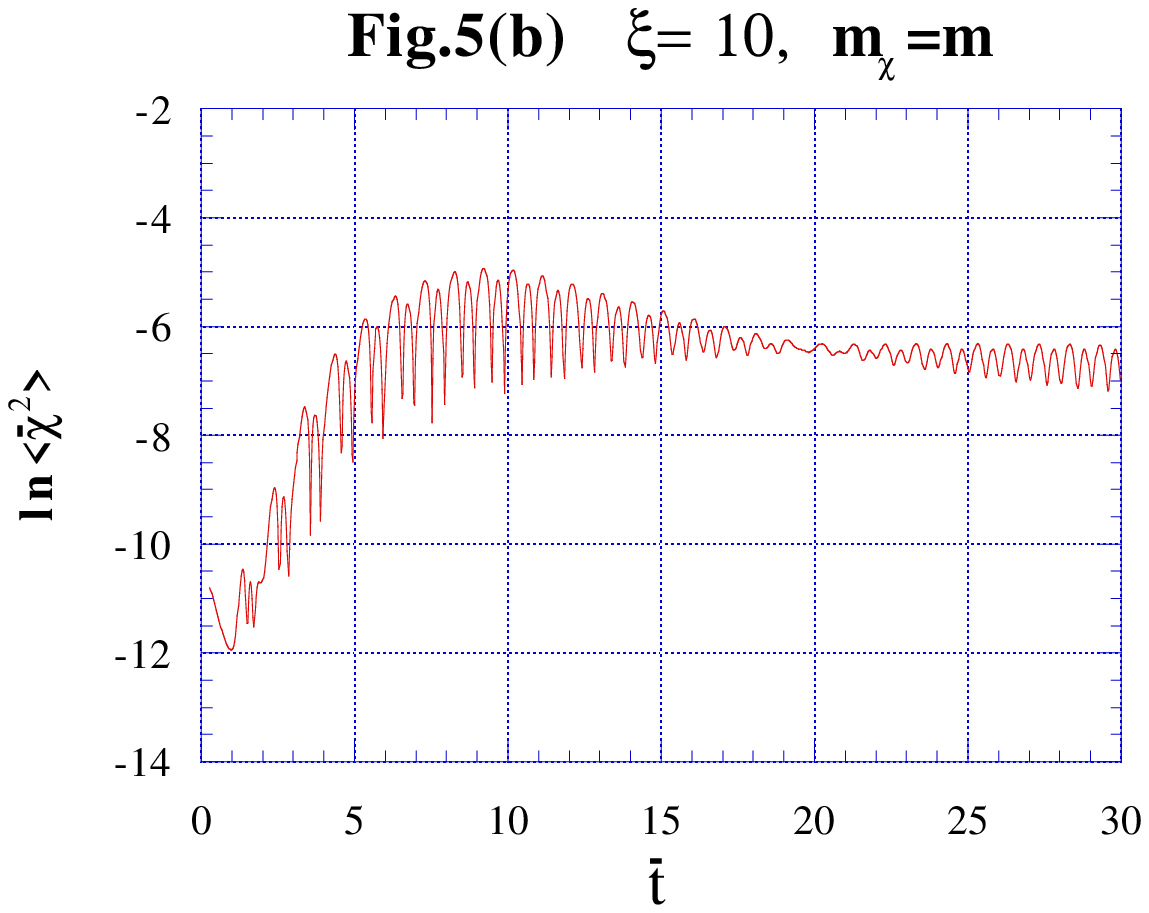}
\end{center}
\end{figure}

\begin{figure}
\begin{center}
\psbox[hscale=0.6,vscale=0.6]{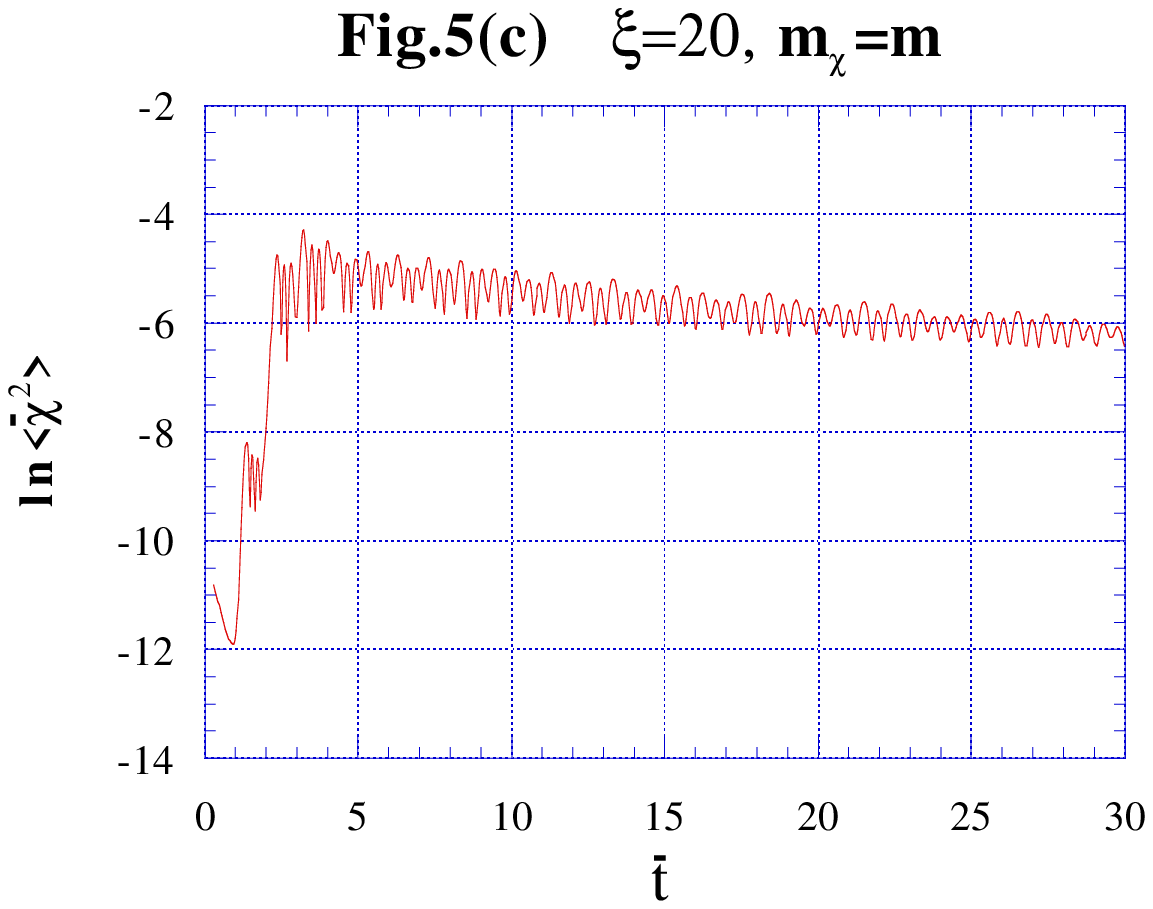}
\end{center}
\end{figure}

\begin{figure}
\begin{center}
\psbox[hscale=0.9,vscale=0.9]{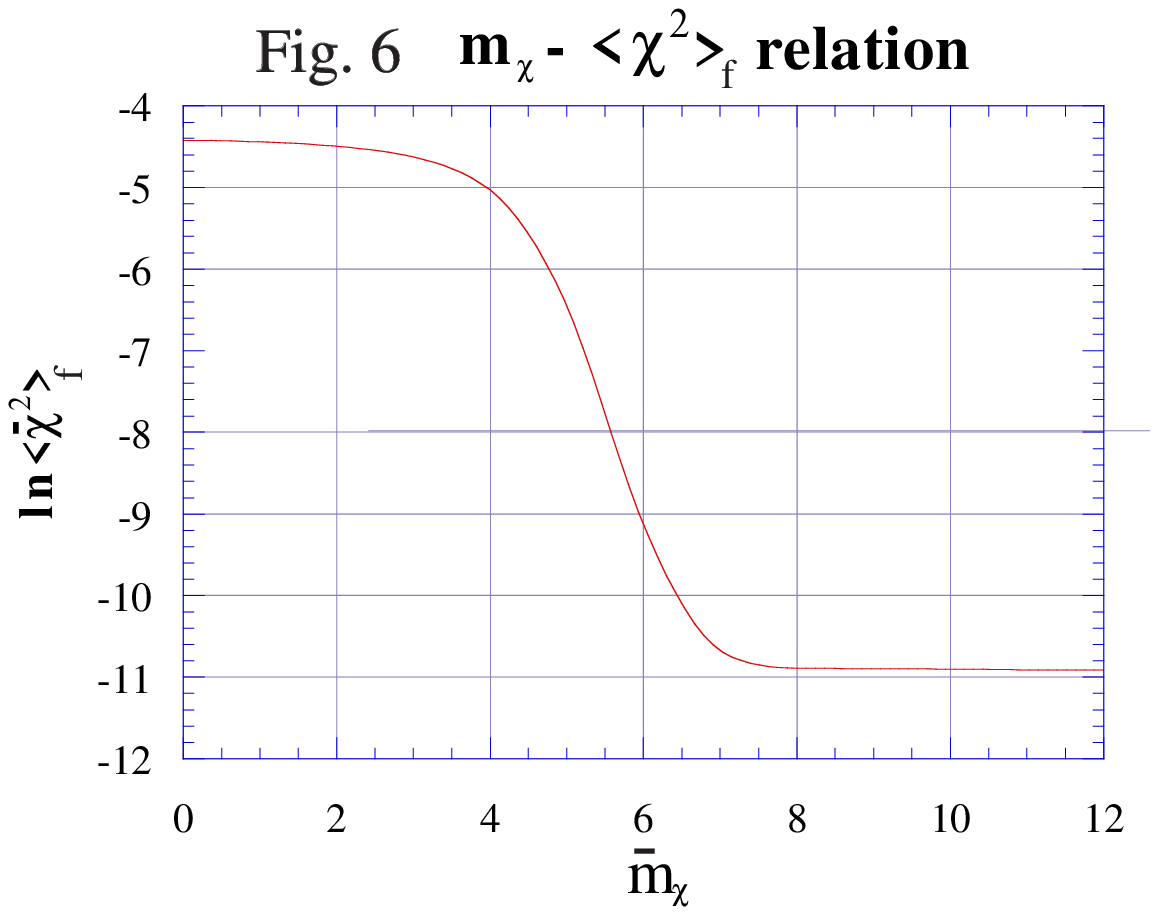}
\end{center}
\end{figure}

\newpage
\begin{figure}
\begin{center}
\psbox[hscale=0.7,vscale=0.7]{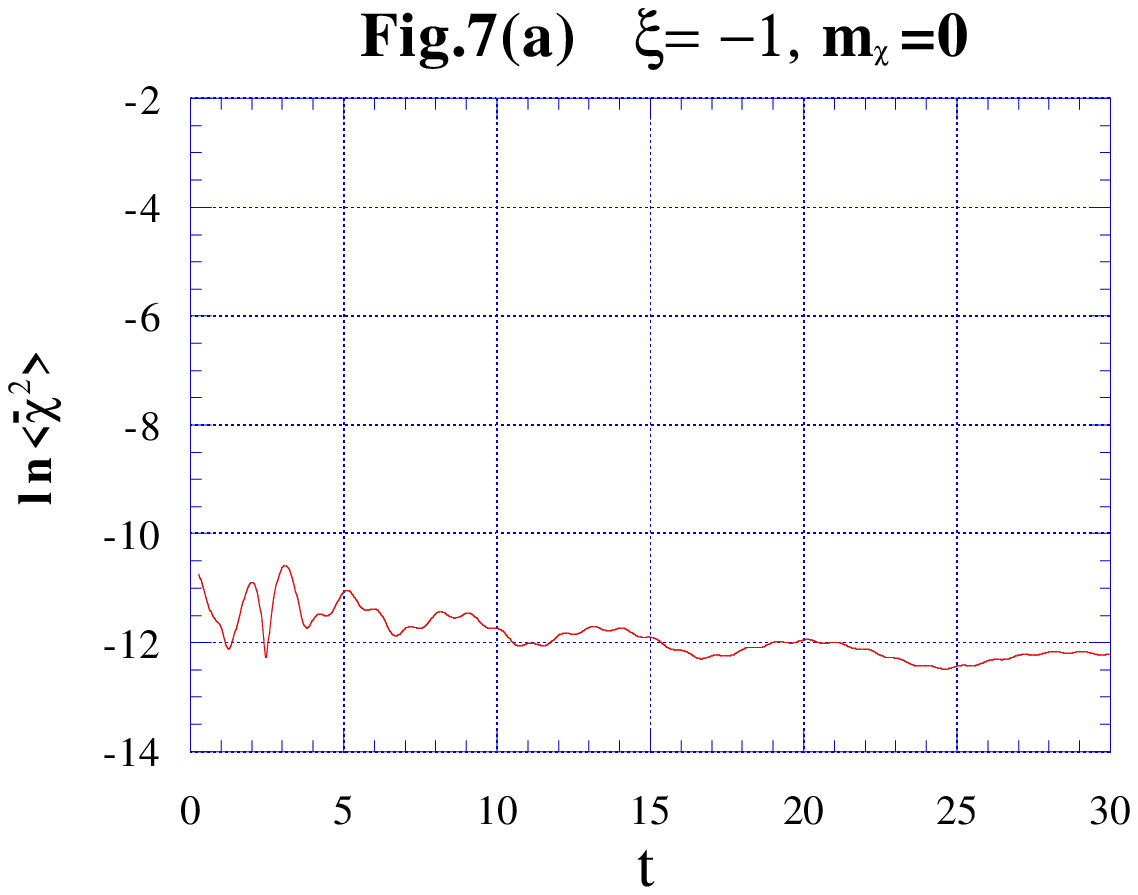}
\end{center}
\end{figure}

\begin{figure}
\begin{center}
\psbox[hscale=0.7,vscale=0.7]{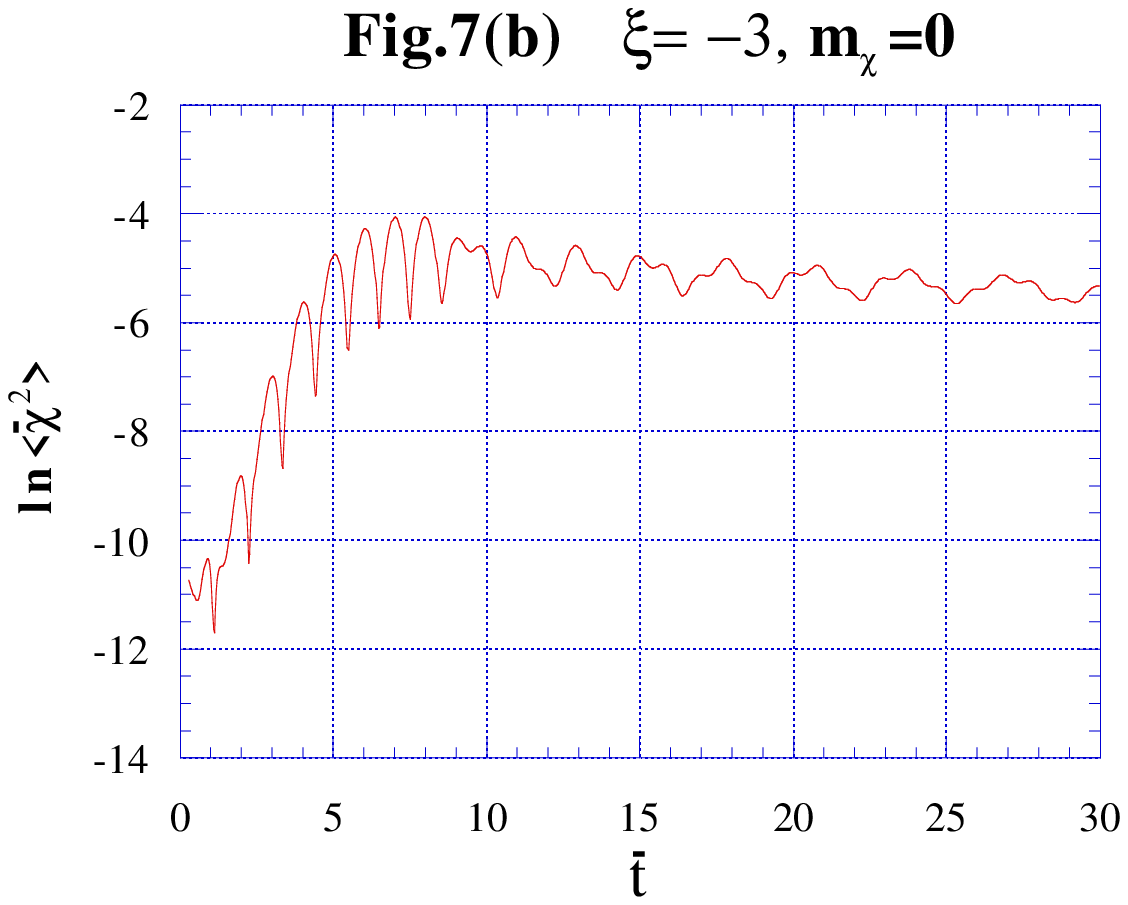}
\end{center}
\end{figure}

\begin{figure}
\begin{center}
\psbox[hscale=0.7,vscale=0.7]{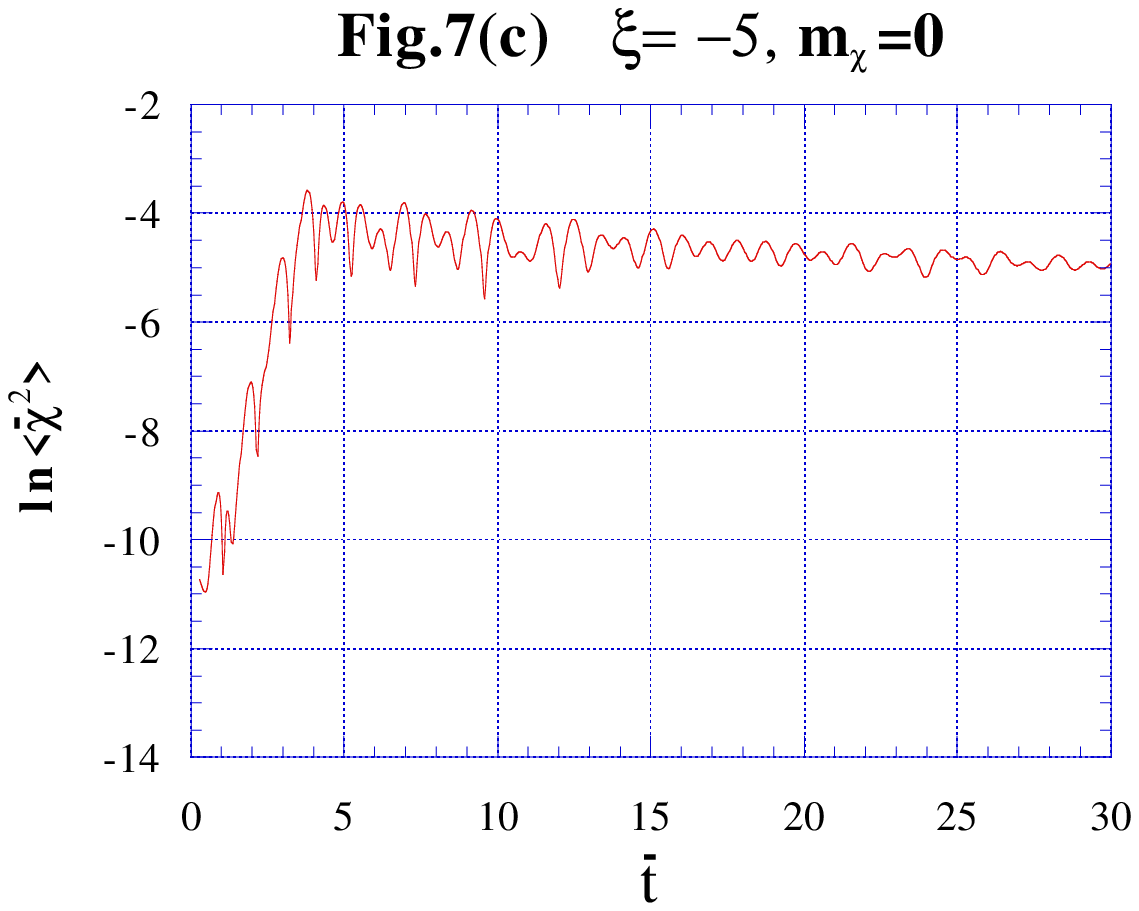}
\end{center}
\end{figure}

\begin{figure}
\begin{center}
\psbox[hscale=0.7,vscale=0.7]{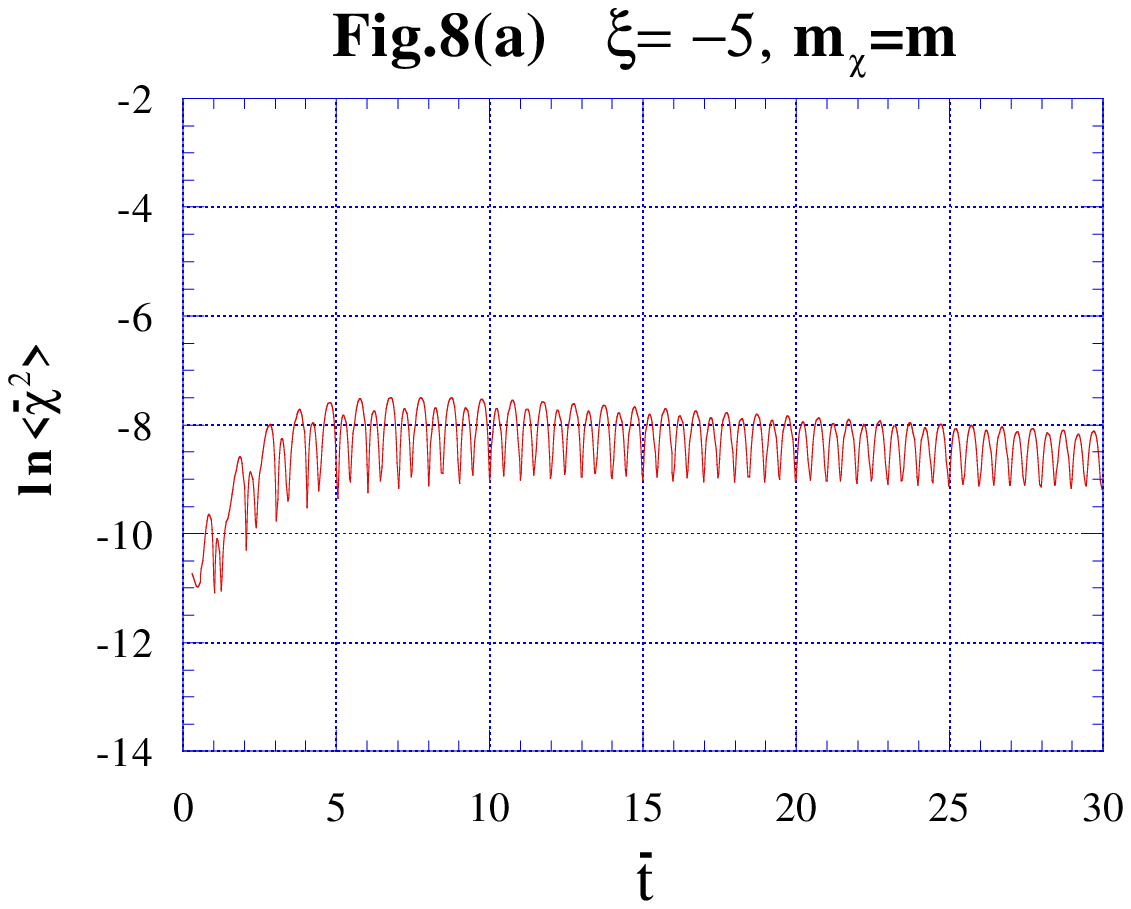}
\end{center}
\end{figure}

\begin{figure}
\begin{center}
\psbox[hscale=0.7,vscale=0.7]{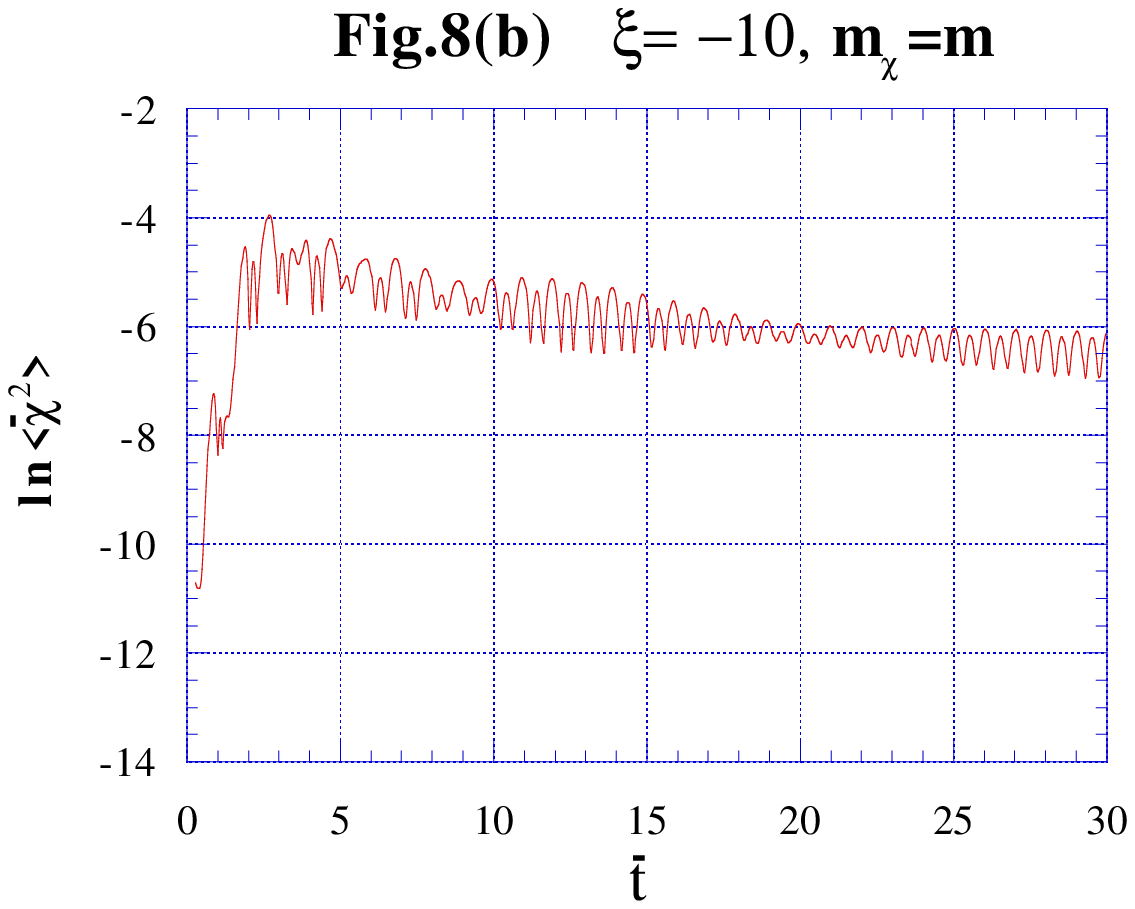}
\end{center}
\end{figure}

\newpage
\begin{figure}
\begin{center}
\psbox[hscale=0.7,vscale=0.7]{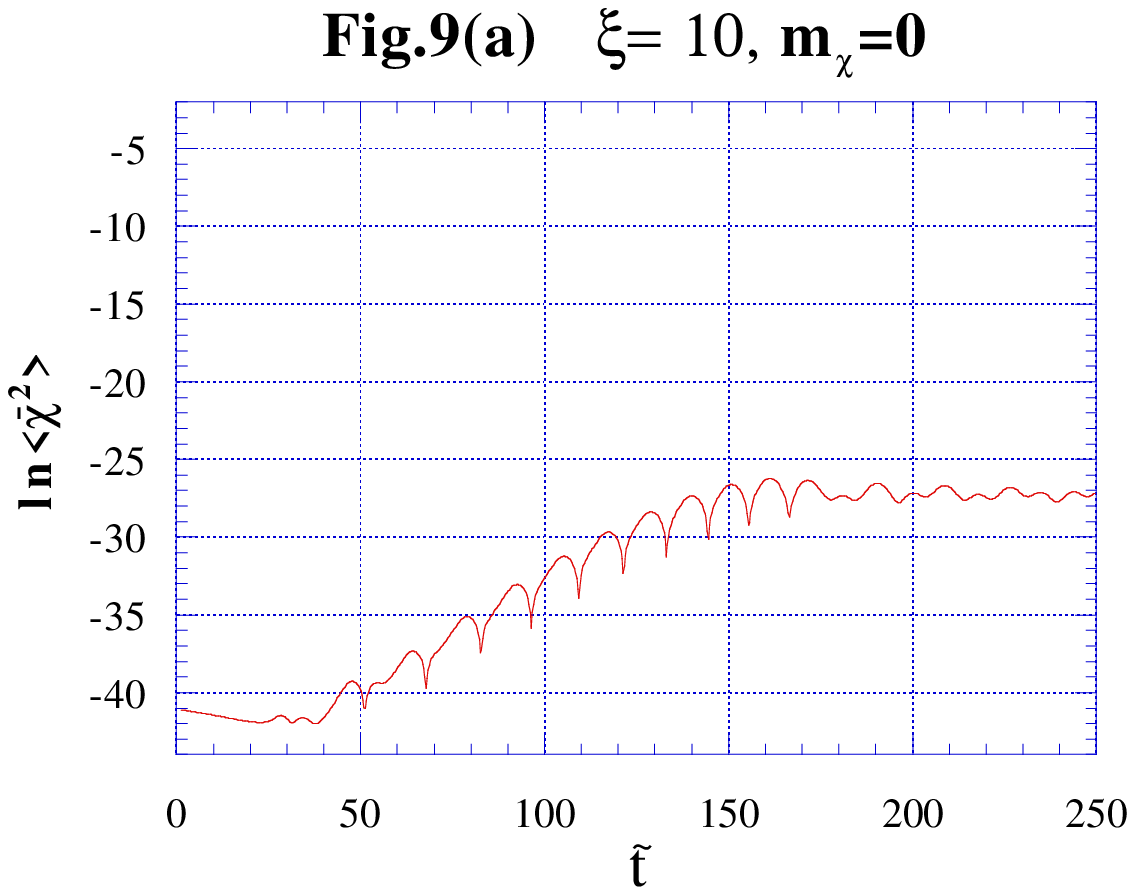}
\end{center}
\end{figure}

\begin{figure}
\begin{center}
\psbox[hscale=0.7,vscale=0.7]{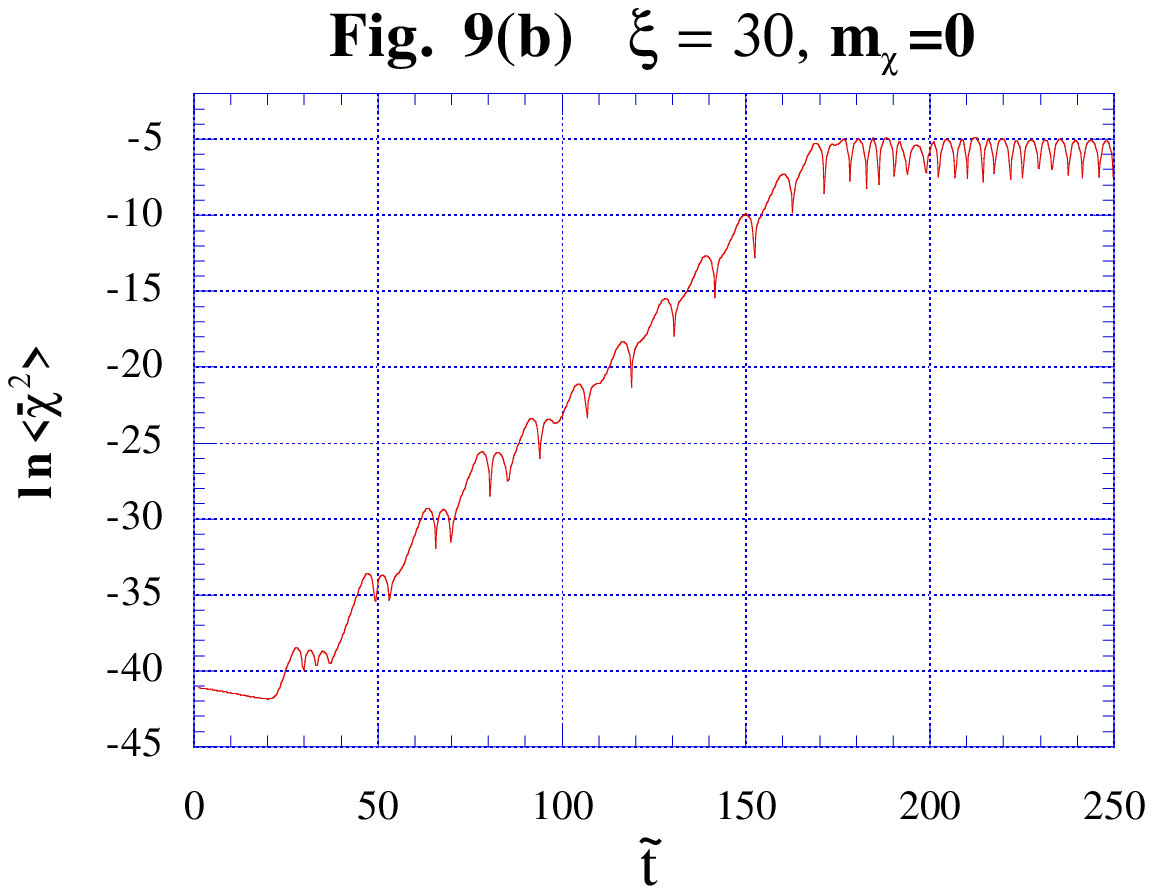}
\end{center}
\end{figure}

\begin{figure}
\begin{center}
\psbox[hscale=0.7,vscale=0.7]{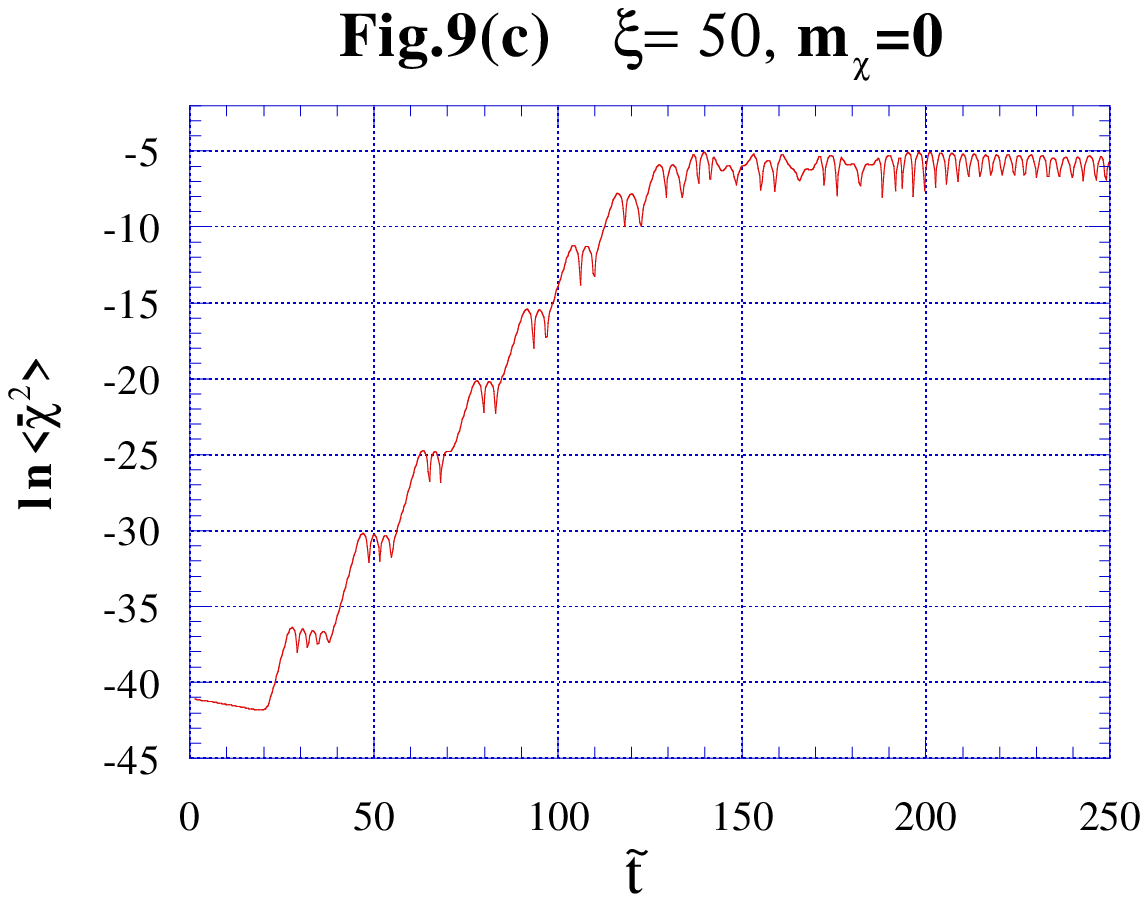}
\end{center}
\end{figure}

\begin{figure}
\begin{center}
\psbox[hscale=0.9,vscale=0.9]{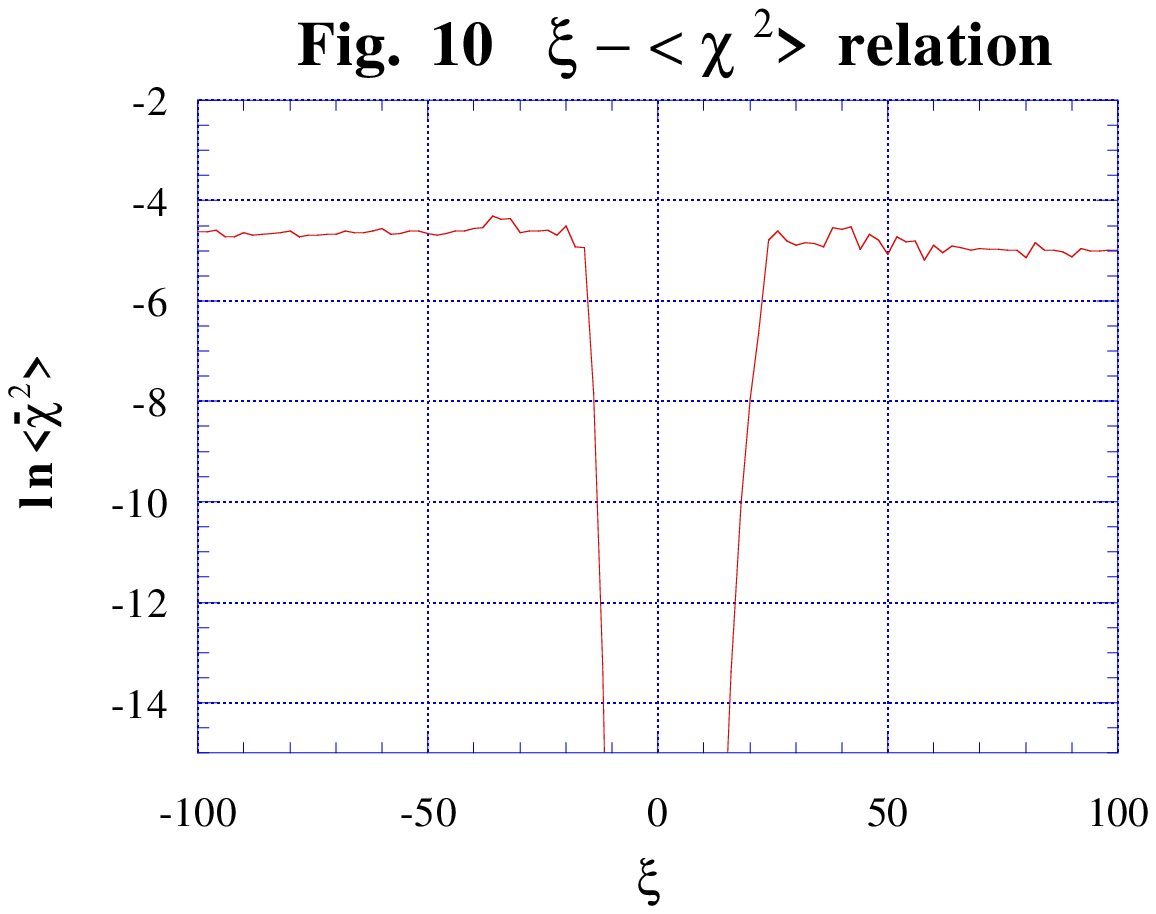}
\end{center}
\end{figure}

\newpage
\begin{figure}
\begin{center}
\psbox[hscale=0.7,vscale=0.7]{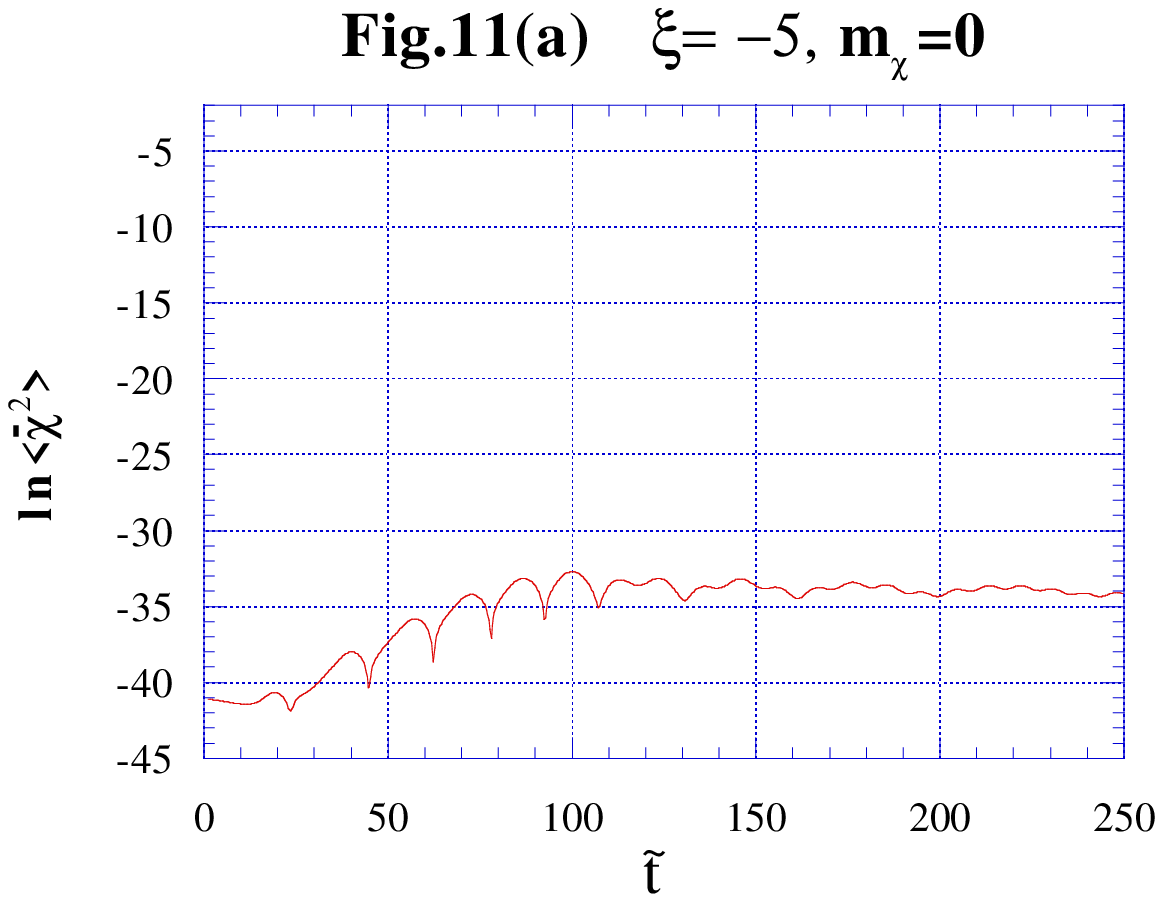}
\end{center}
\end{figure}

\begin{figure}
\begin{center}
\psbox[hscale=0.7,vscale=0.7]{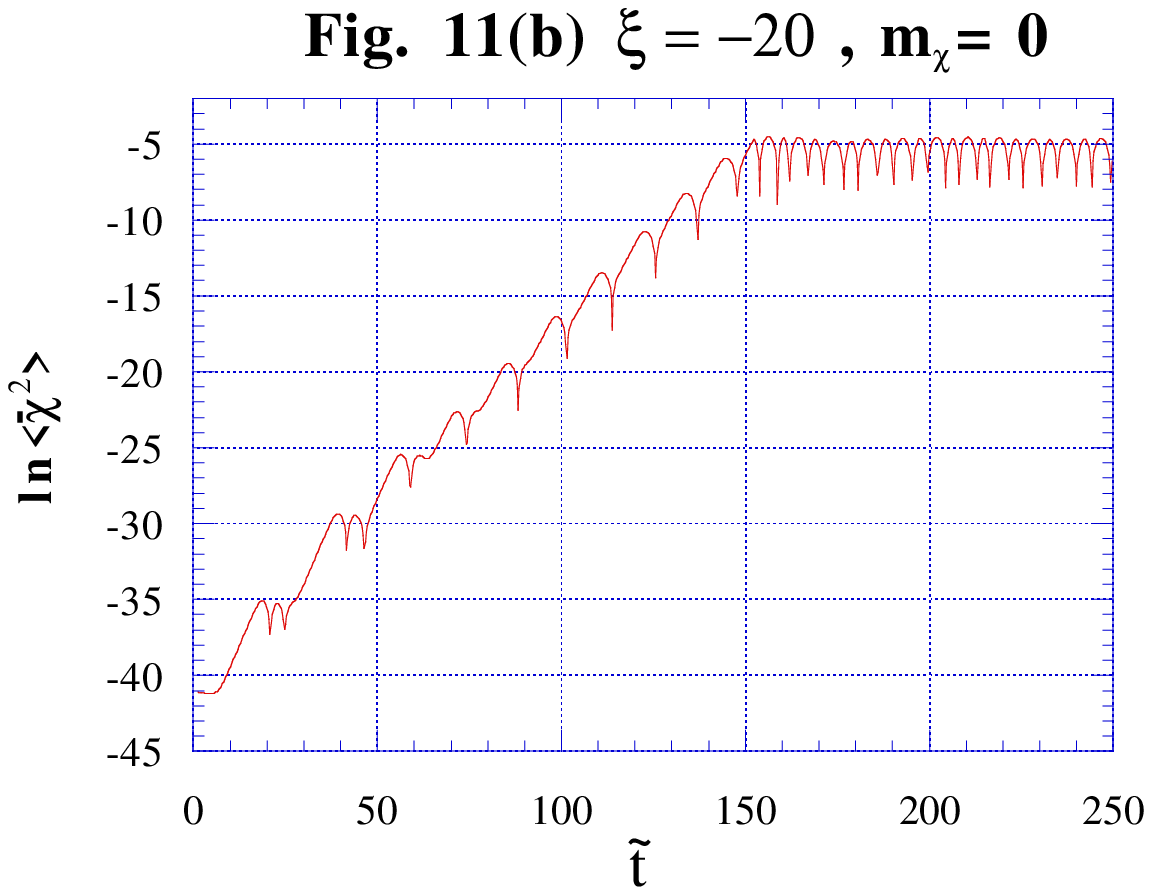}
\end{center}
\end{figure}

\begin{figure}
\begin{center}
\psbox[hscale=0.7,vscale=0.7]{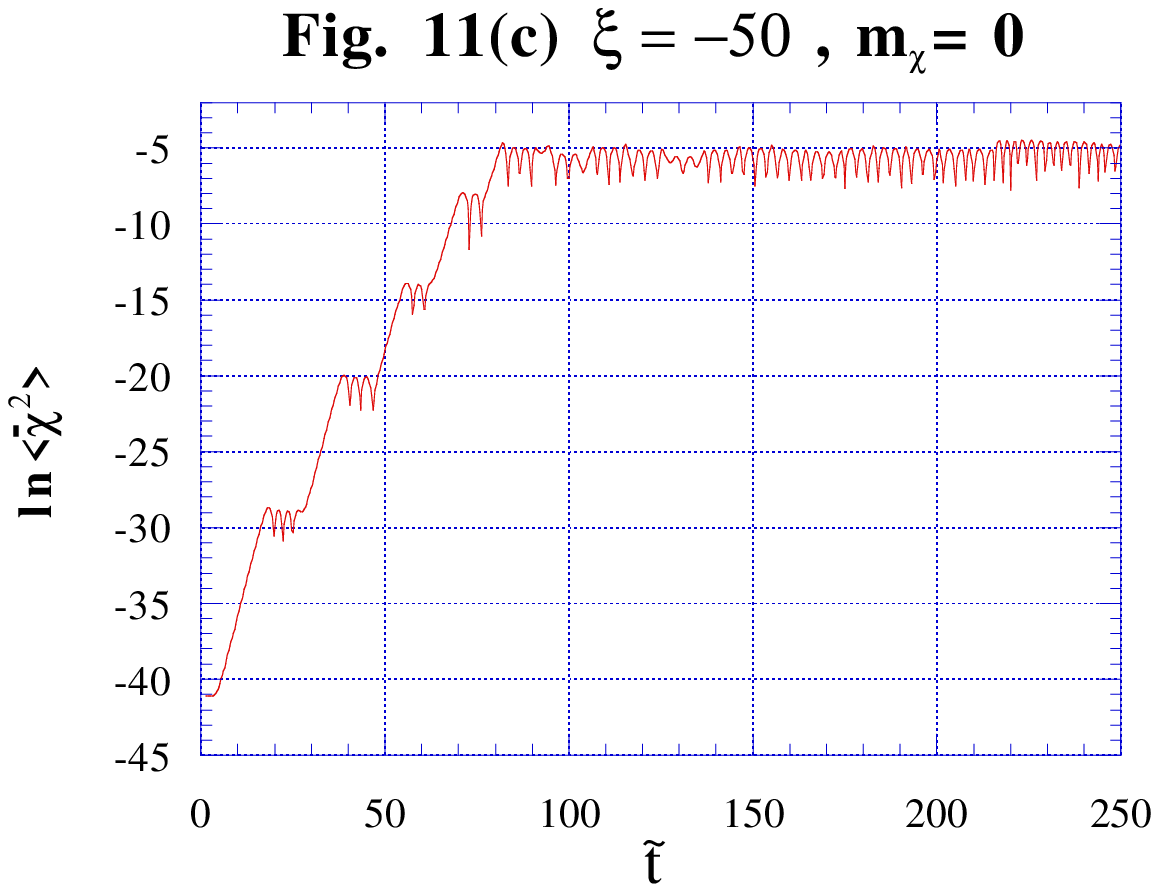}
\end{center}
\end{figure}

\begin{figure}
\begin{center}
\psbox[hscale=0.8,vscale=0.8]{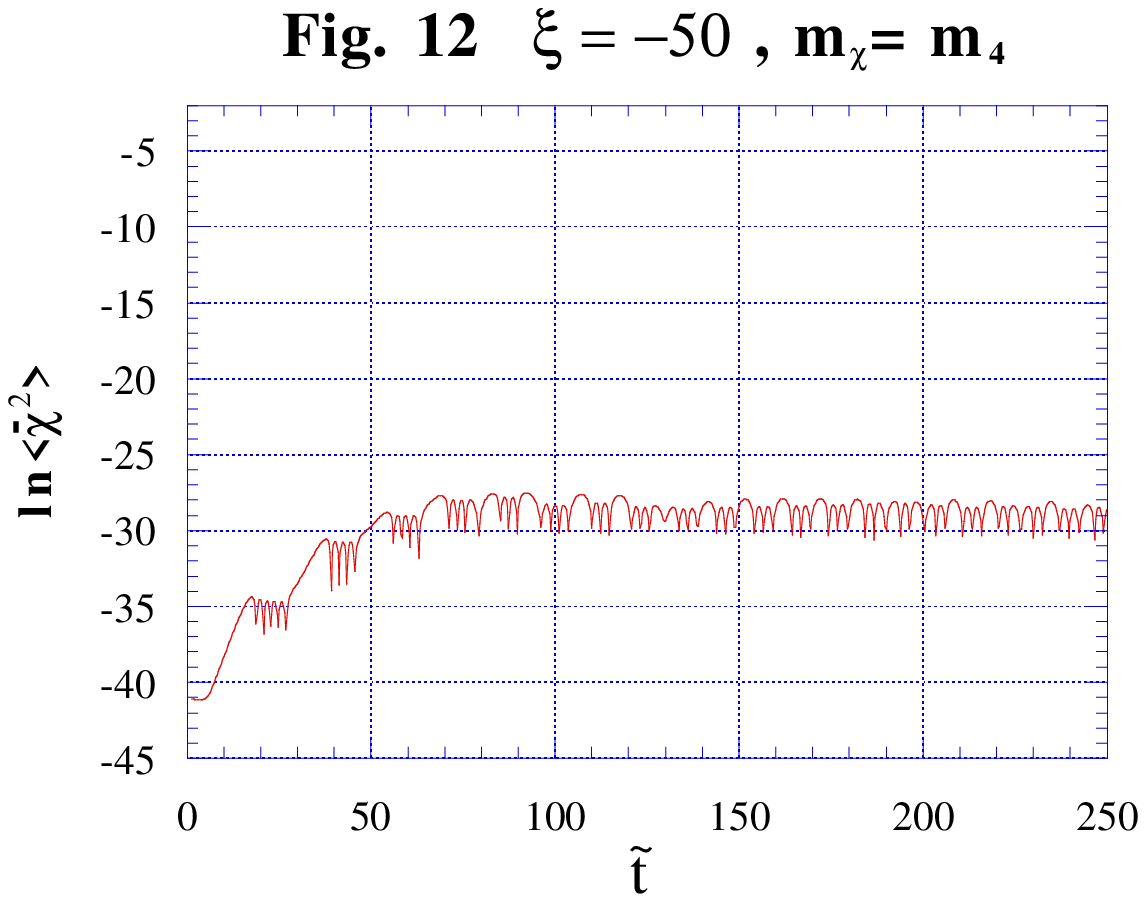}
\end{center}
\end{figure}


\begin{thebibliography}{99}
\bibitem{comment}
Our treatment in this paper is based on investigating
the scalar field dynamics in the Einstein frame
by making use of the mean field approximation.
In order to obtain the precise results, we should
perform the fully non-linear calculation in the original
frame.
\end{thebibliography}
\end{document}